\documentclass[psamsfonts]{amsart}
\usepackage{natbib}
\numberwithin{equation}{section}
\pdfoutput=1
\usepackage{graphics}
\usepackage[dvipdf]{graphicx}
\usepackage{epstopdf}
\usepackage{epsfig}
\usepackage{url}
 
 \usepackage{multirow}
 \usepackage{pstricks}
\usepackage{tikz}
\usetikzlibrary{snakes}
\usetikzlibrary{shapes,snakes}

\begin{document}

\title{Prediction of alpine glacier sliding instabilities: a new hope}

 \author{J\'erome Faillettaz}
 \address{Laboratory of Hydraulics, Hydrology and Glaciology (VAW),
ETH Zurich, CH-8092 Z\"urich, Switzerland.}
\email{faillettaz@vaw.baug.ethz.ch}
 \author{Martin Funk}
  \address{Laboratory of Hydraulics, Hydrology and Glaciology (VAW),
ETH Zurich, CH-8092 Z\"urich, Switzerland.}
 \author{Didier Sornette}
 \address{Department of Management, Technology and Economics, ETH Z\"urich, CH-8092 Z\"urich, Switzerland.}
\address{Department of Earth Sciences, ETH Z\"urich, CH-8092 Z\"urich, Switzerland.}

\date{}
%
%
%


\begin{abstract}
Mechanical and sliding instabilities are the two processes which may lead to breaking off events of large ice masses. Mechanical instabilities mainly affect unbalanced cold hanging glaciers. For the latter case, a prediction could be achieved based on data of surface velocities and seismic activity. The case of sliding instabilities is more problematic. This phenomenon occurs on temperate glacier tongues. Such instabilities are strongly affected by the subglacial hydrology: melt water may cause (i) a lubrication of the bed and (ii) a decrease of the effective pressure and consequently a decrease of basal friction. Available data from Allalingletscher (Valais) indicate that the glacier tongue experienced an active phase during 2-3 weeks with enhanced basal motion in late summer in most years.
In order to scrutinize in more detail the processes governing the sliding instabilities, a numerical model developed to investigate gravitational instabilities in heterogeneous media was applied to Allalingletscher. This model enables to account for various geometric configurations, interaction between sliding and tension cracking and water flow at the bedrock.
We could show that both a critical geometrical configuration of the glacier tongue and the existence of a distributed drainage network were the main causes of this catastrophic break-off. Moreover, the analysis of the modeling results diagnose the
phenomenon of recoupling of the glacier to its bed as a potential new precursory sign announcing the final break-off. This model casts a gleam of hope for a better understanding of the ultimate rupture of such glacier sliding instabilities.
\end{abstract}

\maketitle
\section{Introduction}

Gravity-driven instabilities include landslides, mountain collapse, rockfalls, ice mass break-off and snow avalanches. They pose a considerable risk to mountain communities, real-estate development, tourist activities and hydropower energy generation. Gravity-driven instabilities are the most widespread natural hazard on Earth.
In the US and Europe, they are particularly significant and cause billions of dollars and euros damage each year.
Prediction of such gravity-driven instabilities could be used by policy makers to decide on possible evacuation of the potential dangerous area. Unfortunately, accurate prediction of such phenomena remains a challenge.

In this context, studying glacier break-off is of particular interest because a glacier consists of a unique natural material (ice) where the interface between ice and bedrock is well defined. 
This relative simplicity of the system allows to particularly focus our study on the rupture processes leading to the initiation of the instability.
Two types of glacier instabilities leading to icefalls may be distinguished: (i) mechanical instabilities and (ii) sliding instabilities. 
Mechanical instabilities mainly affect unbalanced cold hanging glaciers (i.e. snow accumulation is mostly compensated with break-off, \citet{Pralong&Funk2006}).
Based on field data combining surface displacement measurements and seismic activity before break-off, \citet{Faillettaz&al2008,Faillettaz&al2011a} showed that a prediction of the event was possible.
The case of sliding instabilities is more problematic. They occur on temperate glaciers (i.e. glaciers that can slide on their beds) where subglacial hydrology plays a major role. Melt water flowing on the glacier bed may influence the glacier dynamics in two ways: it allows (i) a better lubrication of the bed and (ii), if water becomes pressurized, an uplift decouples the glacier from its bed.
In the Alps, only two glaciers are known to have given rise to such sliding instabilities in the 20th century: The glacier du Tour (Mont Blanc, France) in 1949 and the Allalingletscher (Valais, Switzerland) in 1965. 
Up to now, such glacier instabilities are still unpredictable, and the causes of such catastrophic break-off remain unclear.

The present paper is devoted to the study of such instabilities, taking the Allalingletscher as an example. This glacier is indeed of particular interest because it experienced 2 catastrophic break-offs (in 1965 and 2000) and also because a unique set of data was collected since the first catastrophic break-off in 1965.
To address the open questions on the initiation of the instability, we apply a general numerical model developped to investigate gravity-driven instabilities in heterogeneous media.
This model was already successfully applied to a polythermal glacier becoming partly temperate at its bedrock,
i.e. the Altelsgletscher \citep{Faillettaz&al2011b}. Numerical results showed that the final instability is driven by a progressive cold-temperate transition at the ice/bed interface.
The present study significantly differs from the previous case as here the whole glacier is temperate and can slide on its bedrock. 
As subglacial water was shown to drive the sliding instabilities, the initial model was extended to account for subglacial water flow at the bedrock.

After describing the Allalingletscher and analysing available measurements, we apply a general numerical model to this particular gravity-driven instability to assess the causes of the rupture. 
Numerical results are discussed and general criterions leading to the instability are presented.

\section{Allalingletscher}
\subsection{Allalingletscher through the centuries}
Allalingletscher is referred to as early as 1300 in written documents, in connection with the right of pasturage. The glacier is mentioned again and again, through the centuries, mainly in reference to floods caused during stages of glacier advance, and also when it interrupted the path leading to alpine meadows and pass routes to Italy. The hydrological bassin of Allalin (12 $ \rm km^2$, 81 $\%$ glacierized), located near the head of the valley of Saas, consists of a hanging valley, from which the lower part of the glacier tongue extends and flows down the sidewall of the main valley. During the little age, the terminus reached and blocked the river Saaser Vispa leading to the formation of the lake of Mattmark.

\begin{figure}

\begin{tabular}{cc}
\begin{minipage}{10pc}
\noindent{\includegraphics[width=10pc]{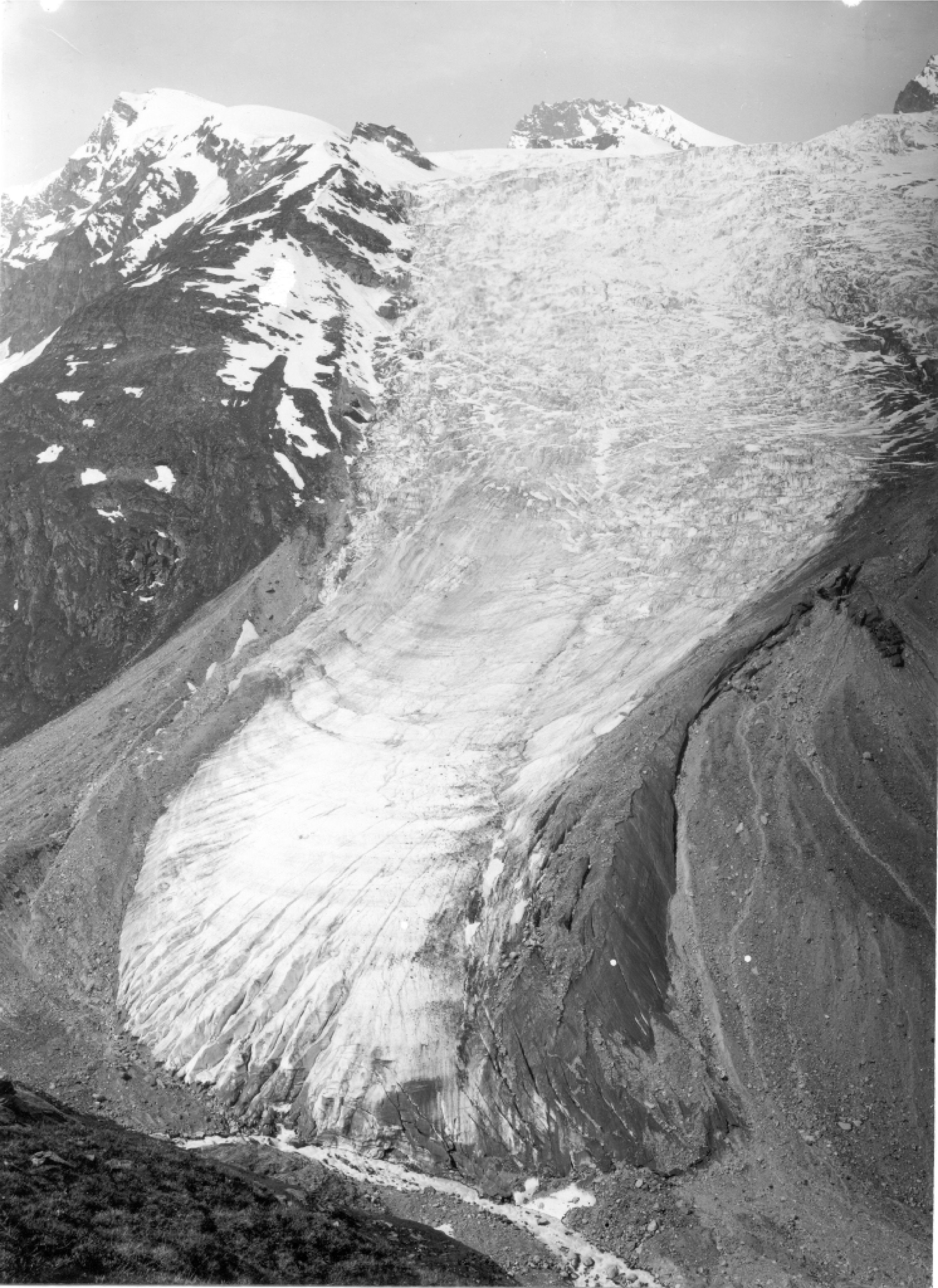}}
(a)
\end{minipage}
&
\begin{minipage}{10pc}
\noindent{\includegraphics[width=9pc]{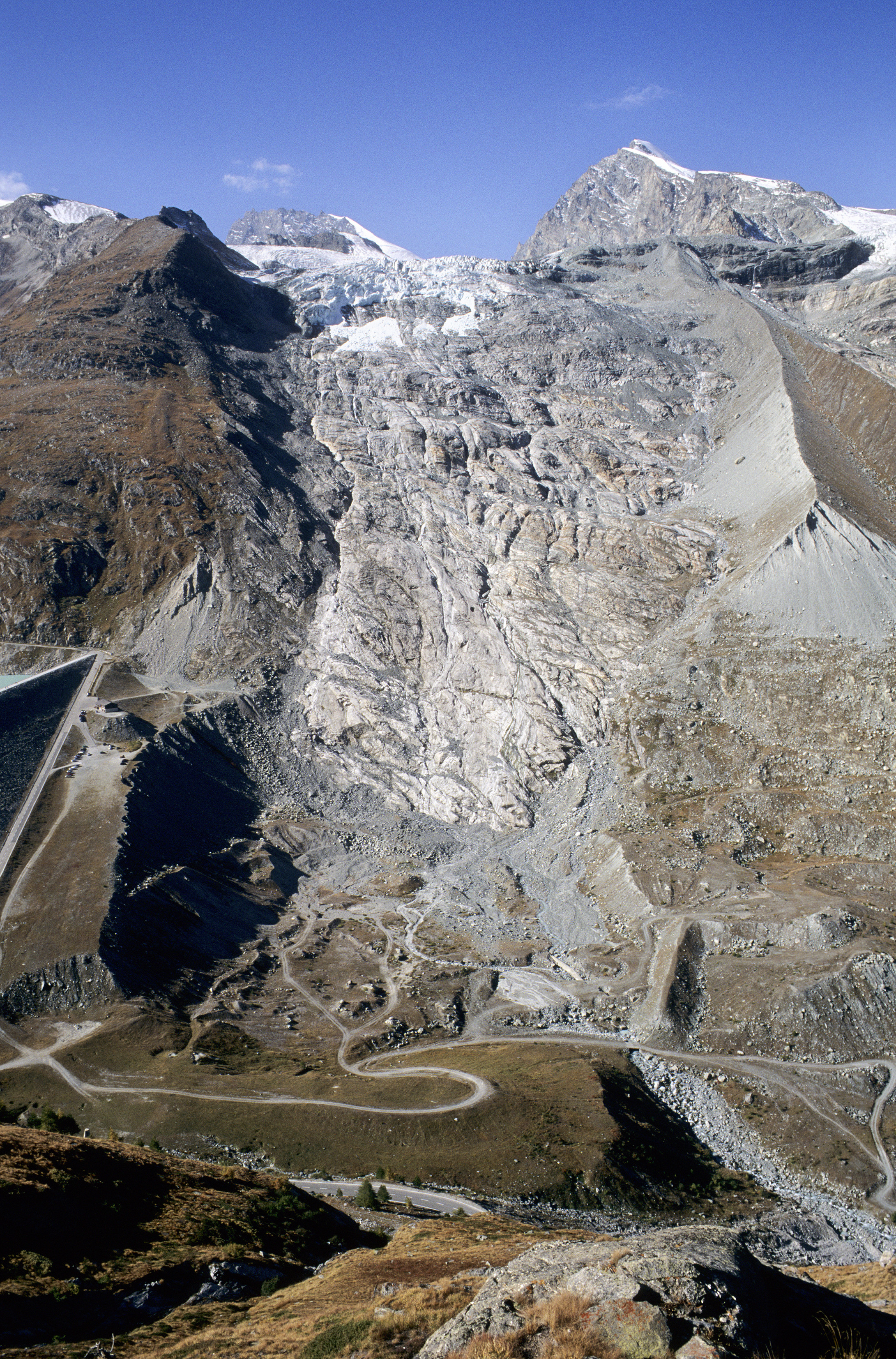}}
(b)
\end{minipage}
\end{tabular}
\caption{Allalingletscher in 1916 (a) and in 2004 (b)}
\label{general situation}
\end{figure}

\begin{figure}

\noindent\includegraphics[width=20pc]{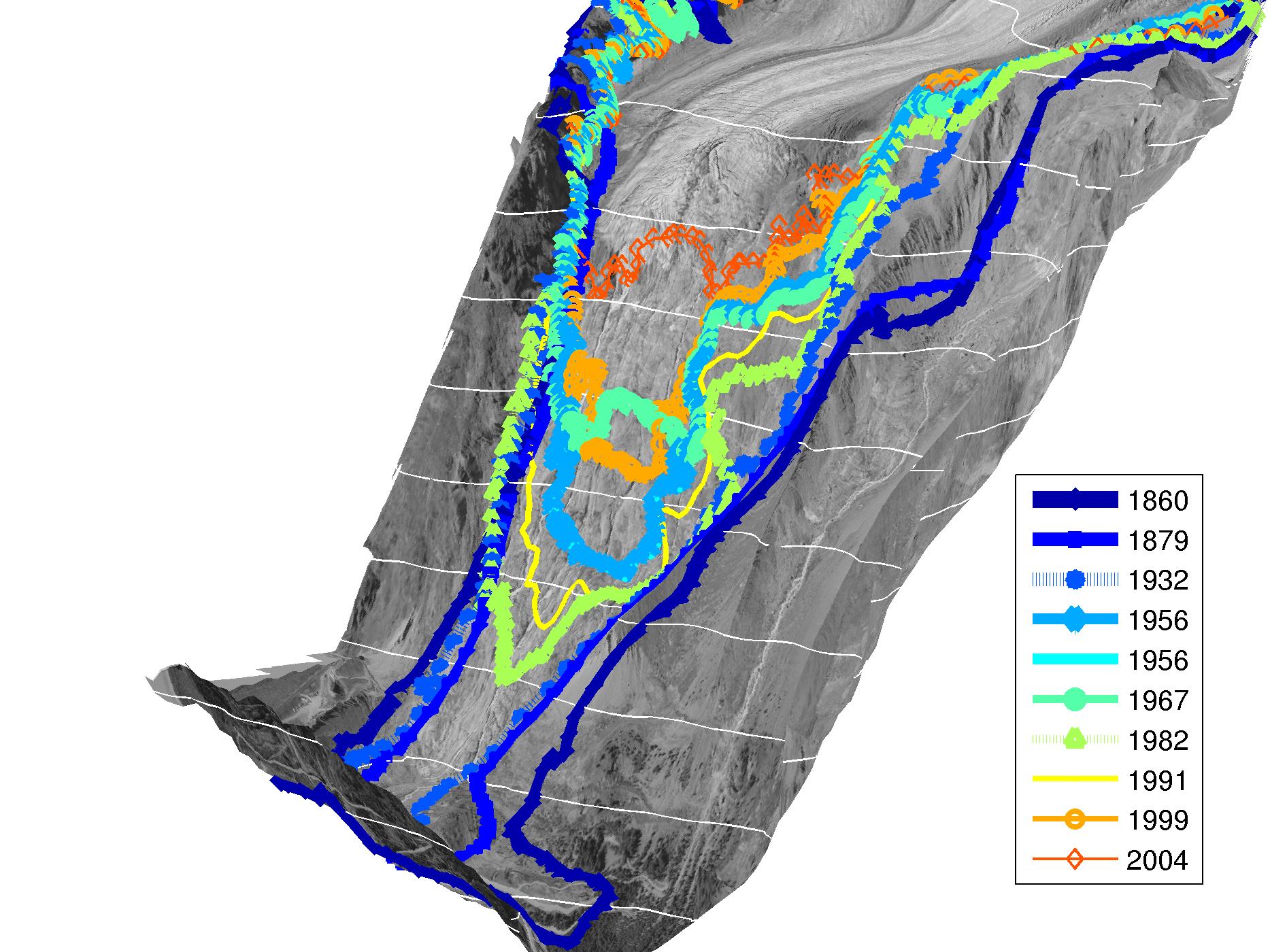}
\caption{Digital Elevation Model and different extensions of the glacier tongue since 1860.}
\label{geometric_evolution}
\end{figure}

The extend of the Allalingletscher has been inferred by L\"utschg (see reference in \citet{Rothlisberger1978}) from references to the size of the lake and to the floods which originated in the outburst of the ice-dammed lake. Combining such information with direct references to the size of the glacier, he assembled a detailed chronicle of the glacier through the centuries, listing large extension of the glacier in 1300 and 1589, and on other occasions during the 17th, 18th and 19th centuries. Maximum known extends occurred in 1822/1823  and 1858. The $\rm 20^{th}$ century maximum in 1923 was smaller, although the terminus still crossed the main river. After that, an almost uninterrupted retreat lasting 31 years from 1923 (Fig. \ref{general situation}) until 1954 occurred. According to the records published in \citet{GlaciologicalReports}, the retreat amounted to about 230 m from 1923 to 1943, 660m from 1943 to 1949, and 90m from 1949 to 1954. The very rapid retreat in the forties was due to the decay of the thinned-down snout on the east-facing rock slope of the lower part of the Saas valley sidewall. Between 1954 and 1964 the terminus has been approximately stationary (Fig. \ref{geometric_evolution}).

The flat area upstream the south lateral morain was used for the construction of an artificial reservoir. Between 1958 and 1967 an earth dam 120m high with a volume of $\rm 10^7 m^3$ was erected immediately upstream the south of the moraine. A major part of the material used in the construction of the dam was obtained from the lateral moraines, while the flat area between the moraines was used for temporary installations, primary in connection with the processing of the earth-fill materials.

\subsection{The ice avalanche of 30 August 1965}
\begin{figure}

\noindent\includegraphics[width=9.5pc]{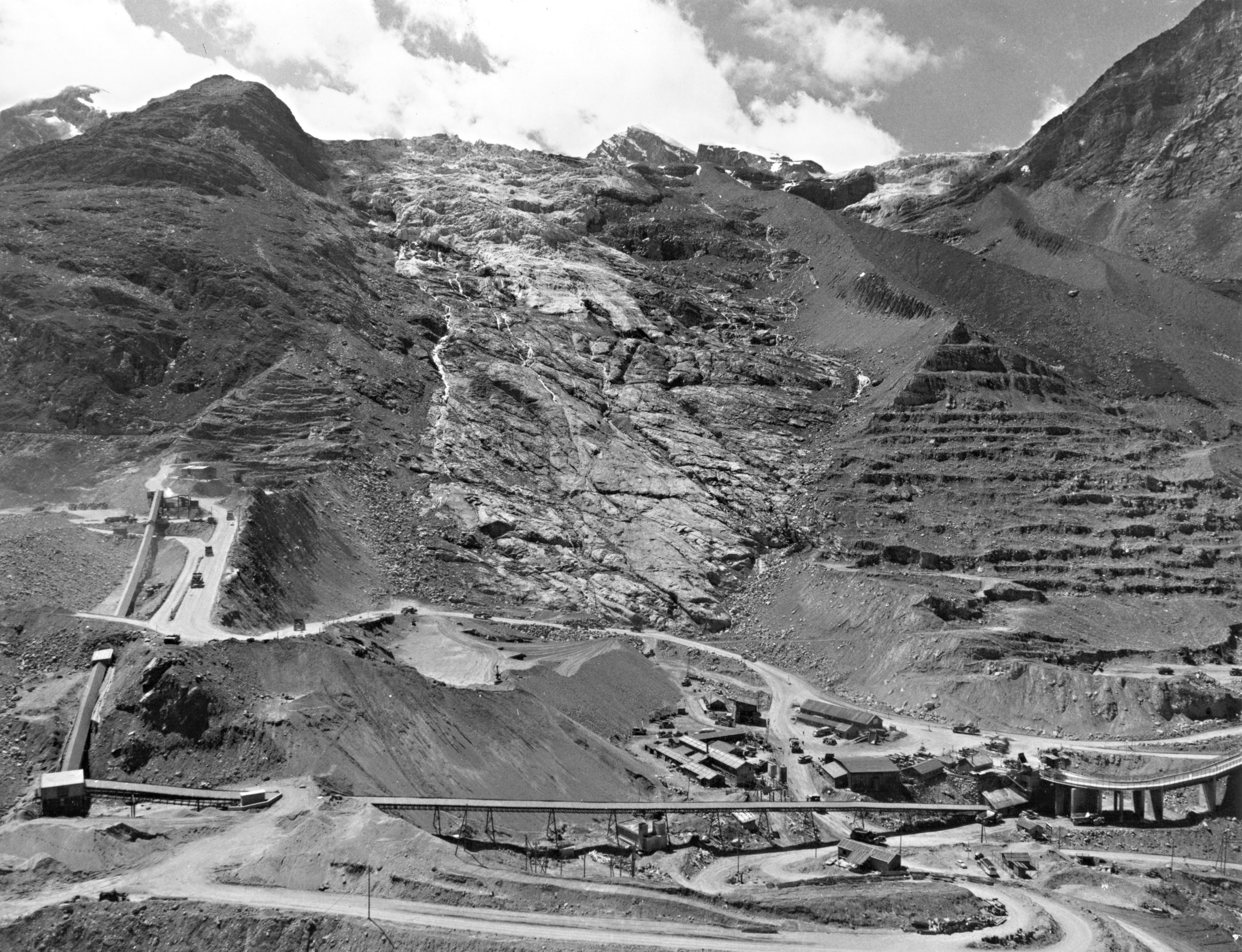}~
\noindent\includegraphics[width=9.5pc]{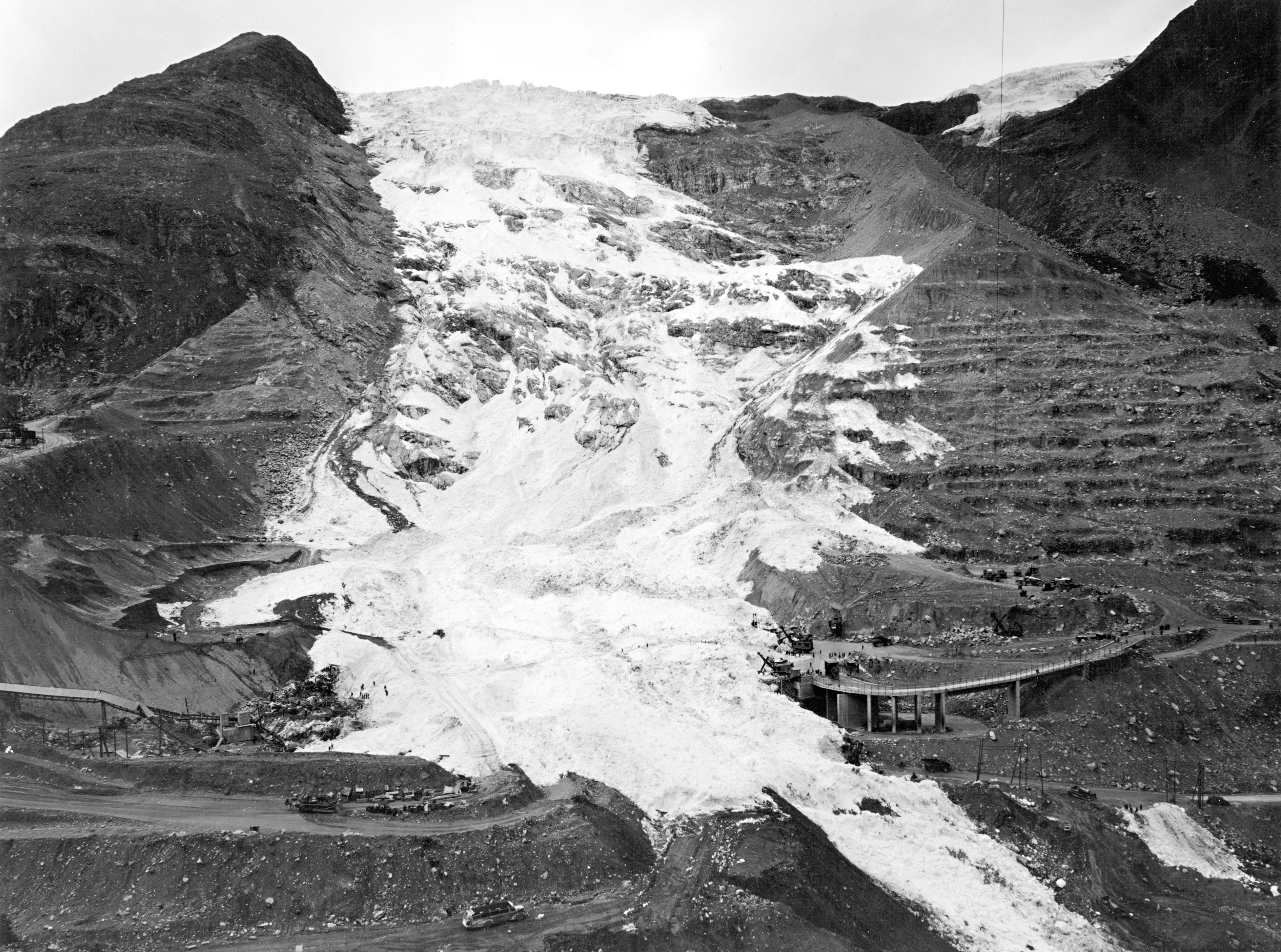}

\caption{The Mattmark catastrophe: 30 August 1965
Breaking off of 2 mio m$^3$ of ice}
\label{catastrophe}
\end{figure}

The Allalin ice avalanche, a landslide-type event, is also referenced to as the Mattmark glacier catastrophe. Approximately~$\rm 2.10^6 m^3$ of ice broke off at the terminus of Allalingletscher, moved down a rock slope of some $\rm 27^o$ over the vertical distance of 400 m and continued for a further 400 m across the flat bottom of the valley, claiming 88 victims at the Mattmark construction site. An overall view of the area shortly after the avalanche is given in Fig. \ref{catastrophe}.

Glaciological investigations showed that the ice avalanche has occurred during a phase of enhanced basal motion as a result of intensive bed-slip of an even larger mass than the one that broke off on August $\rm 30^{th}$ \citep{Rothlisberger1978}. 
Although it seemed that this particular condition was necessary for the lower part of the ice mass to slide off, it does not explain why the ice avalanche occurred in 1965 and not during other active phases either before or after this date. A certain topography of the bed, combined with an unfortunate mass distribution, is believed to have played a major role in the catastrophe.

In terms of glacier variations, the avalanche results in a sudden retreat of the terminus by 400 m in plan view and 220 m in elevation. The avalanche deposit was not counted as part of the glacier in this context, although it took more than a year for the last remnants of the debris to melt completely. The new terminus consisted of a 40 m high ice cliff in the form of a concave crescent-shape arch.

\subsection{Glaciological investigations after the ice avalanche (1965-1978) after \citet{Rothlisberger1978}}

An intensive glaciological study was undertaken immediately after the catastrophe in order to safeguard the rescue operation, and also the construction work when it was resumed later on. Since the termination of the construction of the dam in 1967 a reduced program of measurements has been carried out.

One of the methods of investigation consisted of aerial surveys repeated at short intervals of duration between one to two weeks. The first flight was carried out on 4 September, i.e. 5 days after the fall of the avalanche. 

A detailed monitoring of surface velocities was carried out by means of a theodolite from the rock ridge of Schwarzbergkopf at the south side of the glacier and stakes drilled in the ice fall. One of the practical difficulties was the installation of stable survey markers on this steep and highly crevassed glacier tongue. 

Above the starting zone of the ice avalanche of 1965, a section of medium steepness remained downstream the flat upper part of the tongue. It consisted of two parts, namely an ice-fall with towering seracs separated by deep crevasses followed by a detached ice mass which did not fall down.
This remaining ice mass experienced enhanced motion revealed by several features. Firstly, the ice mass was separated from the glacier by a crescent-shaped depression full of shattered ice. Secondly, numerous sharp-edged, obviously fresh crevasses cut the inner part of the mass involved, which further was more intensely shattered along one edge of the glacier. Thirdly, the talus at the foot of the glacier front was pushed over the edge of a terrasse, a clear sign that the ice cliff was undergoing rapid advance.

In the photographs of the construction site with the Allalingletscher in the background it no fresh crevasses could be observed on August 5 but, by August 30, clear ones already existed. This indicates that the catastrophic event occurred during a phase of enhanced motion, called active phase.

\begin{figure}
\begin{tabular}{@{}c@{}}
\noindent\includegraphics[width=18pc]{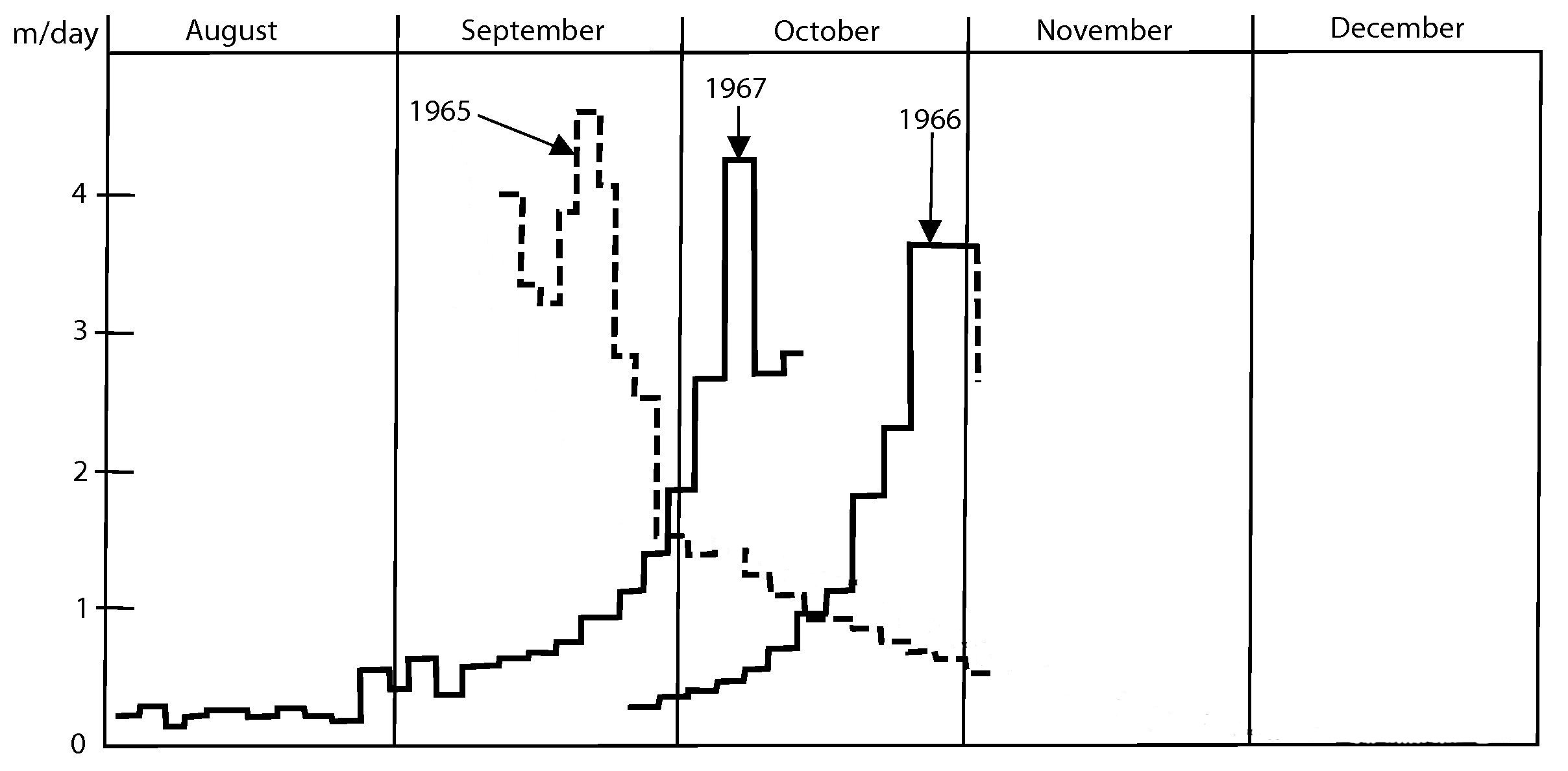}\\
\noindent\includegraphics[width=20pc]{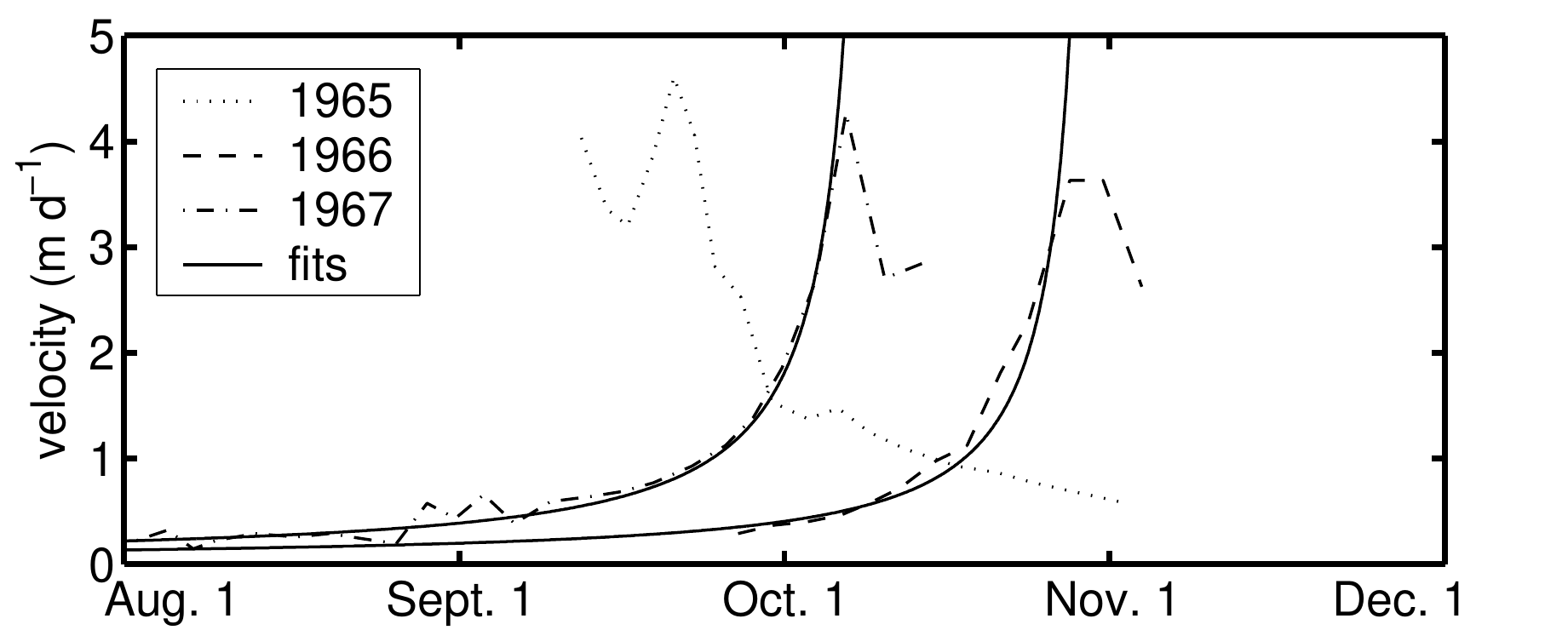}
\end{tabular}

\caption{Evolution of the surface velocity of Allalingletscher tongue between 1965 and 1967.}
\label{velocity1965fit}
\end{figure}

One of the most striking facts illustrated by Fig. \ref{velocity1965fit} is that the timing of the beginning of the active phase was different each year, although its end occurred at the same time in 1965 and 1966. In 1965 the velocity was already fairly high, although still rising, when the survey started. At the beginning of November 1966, the velocity record was interrupted when the work on the dam was stopped because of the high surface velocity; most markers had disappeared when the observations were resumed at the end of the month. In 1967 the survey was discontinued before the active phase ceased completely, but the velocity records of the three years nevertheless gives a unique picture of the active phase. A common feature in the different years is a regular increase and decrease of velocity with time. The regularity of the velocity increase is particularly interesting, since a similar pattern has been observed on other glaciers prior to the breaking-off of a large ice masses; it is noteworthy, however, that the active phases ceased suddenly without formation of any major avalanches similar to that of 1965.

\subsection{Glaciological survey since 1978}
\label{glaciological investigations}
\begin{figure}

\noindent\includegraphics[width=20pc]{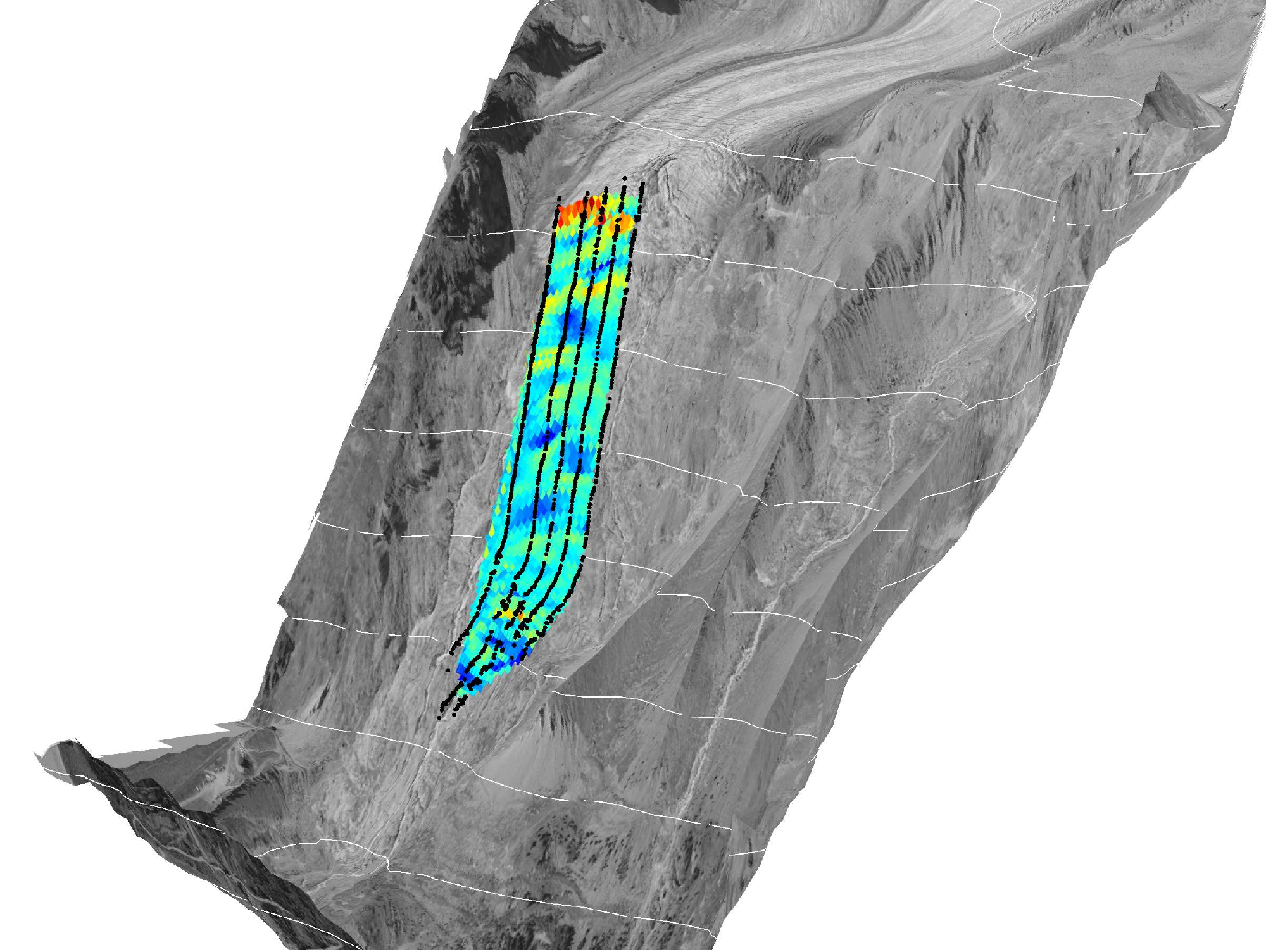}
\caption{Profiles along which glacier thickness was evaluated.}
\label{profil}
\end{figure}

After the intensive glaciological surveys performed by \citet{Rothlisberger1978} after the 1965 event up to 1978, aerial surveys were performed each year. Complete stereophotogrammetry surveys were also performed in 10 years interval to generate a Digital Elevation Model of the glacier surface topography with 25 m of mesh grid (Fig. \ref{timeline}).
Moreover, stereophotogrammetry surveys were performed every year since 1988 to investigate the ice thickness change of the glacier tongue along four different flow lines (see Fig. \ref{profil}).
Mass balance of Allalingletscher was reconstructed with the model GERM (Glacier Evolution Runoff Model) developed by \citet{Huss&al2010}. An enhanced version of this model developed by \citet{Farinotti&al2011} was used to determine daily runoff for Allalingletscher catchment.

\newlength\yearposx
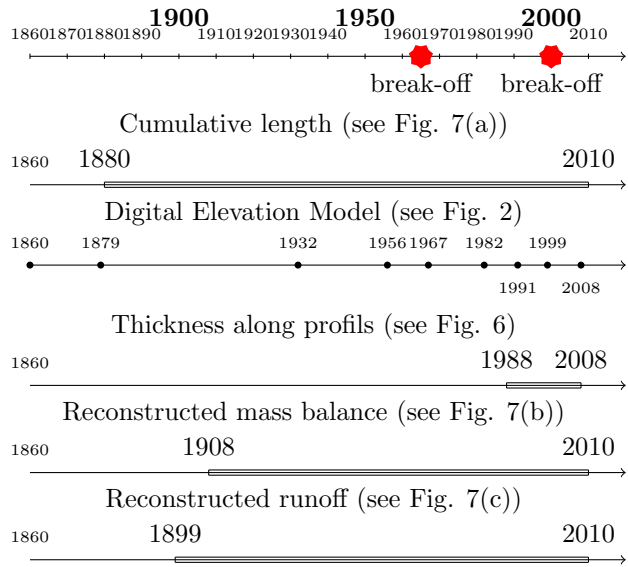
\begin{figure}[htbp]

\begin{tikzpicture}[scale=0.33]

 \foreach \x in {1860,1870,...,2020}{
        \pgfmathsetlength\yearposx{(\x-1860)*0.15cm};
        \coordinate (y\x)   at (\yearposx,0);
        \coordinate (y\x t) at (\yearposx,+3pt);
        \coordinate (y\x b) at (\yearposx,-3pt);
    }
    \pgfmathsetlength\yearposx{(1965-1860)*0.15cm};
    \coordinate (y1965) at (\yearposx,0);

    \draw [->] (y1860) -- (y2020);

    \foreach \x in {1860,1870,...,2010}
        \draw (y\x t) -- (y\x b);

    \foreach \x in {1860,1870,1880,1890,1910,1920,1930,1940,1960,1970,1980,1990,2010}
        \node at (y\x) [above=3pt] {\tiny \x};
        
     \foreach \x in {1900,1950,2000}
        \node at (y\x) [above=8pt] {\bf \x};       

 \node at (y1965) [below=3pt] {break-off};
\draw (y1965) node[star,star points=7,star point ratio=0.8,fill=red]{~};
 \node  at (y2000) [below=3pt] {break-off};
\draw (y2000) node[star,star points=7,star point ratio=0.8,fill=red]{~};
\end{tikzpicture}

Cumulative length (see Fig. \ref{morpho}(a))

\begin{tikzpicture}[scale=0.33]
 \foreach \x in {1860,1880,2010,2020}{
        \pgfmathsetlength\yearposx{(\x-1860)*0.15cm};
        \coordinate (y\x)   at (\yearposx,0);
        \coordinate (y\x t) at (\yearposx,+3pt);
        \coordinate (y\x b) at (\yearposx,-3pt);
    }
     \draw [->] (y1860) -- (y2020);
 \foreach \x in {1880,2010}  {  
\draw[-] (y\x t) -- (y\x b);
\node at (y\x)[above=3pt] {\x}; 

}

\draw [-] (y1880t) -- (y2010t);
\draw [-] (y1880b) -- (y2010b);

\node at (y1860)[above=3pt] {\tiny 1860};
 
\end{tikzpicture}

Digital Elevation Model (see Fig. \ref{geometric_evolution})

\begin{tikzpicture}[scale=0.33] 
    \foreach \x in {1860,1879,1932,1956,1965,1967,1982,1991,1999,2000,2004,2008,2020}{
        \pgfmathsetlength\yearposx{(\x-1860)*0.15cm};
        \coordinate (y\x)   at (\yearposx,0);
        \coordinate (y\x t) at (\yearposx,+3pt);
        \coordinate (y\x b) at (\yearposx,-3pt);
    }
     \draw [->] (y1860) -- (y2020);
       \foreach \x in {1860,1879,1932,1956,1967,1982,1999}{
\node at (y\x) [above=3pt] {\tiny \x};
\fill (y\x) circle (4pt);}
       \foreach \x in {1991,2008}{
\node at (y\x) [below=3pt] {\tiny \x};
\fill (y\x) circle (4pt);}
\end{tikzpicture}

Thickness along profils (see Fig. \ref{profil})

\begin{tikzpicture}[scale=0.33]
 \foreach \x in {1860,1988,2008,2010,2020}{
        \pgfmathsetlength\yearposx{(\x-1860)*0.15cm};
        \coordinate (y\x)   at (\yearposx,0);
        \coordinate (y\x t) at (\yearposx,+3pt);
        \coordinate (y\x b) at (\yearposx,-3pt);
    }
     \draw [->] (y1860) -- (y2020);
 \foreach \x in {1988,2008}  {  
\draw[-] (y\x t) -- (y\x b);
\node at (y\x)[above=3pt] {\x}; 
}

\draw [-] (y1988t) -- (y2008t);
\draw [-] (y1988b) -- (y2008b);
\node at (y1860)[above=3pt] {\tiny 1860};
 
\end{tikzpicture}

Reconstructed mass balance (see Fig. \ref{morpho}(b))

\begin{tikzpicture}[scale=0.33]
 \foreach \x in {1860,1908,2010,2020}{
        \pgfmathsetlength\yearposx{(\x-1860)*0.15cm};
        \coordinate (y\x)   at (\yearposx,0);
        \coordinate (y\x t) at (\yearposx,+3pt);
        \coordinate (y\x b) at (\yearposx,-3pt);
    }
     \draw [->] (y1860) -- (y2020);
 \foreach \x in {1908,2010}  {  
\draw[-] (y\x t) -- (y\x b);
\node at (y\x)[above=3pt] {\x}; 
}

\draw [-] (y1908t) -- (y2010t);
\draw [-] (y1908b) -- (y2010b);
\node at (y1860)[above=3pt] {\tiny 1860};
 
\end{tikzpicture}

Reconstructed runoff (see Fig. \ref{morpho}(c))

\begin{tikzpicture}[scale=0.33]
 \foreach \x in {1860,1899,2010,2020}{
        \pgfmathsetlength\yearposx{(\x-1860)*0.15cm};
        \coordinate (y\x)   at (\yearposx,0);
        \coordinate (y\x t) at (\yearposx,+3pt);
        \coordinate (y\x b) at (\yearposx,-3pt);
    }
     \draw [->] (y1860) -- (y2020);
 \foreach \x in {1899,2010}  {  
\draw[-] (y\x t) -- (y\x b);
\node at (y\x)[above=3pt] {\x}; 
}

\draw [-] (y1899t) -- (y2010t);
\draw [-] (y1899b) -- (y2010b);
\node at (y1860)[above=3pt] {\tiny 1860};
 
\end{tikzpicture}

\caption{Timeline of the available data}
\label{timeline}
\end{figure}

\begin{figure}
\begin{tabular}{c}
\noindent\includegraphics[width=20pc]{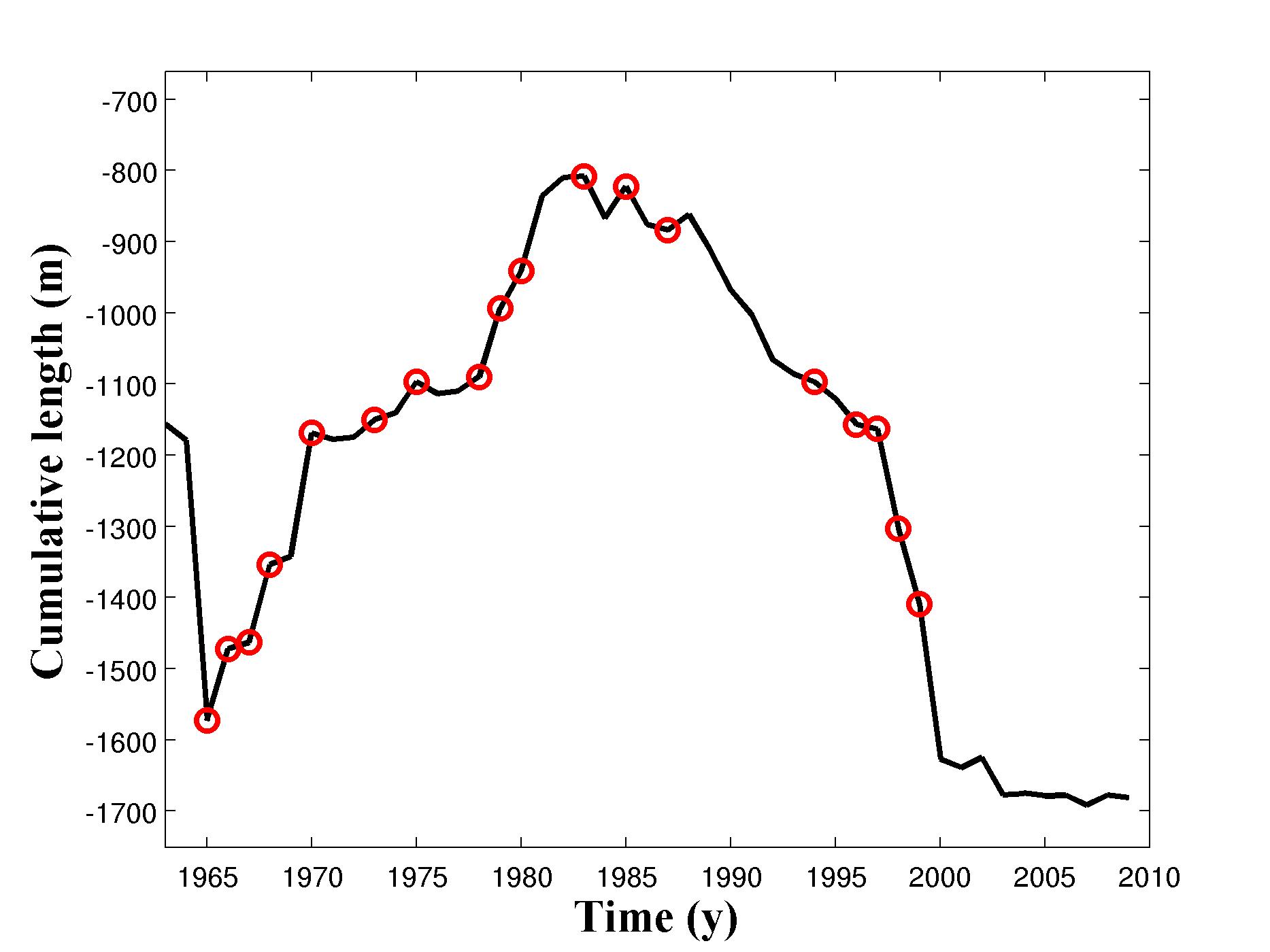} (a)
\\
\noindent\includegraphics[width=20pc]{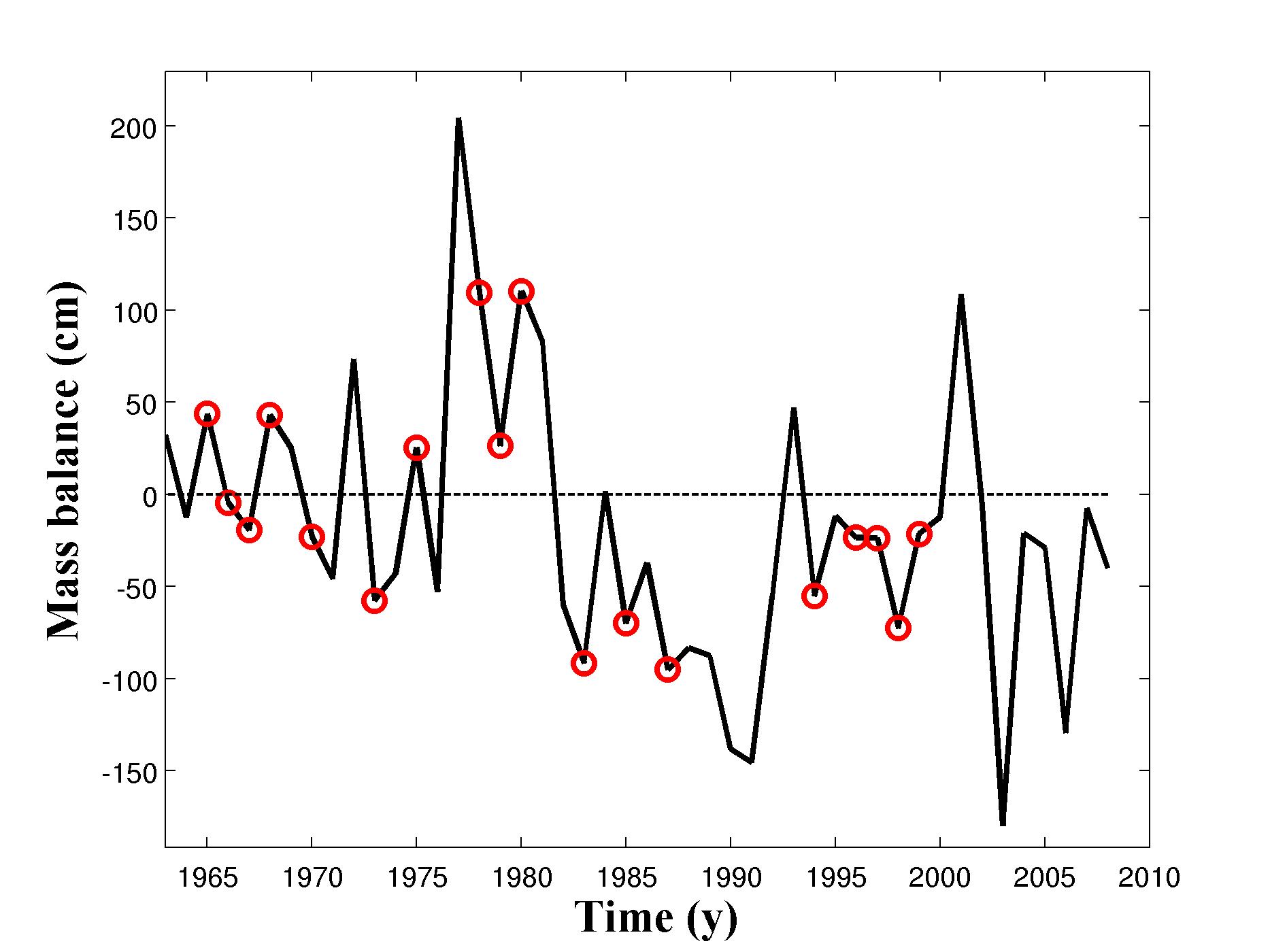} (b)
\\ 
\noindent\includegraphics[width=20pc]{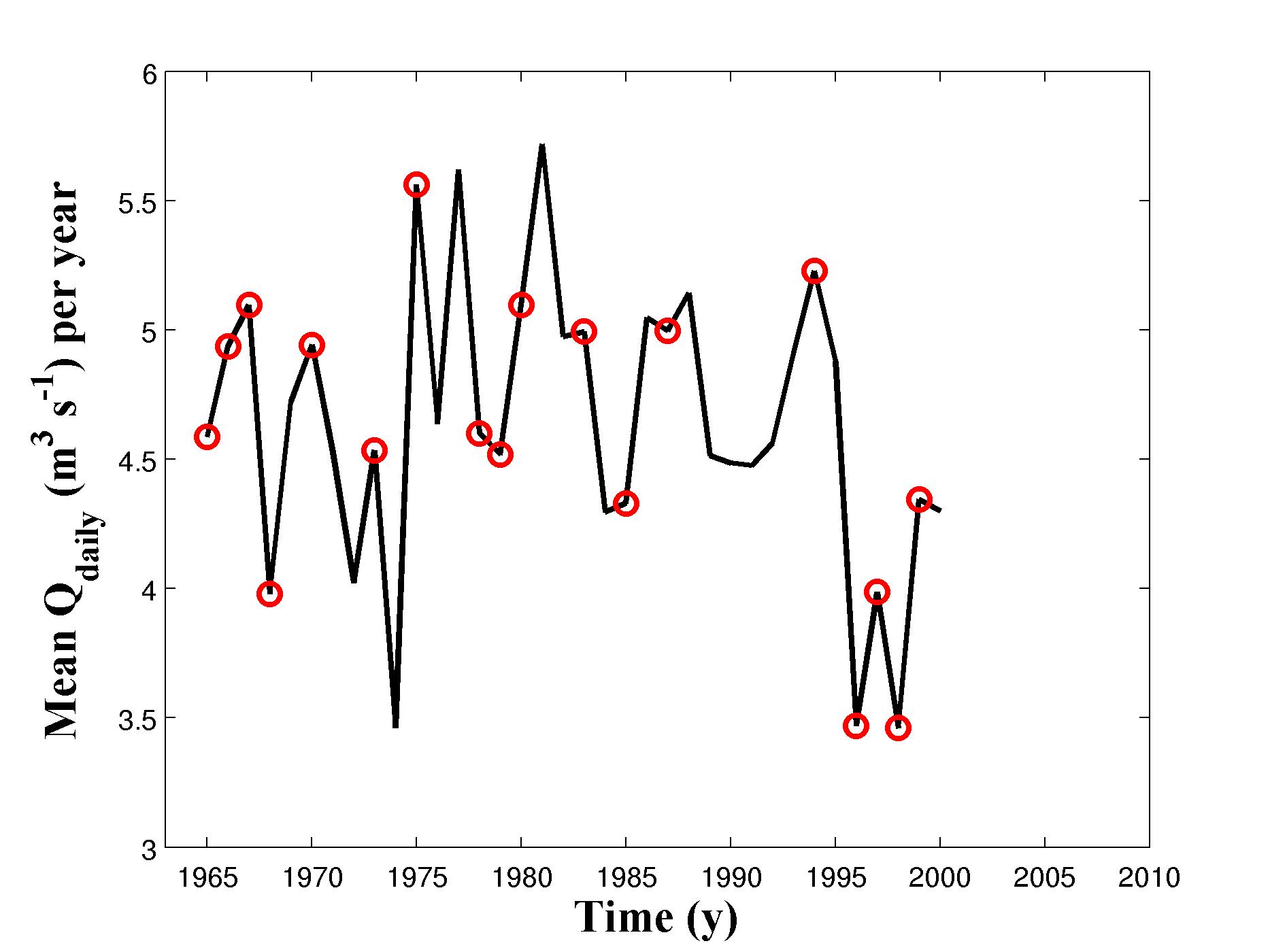} (c)\\

\end{tabular} 
\caption{(a) Cumulative length change of Allalingletscher since 1880. Red dots indicate when active phases occurred. (b) Reconstructed mass balances since 1965. (c) Evolution of the mean daily runoff of subglacial water flow per each year.} 
\label{morpho}

\end{figure}

Thanks to available comprehensive database, it was possible to investigate the geometrical evolution of the glacier tongue, which is shown in Fig. \ref{morpho}(a).
According to this figure, the glacier readvanced rapidly after 1965 and recovered its previous geometry 5 year later. The glacier continued to advance up to 1984. 
Since then, the glacier retreated, with an accelerating rate after 1997.

\begin{figure}

\begin{tabular}{cc}
\noindent\includegraphics[width=10pc]{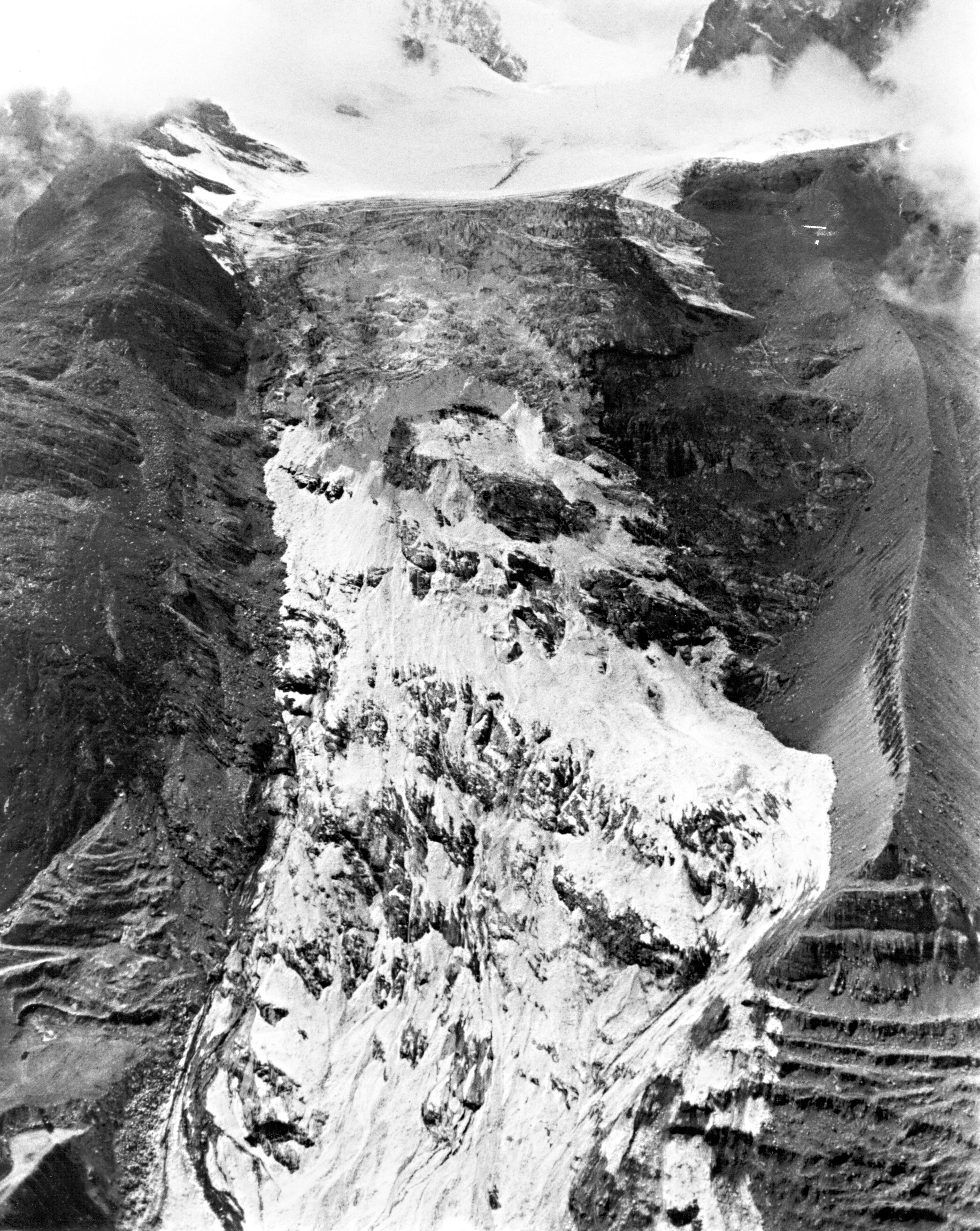}
&
\noindent\includegraphics[width=9pc]{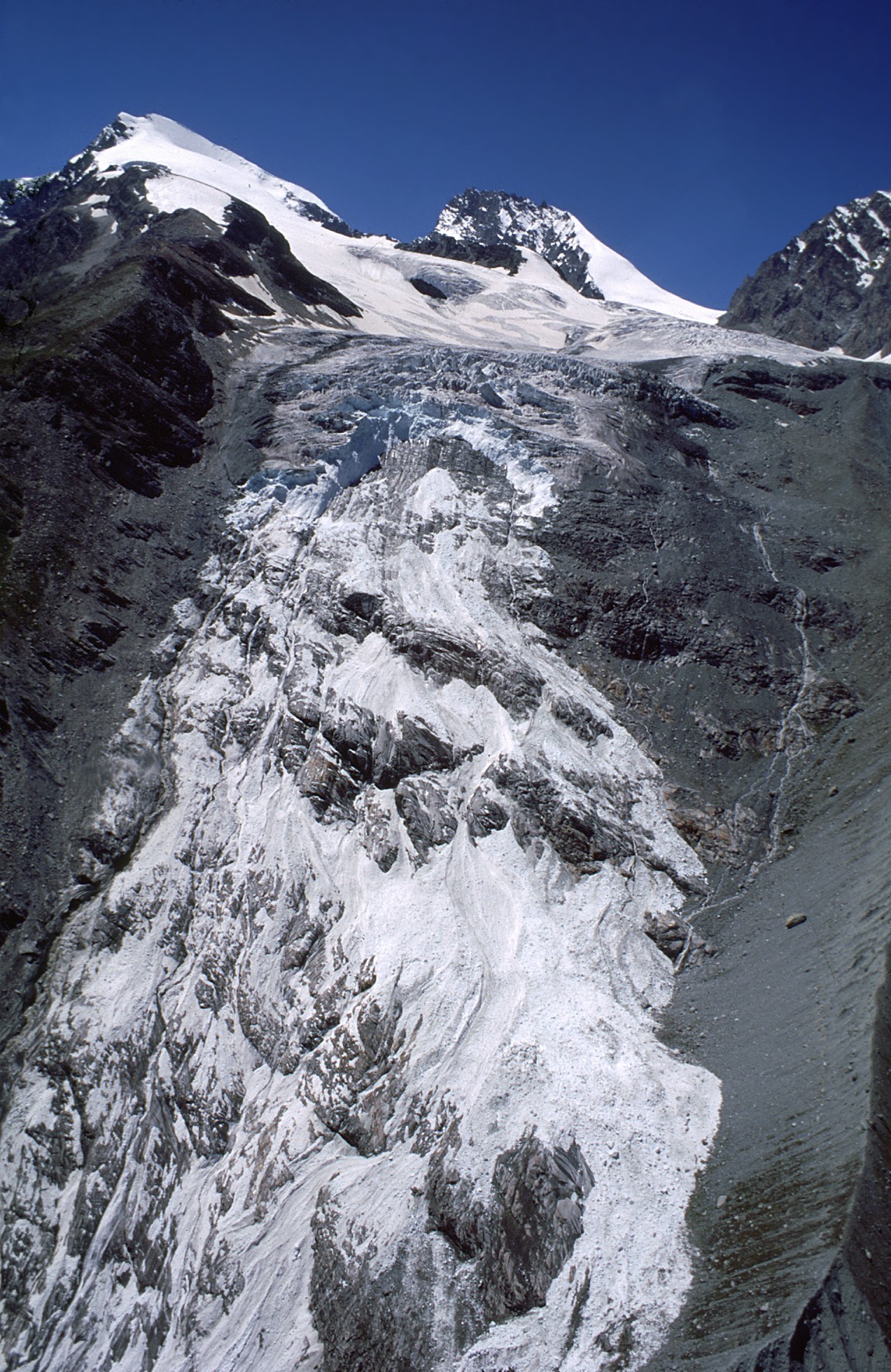}\\

1965&2000\\

\end{tabular}
\caption{Allalingletscher after the 1965 event (2 mio m$^3$ of ice on
  August 30, 1965) and after the 2000 event (1 mio m$^3$ on
  July 31, 2000)}
\label{comp1965_2000}
\end{figure}

In August 2000, the glacier configuration was similar to 1965, and for
security reasons the hazard zone was closed during summer. An ice volume of
1 Million cubic meters broke of, but did not cause any damage (Fig. \ref{comp1965_2000}).
After this event, the glacier was stable, and the terminus stayed more or less at the same altitude.

\subsection{Initiation of active phases}
\label{activephase}
Seasonal changes in surface velocity have been observed in the years after
the catastrophe, and it is now known that Allalingletscher speeded up regularly
every 1-3 years, usually during summer or late autumn. 
In most cases no large release of ice occurred. Apparently the active phase is {\bf necessary but not sufficient} to cause breaking off. 

Figs. \ref{morpho} also emphasizes the years when active phases were observed (red dots).
At a first glance, they seem to appear when large rate of change of the tongue geometry occur, as they are concentrated between 1965 and 1980 (during advance) and between 1994 and 2000 (during retreat).
However, no clear correlation could be found, suggesting another initiation process.
The active phase has always begun during the melt season, and, perhaps even more significantly, has stopped at the beginning of the winter. This seems to indicate that meltwater plays a major role in the processes triggering the active phases. 
However, the initiation of such phenomenon does not seem to be correlated with the total amount of water flowing each year under the glacier (Fig. \ref{morpho}(c)).

\begin{figure}

\begin{tabular}{@{}c@{}r@{}}
 \begin{minipage}{38pc}
\begin{tabular}{@{}c@{}c@{}c@{}c@{}}
\noindent\includegraphics[width=8pc]{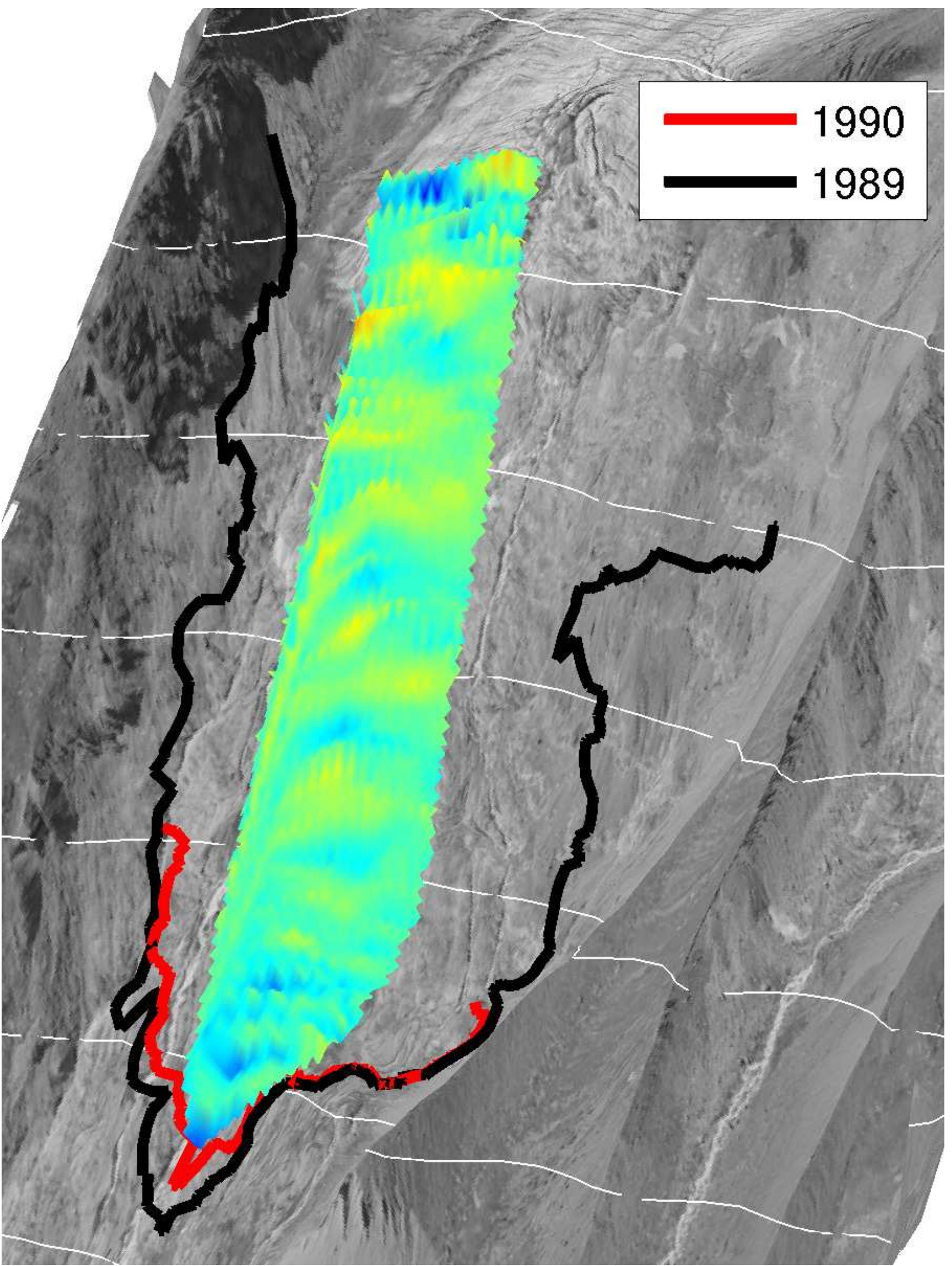}(a)&
\noindent\includegraphics[width=8pc]{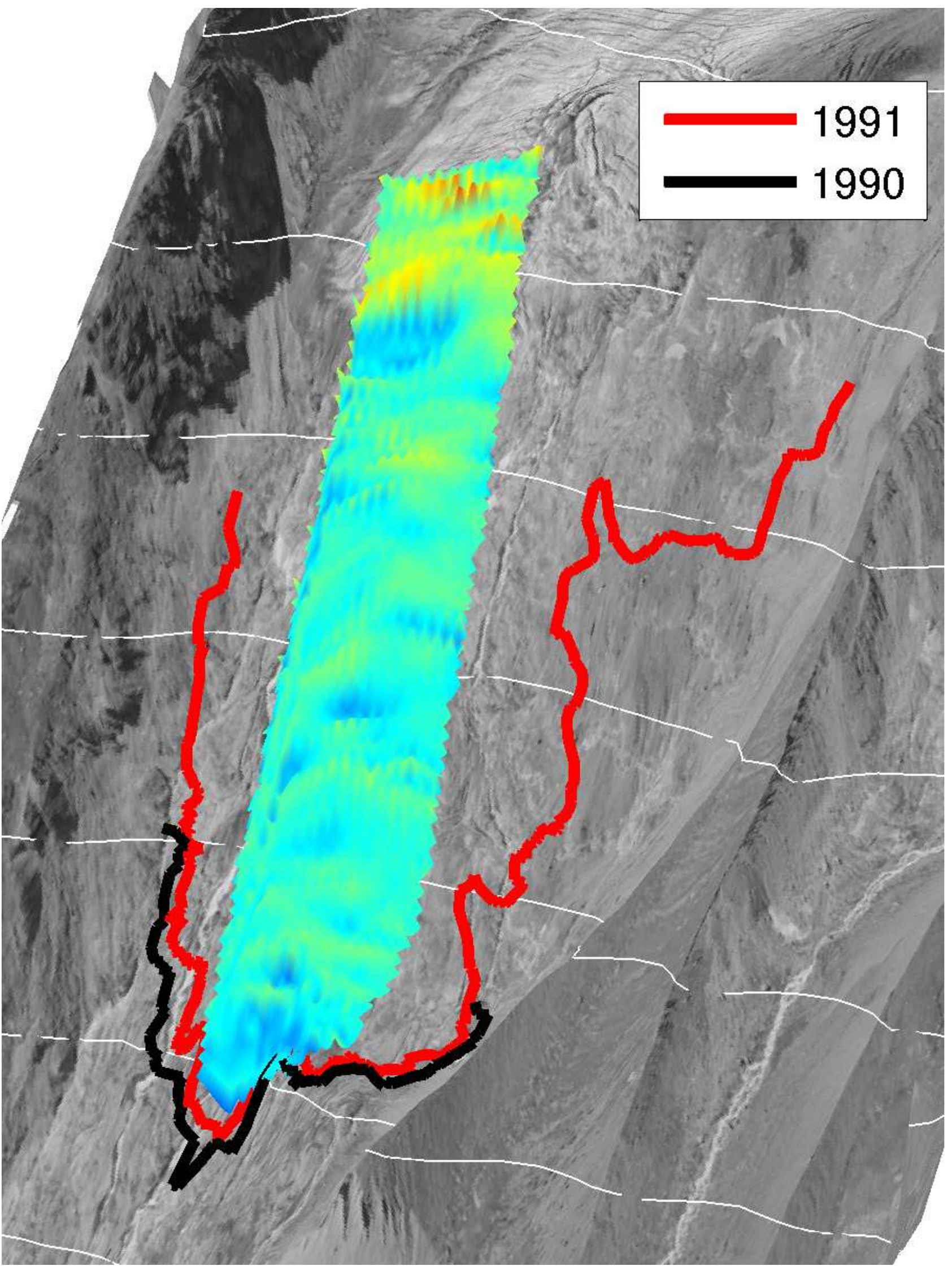}(b)&
\noindent\includegraphics[width=8pc]{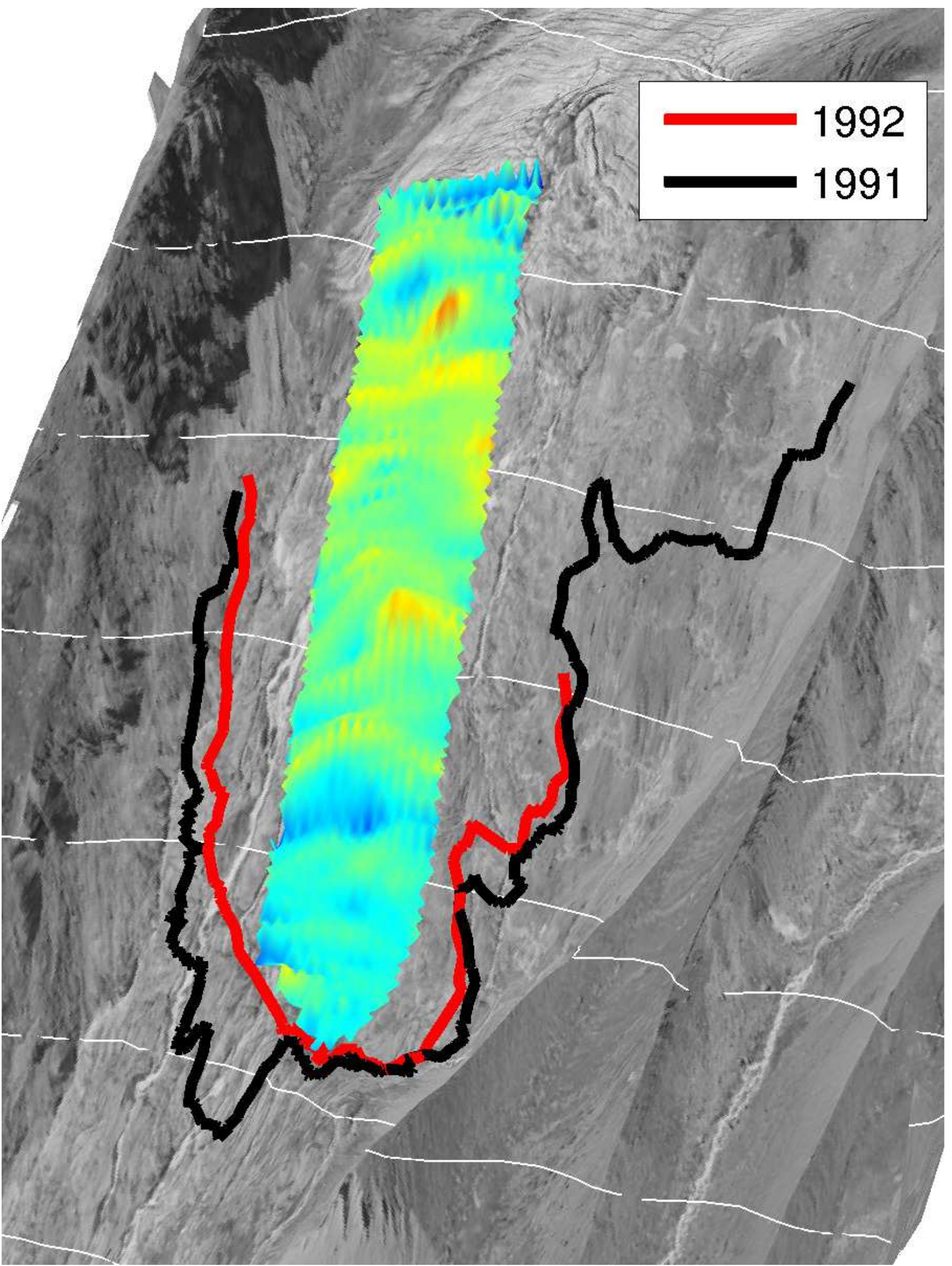}(c)&
\noindent\includegraphics[width=8pc]{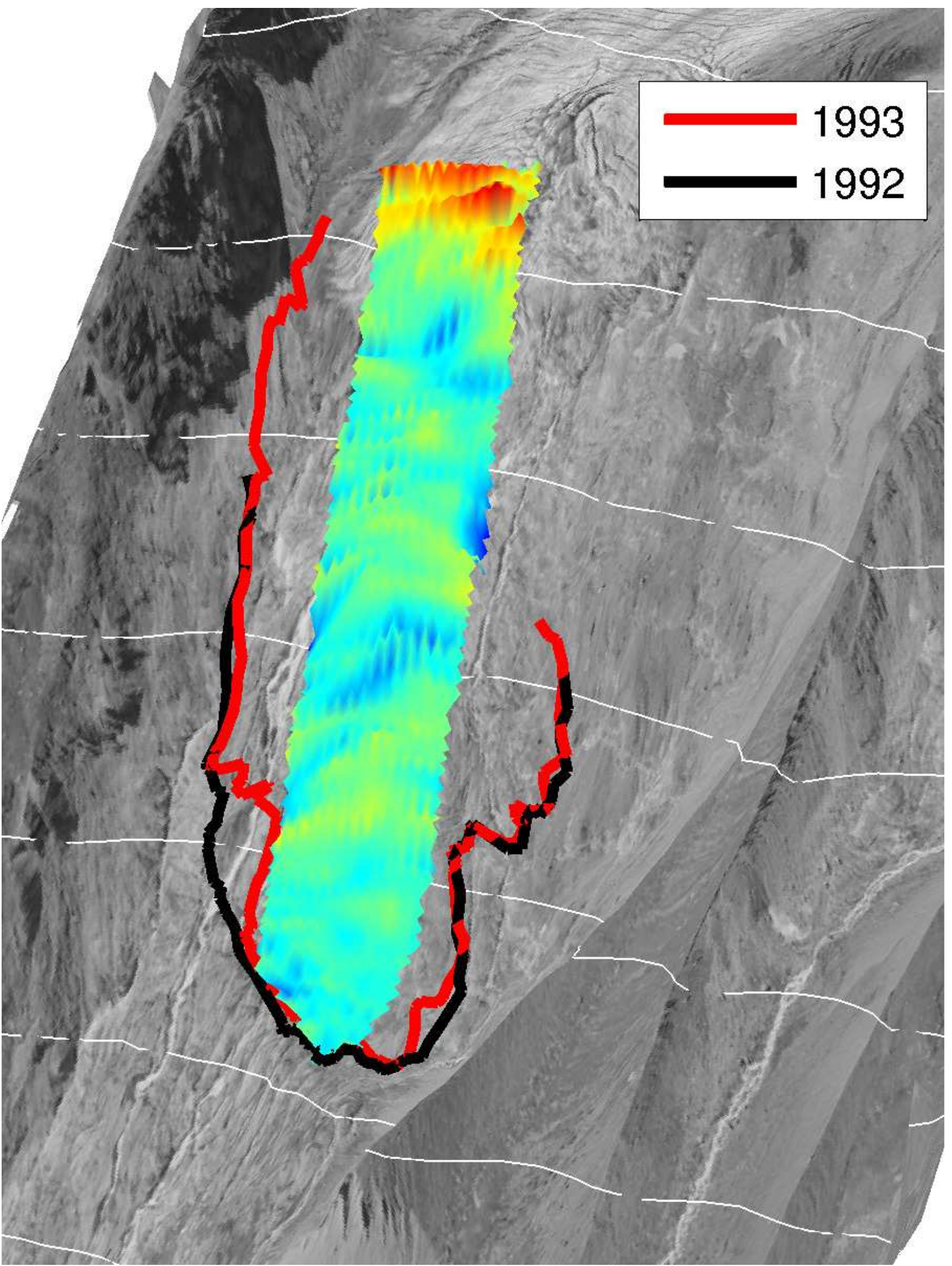}(d)\\
\noindent\includegraphics[width=8pc]{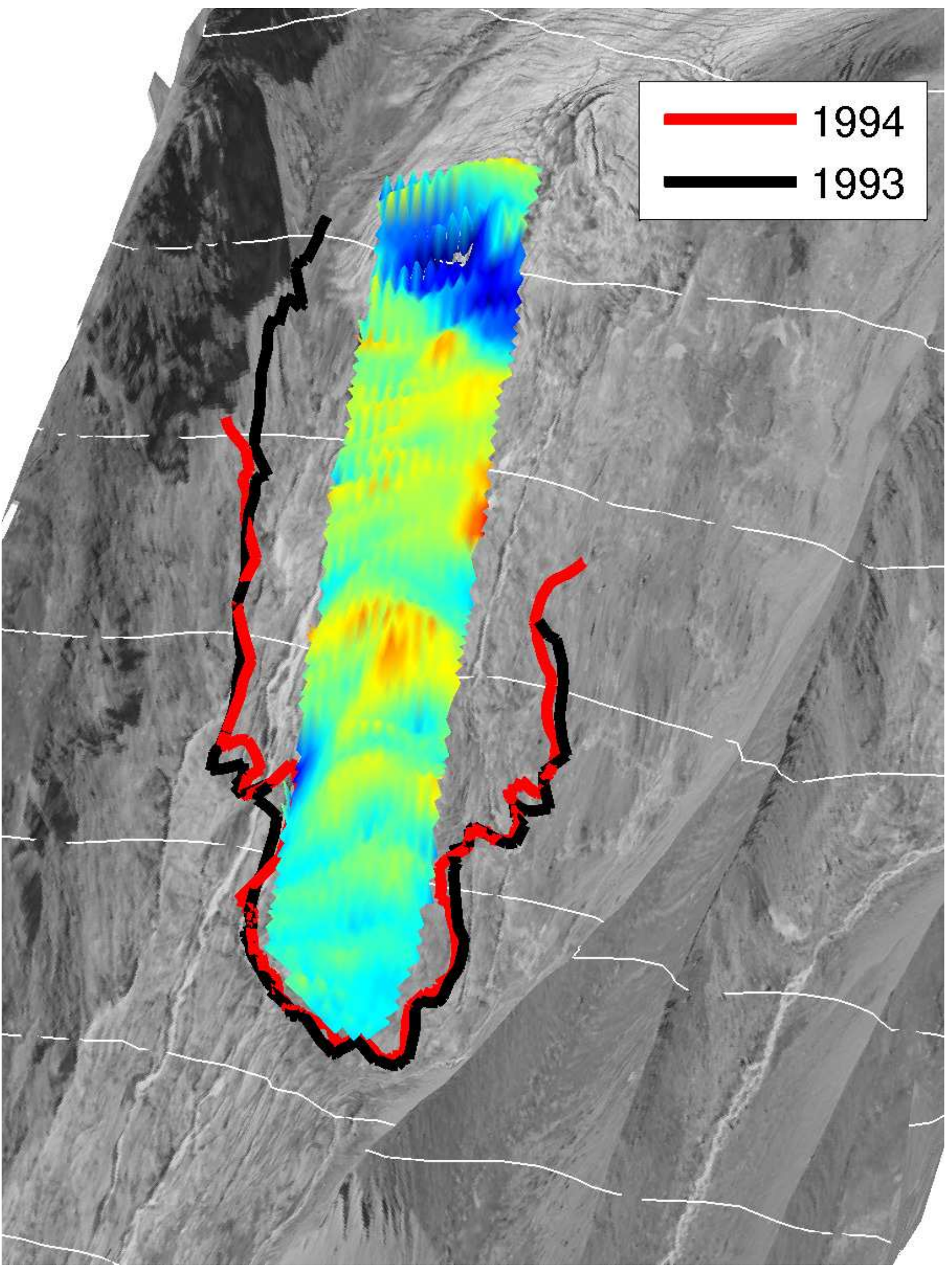}(e)&
\noindent\includegraphics[width=8pc]{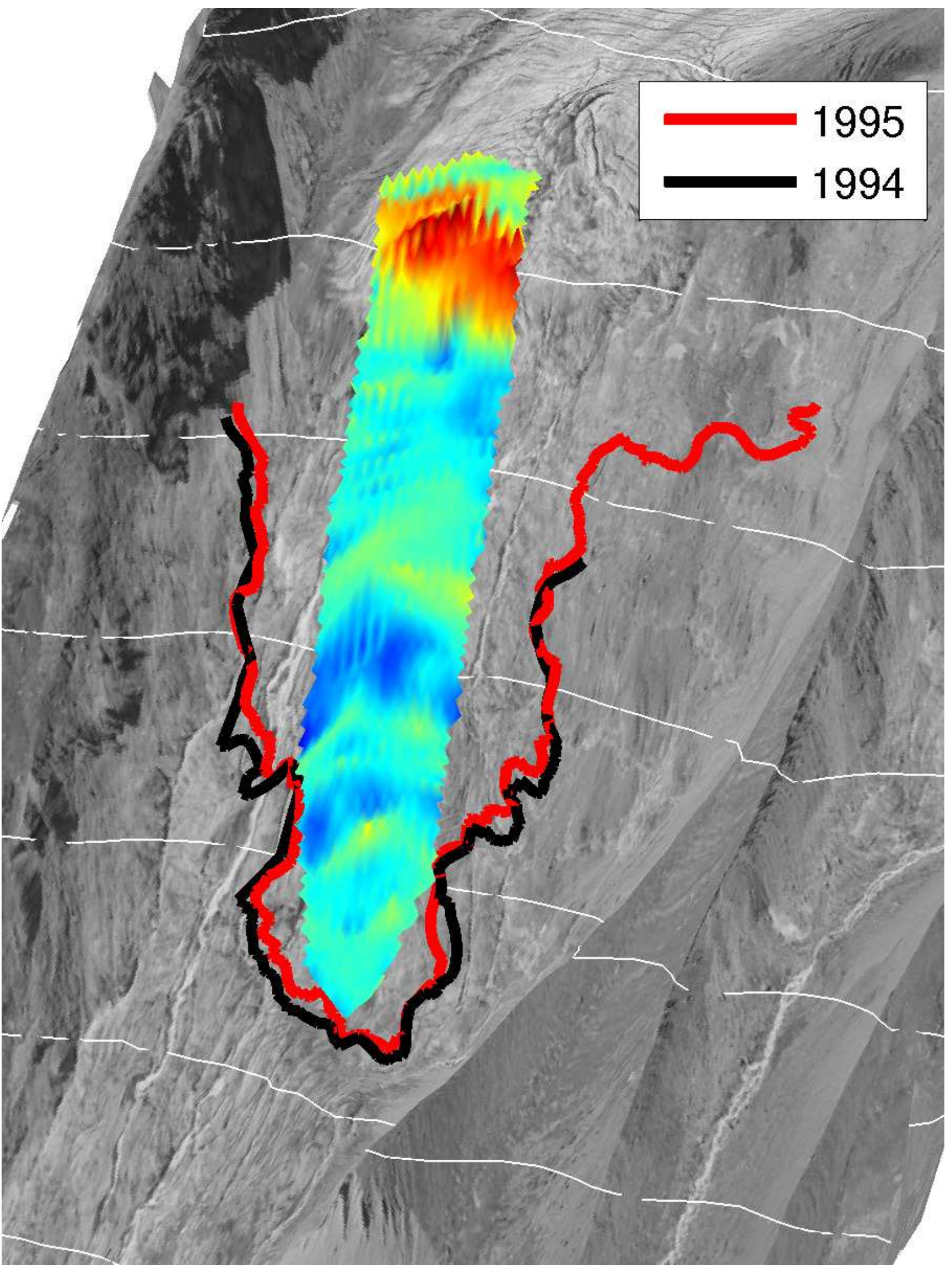}(f)&
\noindent\includegraphics[width=8pc]{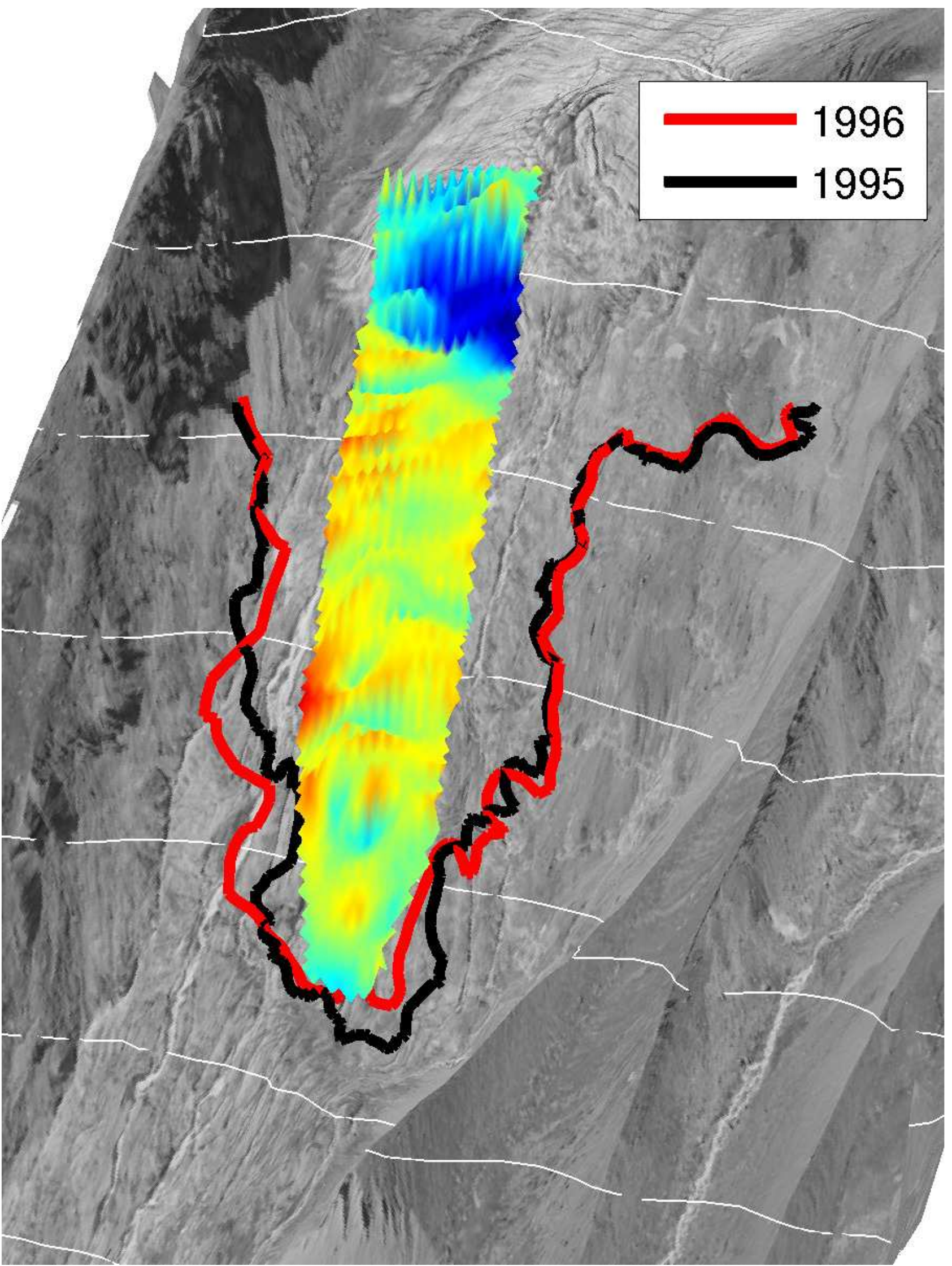}(g)&
\noindent\includegraphics[width=8pc]{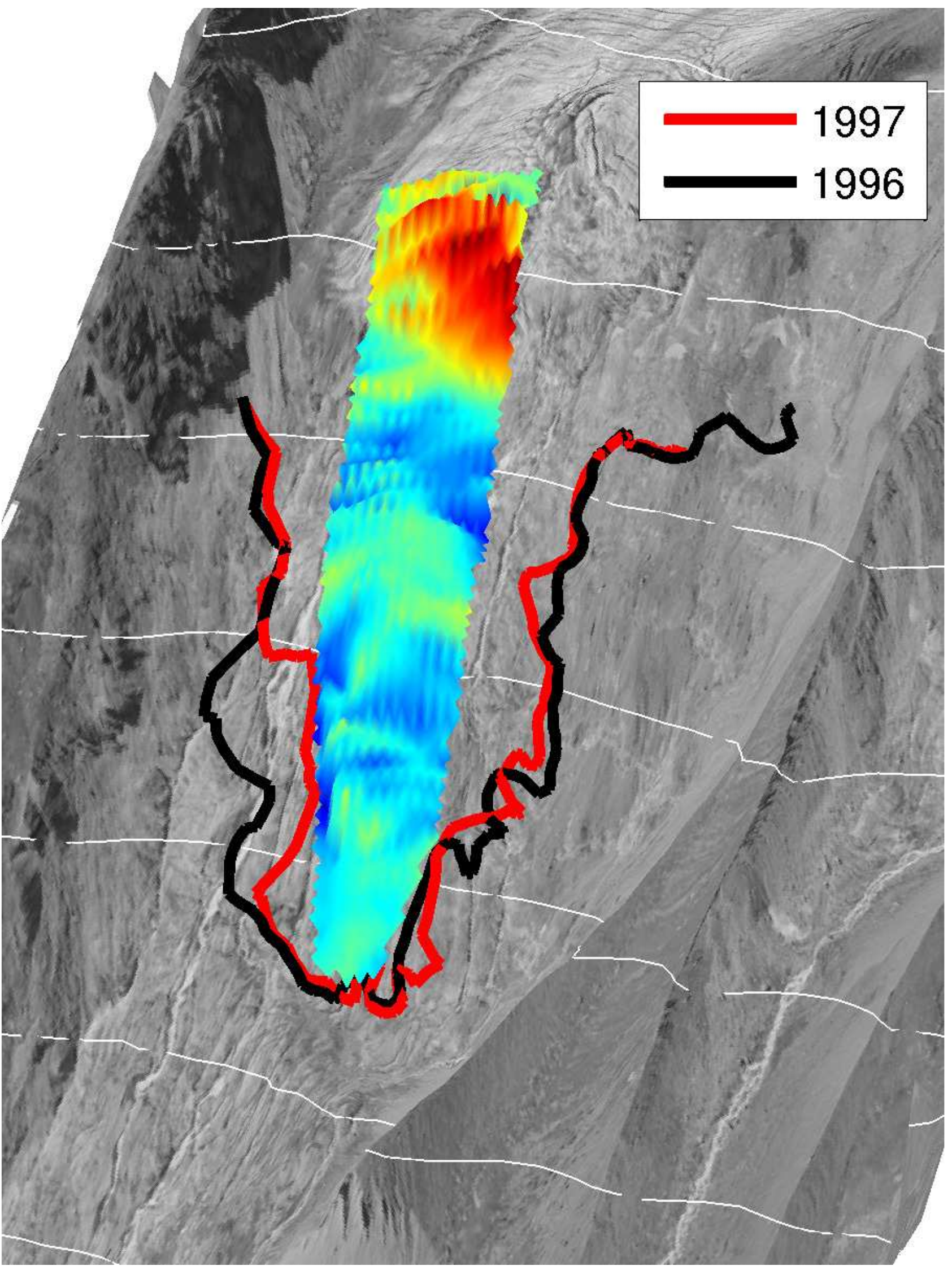}(h)\\
\noindent\includegraphics[width=8pc]{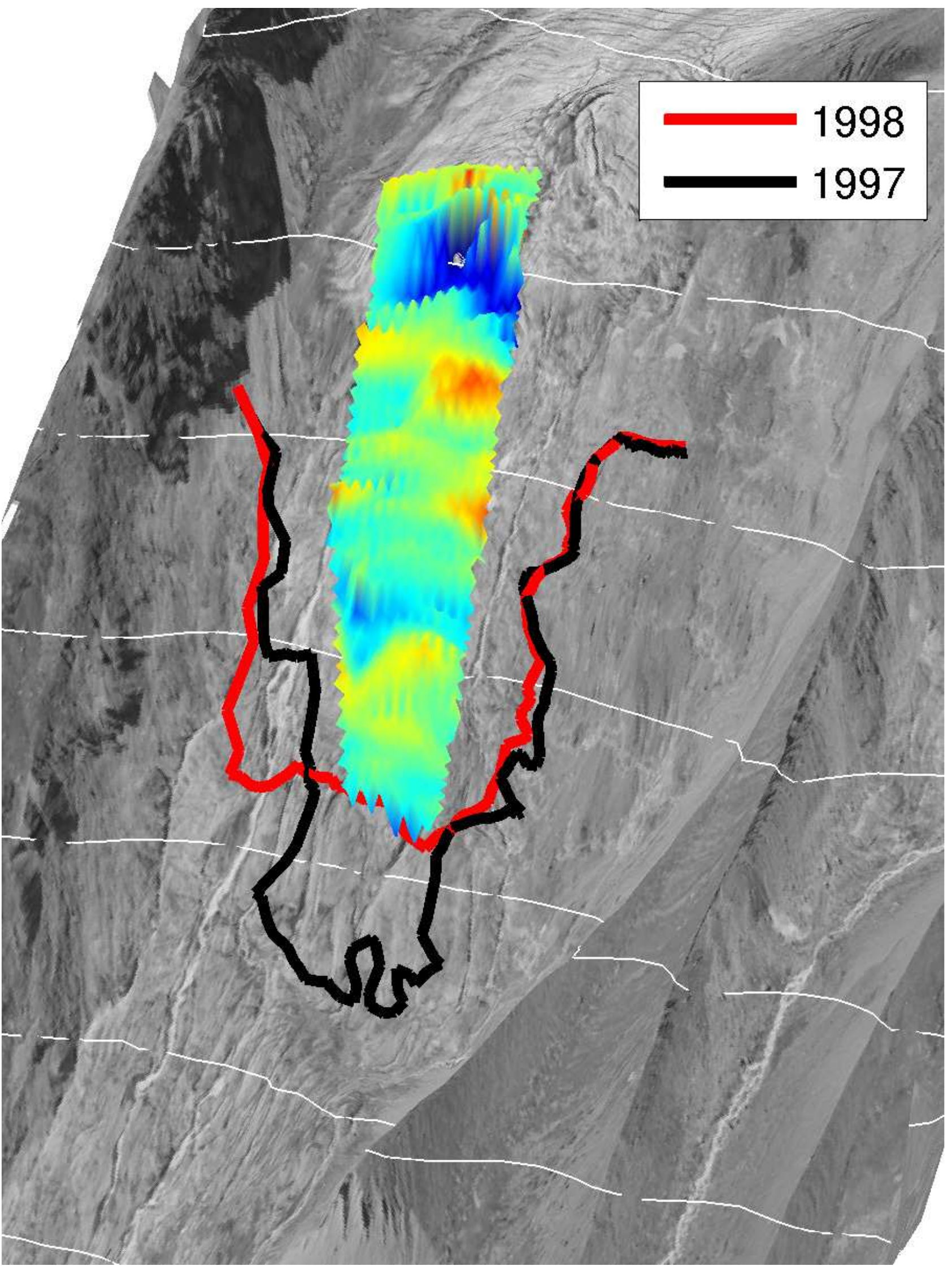}(i)&
\noindent\includegraphics[width=8pc]{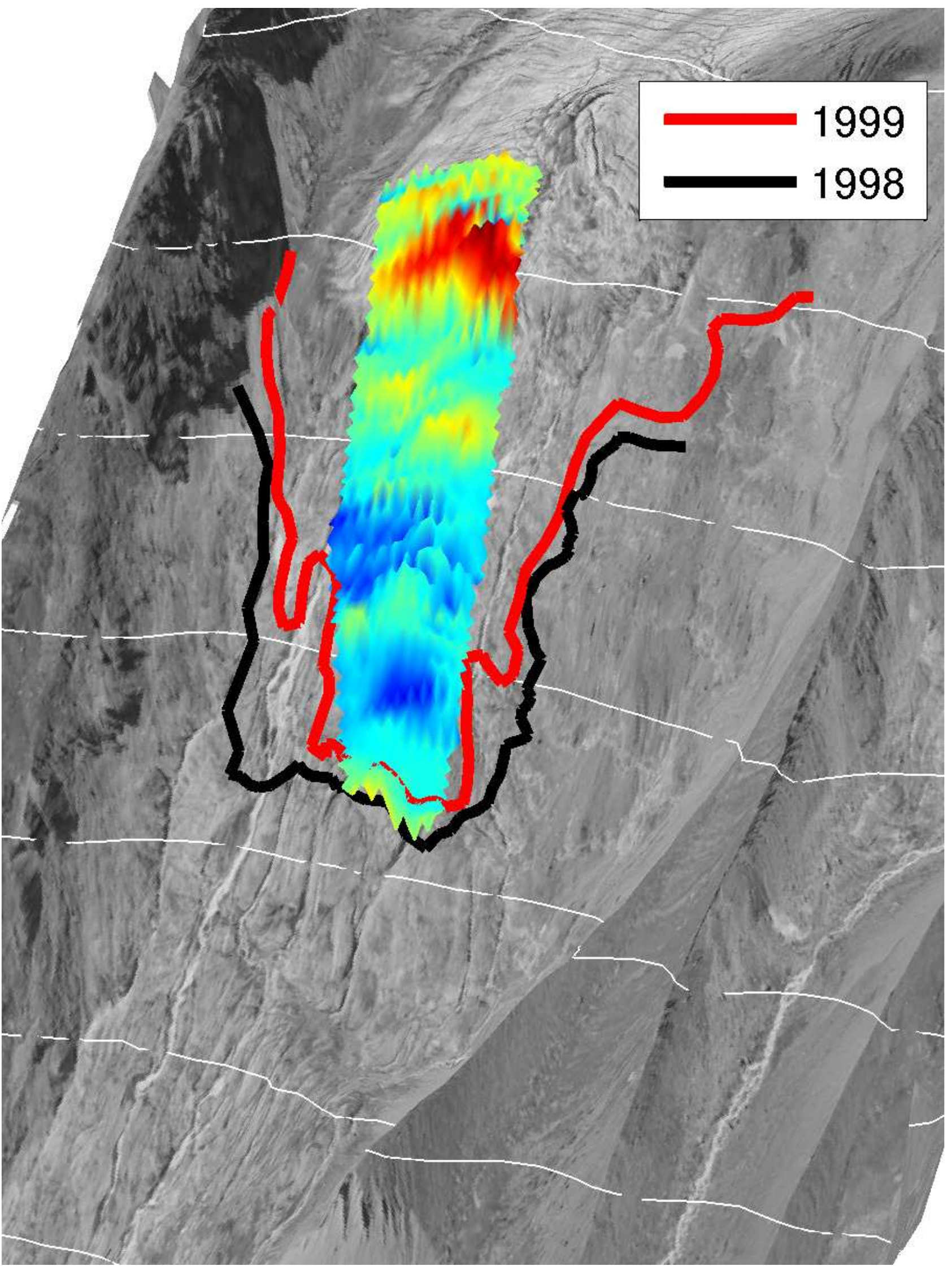}(j)&
\noindent\includegraphics[width=8pc]{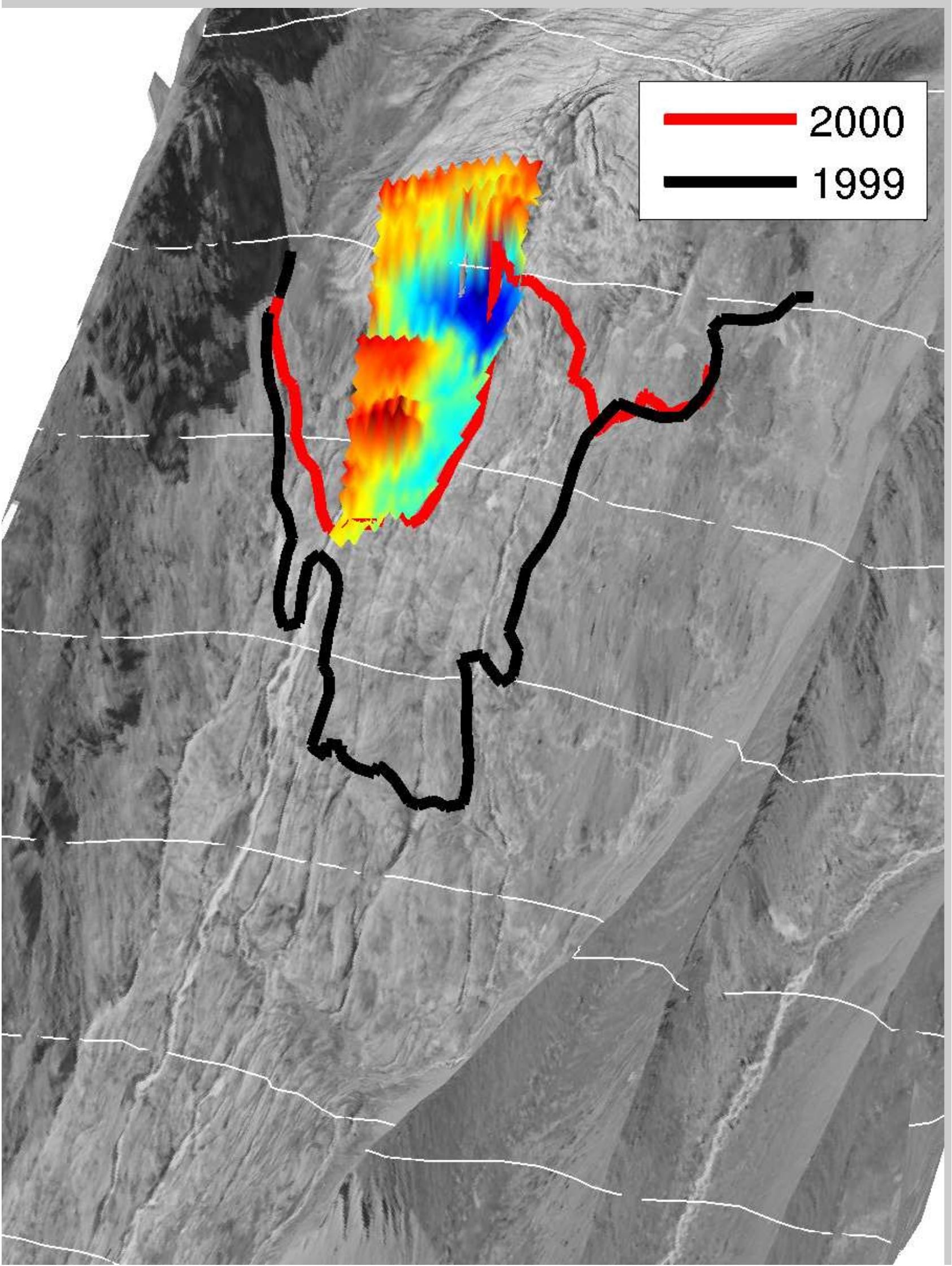}(k)&
\noindent\includegraphics[width=8pc]{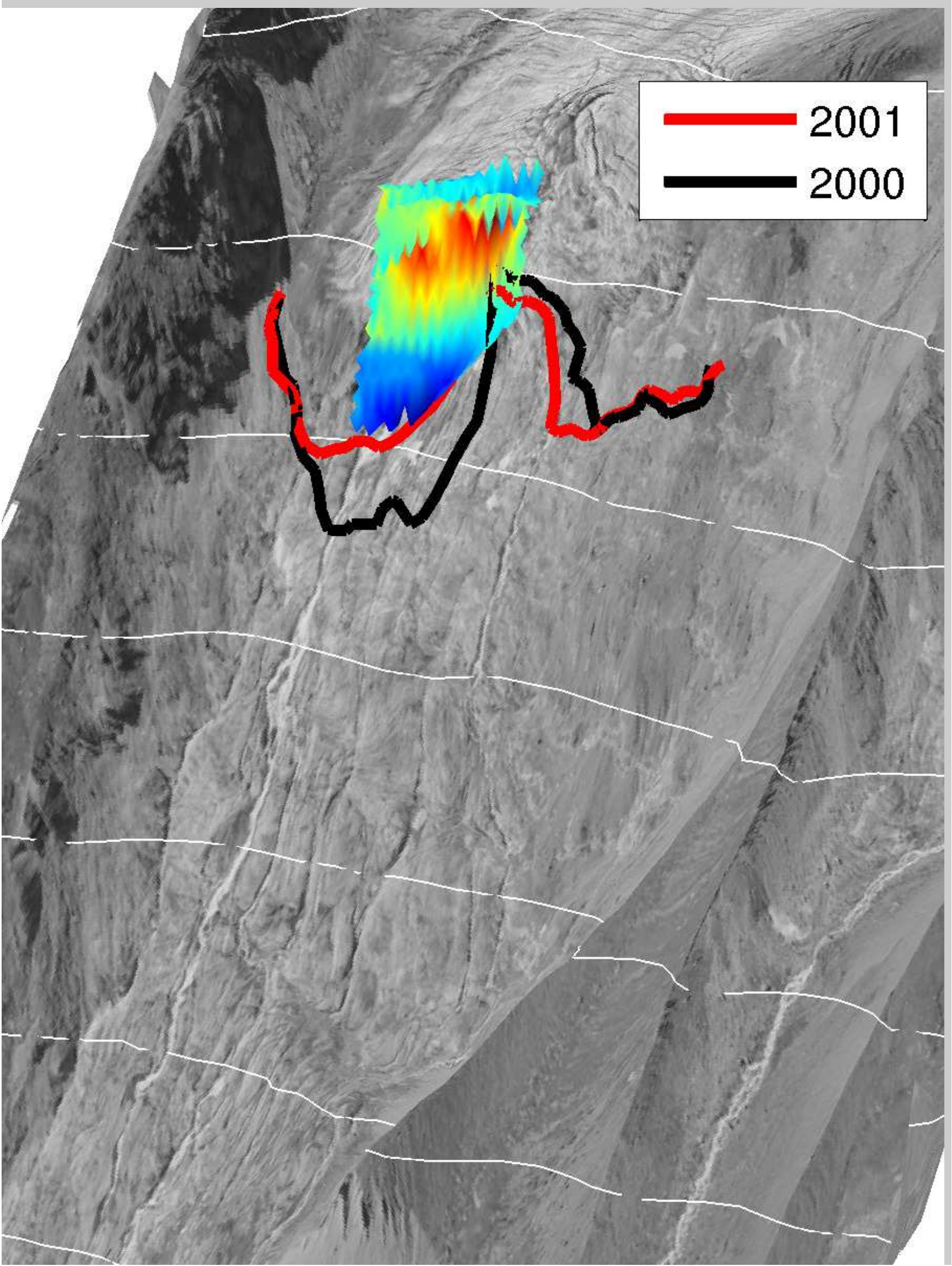}(l)\\
\end{tabular}
\end{minipage}
&
\begin{minipage}{1pc}
\noindent\includegraphics[height=30pc]{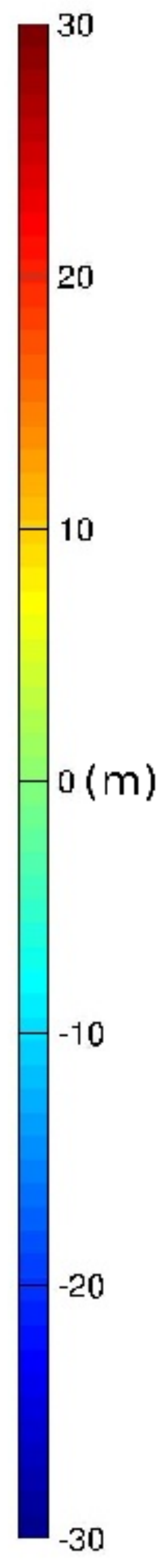}
\end{minipage}

\end{tabular}

\caption{\label{diffprofil} Difference of thickness of the glacier tongue between two successive years from 1989 to 2000. Spatial extension of the tongue is also shown.}

\end{figure}

Ice thickness changes at the tongue was analyzed every year since 1990 by interpolating the measured thicknesses along the 4 profiles given in Fig. \ref{profil}.
Fig. \ref{diffprofil} shows the difference in tongue thickness from one year to the next.

Fig. \ref{morpho}(a)  shows that, during this period, active phases occurred in 1994, 1996, 1997, 1998 and 1999.
Interestingly, between 1990 and 1993, the geometrical extension of the glacier tongue was stable and the thickness of the tongue experienced a small homogeneous thinning. No active phases were detected during this period. 
Then, in 1993, the glacier started a progressive retreat and a thickenning of the upper part of the tongue could be observed. This initial ''pulse'' is due to the positive mass balance in 1993 (Fig. \ref{morpho}(b)). 
In 1994, the tongue became thinner in the upper part while a thickening was observed behind the terminus. An active phase was detected in 1994. 
Since then, a ''oscillating behavior'' was initiated, where ice mass is first accumulated in the upper part of the tongue and then transferred to its terminus the following year.
Such oscillating behavior and active phases seems to be correlated.

 \subsection{Summary of the Allalingletscher break-off}
 
 \begin{enumerate}
 \item There are two known events: 31.8.1965 and 31.7.2000.
 \item The glacier tongue is {\bf temperate}: the glacier is sliding on its bedrock.
 \item A regular speed up of the glacier tongue during 2-3 weeks has been observed almost every year after 1965 between July and October.
 \item Subglacial hydrology plays a major role on the instability.
 \item The speed up of the glacier tongue is only a {\bf necessary condition} for 
 the breaking off.
 \item A critical mass distribution within the glacier tongue is probably a key factor for the instability.
 \end{enumerate}

To investigate in more details the causes of this instability, we reanalysed this event by applying a new numerical model designed for describing natural gravity-driven instabilities \citep{Faillettaz&al2010}. This model allows us to test the different hypothesis proposed previously to explore the possible causes of the break-off of this glacier.

\section{Numerical modeling}
\label{numdes}

\subsection{Model description}
\label{shortdes}

We use a model describing the progressive maturation
of a mass towards a gravity-driven instability, which combines
basal sliding and cracking. Our hypothesis is that
gravity-driven ruptures in natural heterogeneous materials are characterized by a common
triggering mechanism resulting from a competition between frictional sliding
and tension cracking. Heterogeneity of material properties and dynamical
interaction of damage and cracks along the sliding layer seem to have a significant influence on the global behavior and have to be modeled.

\begin{figure}
\noindent{
\includegraphics[width=20pc]{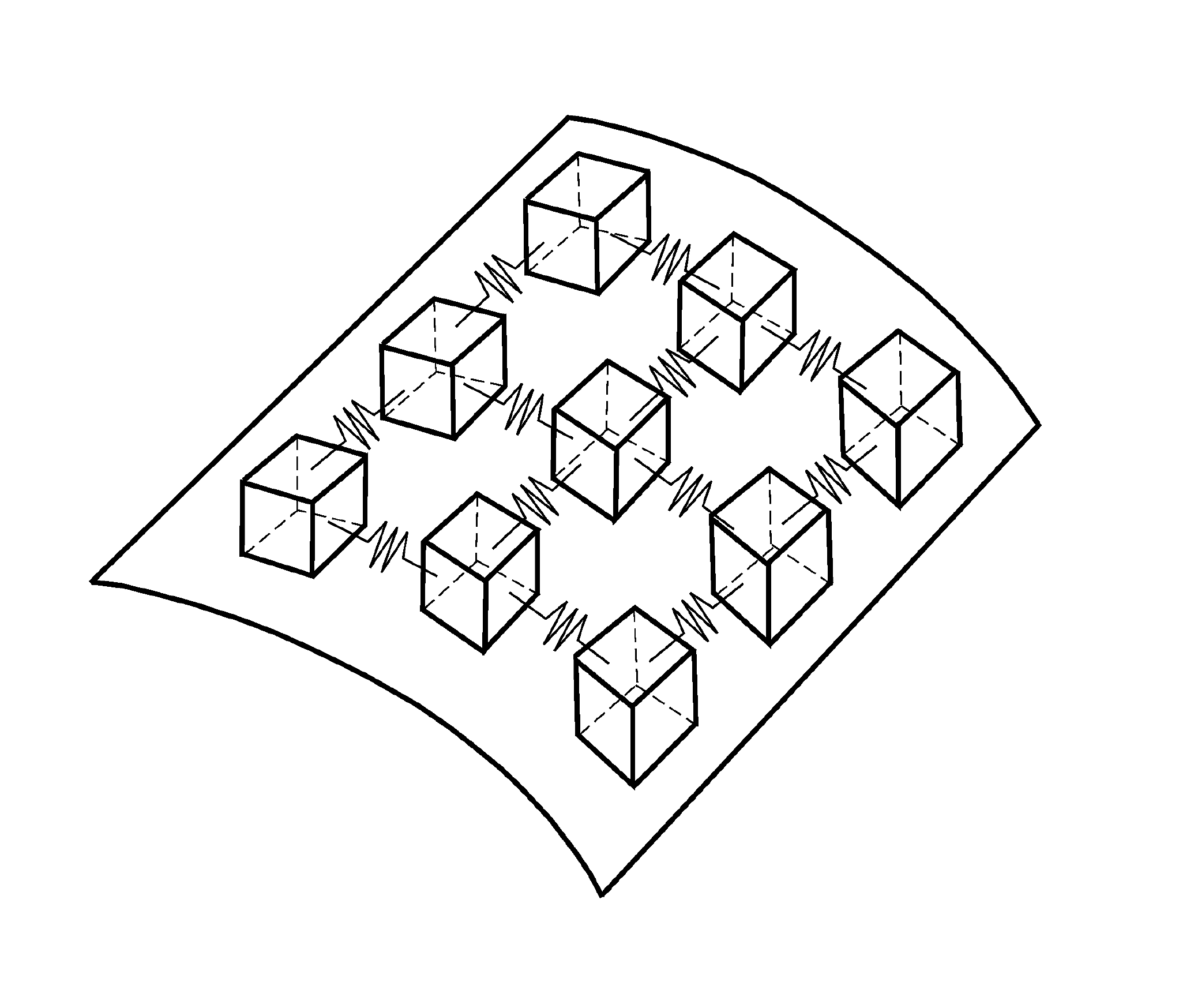}}
\caption{\label{blocks} Illustration of the model consisting of spring-blocks resting on an inclined plane.
The blocks lie on an inclined curved surface and gravity is the driving force. Only
a small subset of the spring-block system is shown here.
}

\end{figure}

This numerical model is based on the discretization
of the natural medium in terms of blocks and springs forming a two-dimensional network
sliding on an inclined plane (Fig. \ref{blocks}. Each block, which can slide, is connected to its
four neighbors by springs that can fail, depending on the history of displacements and damage. We develop physically realistic models describing
the frictional sliding of the blocks on the supporting surface and the tensile
failure of the springs between blocks proxying for crack opening. Frictional sliding is modeled with a state-and-velocity weakening friction law with
threshold. 
This means that solid friction is not used as a
parameter but as a process evolving with the concentration of deformation and
properties of sliding interfaces.
Crack formation is modeled with a time-dependent cumulative damage
law with thermal activation including stress corrosion. In order to reproduce cracking
and dynamical effects, all equations of motion (including inertia) for each
block are solved simultaneously.

The present model improves the multi-block model of \citet{Andersen&al1997}
and \citet{Leung&Andersen1997} in two ways. First, we use a
state-and-velocity weakening friction law instead of 
a constant (or just state- or velocity-weakening) solid friction
coefficient. Second, rather than a static threshold for the spring failures, we model the 
progressive damage accumulation
via stress corrosion and other thermally activated processes aided by stress.
Both improvements make the numerical simulations significantly
longer but present the advantage of embodying rather well
the known empirical phenomenology of sliding and damage processes.
Adding the state and velocity-dependent friction law and time-dependent
damage processes allows us to model rather faithfully the 
interplay between sliding and cracking between blocks and the overall
self-organizing of the system of blocks \citep{Faillettaz&al2010}. 

The geometry of the system of blocks interacting via springs and with 
a basal surface is depicted in Figure~\ref{blocks}. 
To sum up, the model includes the following characteristics:
\begin{enumerate}
\item Frictional sliding on the ground,
\item Heterogeneity of basal properties,
\item Possible tension rupture by accumulation of damage,
\item Dynamical interactions between blocks along the sliding layer,
\item Geometry and boundary conditions, and
\item Interplay between frictional sliding and cracking.
\end{enumerate}

We describe the two key processes in the model, the friction and
damage laws, that are applied to blocks and bonds respectively.

\subsection{Friction law between the discrete blocks and the basal surface (after \citet{Faillettaz&al2011b}}
\label{frictionlawchap}

Guided by the rate-state friction character of rock \citep{Dietrich1994} and ice \citep{Fortt&Schulson2009,Lishman&al2011}, a rate- and state-dependent friction law 
seems also to be adequate to describe 
ice-bedrock friction.
The version of the rate/state-variable constitutive law, currently most
accepted as being in reasonable agreement with experimental 
data on solid friction, is known as the Dieterich-Ruina law \citet{Dietrich1994}:

\begin{equation}
\mu(\dot{\delta}, \theta) = \mu_0 + A \ln {\dot{\delta} \over \dot{\delta}_0} + B \ln 
{\theta \over \theta_0}~,
\label{Dietrich-Ruina}
\end{equation}

where the state variable $\theta$ is usually interpreted as the
proportional to the surface
of contact between asperities of the two surfaces. The constant $\mu_0$ is the
friction coefficient for a sliding velocity $\dot{\delta}_0$
and a state variable $\theta_0$. $A$ and $B$ are coefficients depending on material properties.

The time evolution of the state variable $\theta$ is described by
\begin{equation}
{{\rm d}\theta \over {\rm d}t} = 1 - {\theta \dot{\delta} \over 
D_c}~, 
\label{qwertyu}
\end{equation}
where $D_c$ is a characteristic slip distance, usually interpreted as
the typical size of asperities.
The friction law (\ref{Dietrich-Ruina}) with (\ref{qwertyu}) accounts for
the fundamental properties of a broad range of surfaces in contact, 
namely that they strengthen (age)
logarithmically when aging at rest, and tend to weaken (rejuvenate) when sliding.

The primary parameter that determines stability, $A-B$, is a
material property. Physically, $A-B$ indicates the sensitivity of the friction coefficient to velocity change: a negative value indicates velocity-weakening leading to an unstable slip; a positive value indicates velocity-strengthening leading to a stable slip.
\citet{Faillettaz&al2010} showed that 
the case $A=B$ is of special interest because it retrieves the main
qualitative features of the two classes, and also because, 
empirically, $A$ is very close to $B$. Assuming 
$A=B$ is therefore reasonable and ensures more robust
results.
In Appendix A we present how the equation for the critical time $t_f$ was obtained. $t_f$ indicates the transition from 
a subcritical sliding to the dynamical inertial sliding. 
It is given in this case \citep{Faillettaz&al2010} by
\begin{equation}
\label{frictionlaw}
t_f\;=\;\frac{\theta_0}{\exp(\frac{\mu-\mu_0}{A})-1}~,
\end{equation}
for $\mu > \mu_0$; Note that $t_f \to \infty$ for $\mu \leq \mu_0$,
where $t_f$ is the time when the block starts sliding, $\mu_0$ is a constant
friction coefficient, $A$ is a constant parameter depending on material
properties and $\theta_0$ is the state parameter at steady state. The parameter
$\mu$ is evaluated for each block with the definition of the solid friction $\mu~=~\frac{T}{N}$ where T is the tangential force 
determined from the position of its
connected neighbors and N is the normal component of the weight of the block.
The state parameter $\theta_0$ in Eq. \ref{frictionlaw} is given (see Appendix A) by
\begin{equation}
\label{theta0}
\theta_0\;=\;{D_c \over \dot{\delta}_0},
\end{equation}
where $\dot{\delta_0}$ is generally interpreted as the initial low velocity
of a sliding mass before it starts to accelerate towards its
dynamical instability and $D_c$ can be interpreted as a characteristic slip distance over which
different asperities come in contact.

Three parameters have to be correctly evaluated to model the frictional processes within
the glacier: $\mu_0$, A and $\theta_0$.
An increasing friction coefficient with increasing sliding velocity (velocity-strengthening) at low velocities has been observed for polished ice-on-ice \citep{Kennedy&al2000,Montagnat&Schulson2003}, ice-on-granite \citep{Barnes&al1971} and ice-on-ice along Coulombic fault \citep{Fortt&Schulson2009}.
A decreasing friction coefficient with increasing sliding velocity (velocity-weakening) at high velocities has been observed as well \citep{Barnes&al1971,Kennedy&al2000}. The transition between these two regimes occurs at a velocity of $10^{-5} \rm m s^{-1}$ \citep{Fortt&Schulson2009}.
These two different behaviors are generally explained by two physical mechanisms
depending on the sliding velocity regimes:
\begin{itemize}
\item[(i) ] The first mechanism is the water lubrification mechanism (produced by frictional heat at the
sliding surface) working at sliding velocities above roughly 0.01 $\rm m~s^{-1}$. The water lubrification mechanism is characterized by the low viscous
resistance of a water film produced by frictional heat at the sliding interface
\citep{Maeno&Arakawa2004,Barnes&al1971,Kennedy&al2000,Montagnat&Schulson2003}. 
\item[(ii) ] The second mechanism is the adhesion and plastic deformation of
  ice at the friction interface, which is present at velocities lower than
  roughly 0.01 $\rm m~s^{-1}$ \citep{Kennedy&al2000,Montagnat&Schulson2003,Maeno&Arakawa2004}.
 \end{itemize}

By analogy with rock physics, a rate- and state-dependent friction model for fresh and sea ice was proposed recently \citep{Fortt&Schulson2009,Lishman&al2011}.
Such a model enables to explain the change in the friction coefficient when varying sliding velocity along Coulombic shear fault or slip history.
The experiments performed by \citet{Fortt&Schulson2009} at -10$^\circ$C on sliding along Coulombic shear faults in ice suggests that the ice-ice friction coefficient is velocity-dependent and varies between 0.6 to 1.4 depending on the applied sliding velocity along the fault. As these last results are the closest to our case, we used a mean value of $\mu_o^{ice-ice}$ equal to 0.8 in our calculation.

Once the block slides, the dynamics
is controlled by a kinetic friction coefficient, which is in general
smaller than the static coefficient $\mu_o^{ice-ice}$.
For low temperatures, \citet{Kennedy&al2000} and \citet{Fortt&Schulson2009} found a relatively
constant friction coefficient around $0.6$ for sliding
velocities below $10^{-4} \rm ms^{-1}$, and rapidly decreasing values down to $0.1$
at velocities above $10^{-4} \rm ms^{-1}$ . 
Because velocities above $10^{-4} \rm ms^{-1}$ are not expected in our model
describing the nucleation phase to the catastrophic rupture, we thus 
assume a kinetic friction coefficient $\mu_d=0.6$.  

Next, we turn to the evaluation of the coefficient $A$ (and $B$ since we have
made the simplifying assumption that $B=A$)  in the
rate- and state-dependent friction law (Eq. \ref{Dietrich-Ruina}). 
Laboratory experiments suggest that $A$ is smaller
than $\mu_0$ by typically one 
and sometimes up to two orders of magnitude for rock \citep{Ohmura&Kawamura2007,
Scholz2002,Scholz1998}, ranging from 0.01 to 0.2.
A recent experiment on saline ice suggests however that $A$ could be significantly higher for ice (i.e. $A \approx 0.3$) \citep{Lishman&al2011}. 
As we do not have access to 
strong experimental constraints and
since the friction law should be valid for the ice-ice and the ice-bedrock friction,
 we choose $A \approx 0.1$,
corresponding to one-tenth of the static friction coefficient, which is the upper limit given by \citet{Ohmura&Kawamura2007} for rock. 

The last parameter to be determined is $\theta_0$ in Eq. \ref{theta0}.
In the case of a glacier, the sliding velocity is 
typically of the order of centimeters per day. Therefore a roughly correct estimation is $\dot{\delta_0} \approx 1 \;\rm cm.d^{-1}$.
$D_c$ can be interpreted as a characteristic slip distance over which
different asperities come in contact. It is
difficult to evaluate this value. The recent seismological literature
reports $D_c$ to lie in the range of tens of centimeters to meters for earthquakes
\citep{Mikumo&al2003,Zhang&al2003}. We arbitrary choose
$D_c \simeq 1 \;\rm m$.
Finally, inserting this value in equation (\ref{theta0}), we obtain $\theta_0=100 \;\rm days$. 

To account for the heterogeneity and
roughness of the sliding surface, the state variable $\theta_i$ is reset to a
new random value after the dynamical sliding stops. This random value should not be
choosen too low in order to prevent a block which just stopped sliding from switching
immediately in a new dynamical phase (i.e. $t_f = 0$). Thus, we assign 
$\theta_i=\nu . \theta_0$ with $\nu$ uniformly distributed between 0.5 and 1.5.

Finally, the possible water lubrication of the bedrock leading to a progressive decrease of the ice-rock frictional resistance is modeled using a progressive decrease of the friction coefficient $\mu_0$ at the interface between the simulated ice block and the basal surface (see section \ref{improvement}).

\subsection{Creep law}
\label{creeplaw}
As explained in section \ref{shortdes}, bonds are modeled as
linear springs in parallel with an Eyring dashpot. The springs
transmit the forces associated with the relative displacements
of the blocks. The spring stiffness has also to be evaluated in order
to reflect the elastic property of the bulk ice mass.
In continuous elasticity, Hooke's law of elasticity relates stress $\sigma$ and
strain $\epsilon$ via Young's modulus $E$: $\sigma=E \epsilon$.
This leads to the expression: $\sigma_{bulk}=E_{ice} \frac{\delta L}{L}$.
This stress is applied to a side surface of the block $S=L \times H$ (where $H$ corresponds to the height and L the length of the surface), leading to an equivalent
force in the bulk equal to $F_{bulk}=E_{ice} \frac{\delta L}{L}\;S$.
A linear spring is subjected to forces given by $F_{bond}=K_{bond}\;\delta L$.
In order to find an equivalent behavior, these two forces have to be of the
same order, leading to a spring stiffness given by $K_{bond}=E_{ice} H$.
Usually, values for $E_{ice}$ are reported to be 9 GPa \citep{Petrovic2003,Petrenko&Whitworth1999}. 
However, there is a disagreement of an order of magnitude
between measurements of $E$ in laboratory (9 GPa) and from field observation
($\approx 1 \rm GPa$) as argued by \citet{Vaughan1995}.

Depending on the applied stress and the time scale of interest, ice has either a linear viscous or a non-linear viscous behavior.  
In glaciers, ice creep is usually described with a non-linear viscous rheology
called Glen's flow law (see \citet{Hutter1983} and references therein). This law relates,
in steady-state conditions, strain rate and stress in the secondary creep regime. 
It is thus not possible with this law to describe the tertiary creep and the
rupture of a bond. 
However, \citet{Hutter1983} states that the so-called Prandtl-Eyring flow model shows a compatible behavior with Glen's flow law at low stresses. 
Non-linear viscous behavior is introduced in our model with an Eyring dashpot in parallel with a linear
spring of stiffness E (which is analogous to the Prandtl-Eyring flow model), following \citet{Nechad&al2005}. Its deformation e is governed by the
Eyring dashpot dynamics and reads
\begin{equation}
\label{eyring}
\frac{de}{dt}=K \sinh(\beta s_{\rm dashpot}),
\end{equation}
where the stress $s_{\rm dashpot}$ in the dashpot is given by 
\begin{equation}
s_{\rm dashpot}={s \over 1-P(e)} -Ee~.
\label{kthnmb}
\end{equation}
Here, $s$ is the total stress applied to the bond and $P(e)$ is the 
damage accumulated within the bond during its history
leading to a cumulative deformation $e$.
$P(e)$ can be equivalently interpreted as the fraction of representative
elements within the bond which have broken, so that the applied stress $s$
is supported by the fraction $1-P(e)$ of unbroken elements.
Following \citet{Nechad&al2005}, we postulate the following dependence
of the damage $P(e)$ on the deformation $e$:
\begin{equation}
P(e)=1- \Bigg( \frac{e_{01}}{e+e_{02 }}\Bigg) ^\xi~,
\label{mgm,tbl;}
\end{equation}
where $e_{01}, e_{02 }$ and $\xi$ are 
constitutive properties of the bond material.

Finally, by combining the previous equations, \citet{Faillettaz&al2010} ended up with a creep model that computes the critical time (i.e. failure of the bond) as a function of the stress experienced by the bond {\em s} given by:
\begin{equation}
\label{tcs}
t_c =\left\{ 
\begin{array}{ll}
{1 \over K} \;\exp(-\gamma \;s) & \textrm{if } s > s^\star\\
\to \infty & \textrm{if }  s \leq s^\star
\end{array} \right.
\end{equation}
where
\begin{equation}
\gamma=\frac{\beta e_{02}^\xi}{e_{01}^\xi},
\label{hy3ghte}
\end{equation}
and
\begin{equation}
s^\star=E\;\;\Big(\frac{e_{01}}{\xi}\Big)^\xi\;\;\Big(\frac{\xi-1}{e_{02}}\Big)^{\xi-1}.
\label{sstar}
\end{equation}
 Creep properties are defined by the parameters
$K$, $\beta$, $e_{01}$, $e_{02}$ and $\xi$, that we need to fix for
our simulations. 

We need to find the most appropriate
parameters to describe creep behavior of ice.
The ice of natural glaciers has a complex polycrystalline structure composed of
crystals of different sizes. 
As Equation \ref{mgm,tbl;} shows, a fraction  $1-(e_{01}/e_{02})^\xi $ of all
present representative
elements undergo abrupt failure immediately after the stress is
applied. Since ice does not show significant damage
immediately after being loaded, a reasonable assumption is
$e_{01}=e_{02}$.
The relative heterogeneity of the material is introduced with the parameter $\xi$.
Despite its complex structure, ice is a
fairly homogeneous material compared to fiber matrix composite. The
more homogenous a material,  the greater $\xi$. 
In the following, we set $\xi~=~10$ (which
means that ice is not very heterogeneous, for example, \citet{Nechad&al2005} used $\xi \approx 2$ for fiber matrix composite).
Moreover, as $e_{01}=e_{02}$, $\xi$ has just an influence on the parameter $s^{\star}$, i.e. the critical stress above which damage starts. 

The other parameters describing the deformation of the Eyring dashpot
under an applied stress are $\beta$ for the non-linear term and $K$ for the
linear one. 
It is difficult to infer such parameters for ice.
\citet{Nechad&al2005} used for fiber matrix composite $\beta = 50 ~10^{-9}\; \rm Pa^{-1}$ and $K=10^5 \; \rm s^{-1}$. As ice is significantly weaker, we arbitrary choose
$\beta\;=\;10^{-7}\; \rm Pa^{-1}$ and $K\;=\;10^{-3}\; \rm s^{-1}$. With a tensile strength of ice equal to 1 MPa, we obtain from Eq. \ref{eyring} $\frac{de}{dt} = 10^{-4} \rm s^{-1}$, which is coherent with the behavior of polycrystalline ice \citep{Schulson&Duval2009}.

\subsection{Geometric parameters}
\label{geompar}
We have to consider the geometric input parameters for modeling 
Allalingletscher.
The glacier is discretized into a system of regular cubic blocks.
As we saw in section \ref{glaciological investigations}, spatial extension of the glacier tongue evolved a lot during the last century. Digital elevation models (DEM) of the tongue for different years are available (see Fig. \ref{timeline}). In this study, we perform simulations for two extreme cases, one where the extension of the glacier tongue was maximum i.e. in 1982 and the other where the glacier broke-off i.e. in 2000.
Since the glacier has completely retreated from the steep part, a detailed bedrock topography of the former glacier tongue could be obtained by a photogrammetric processing of recent aerial photographs (2008).

In order to obtain a realistic description of the damage and fragmentation process
that may develop in the ice mass, we need a sufficiently large number
of blocks. 

As a compromise between reasonable sampling
and numerical speed, we use a model composed of 20 m x 20 m blocks.

\begin{figure}

\noindent\includegraphics[width=20pc]{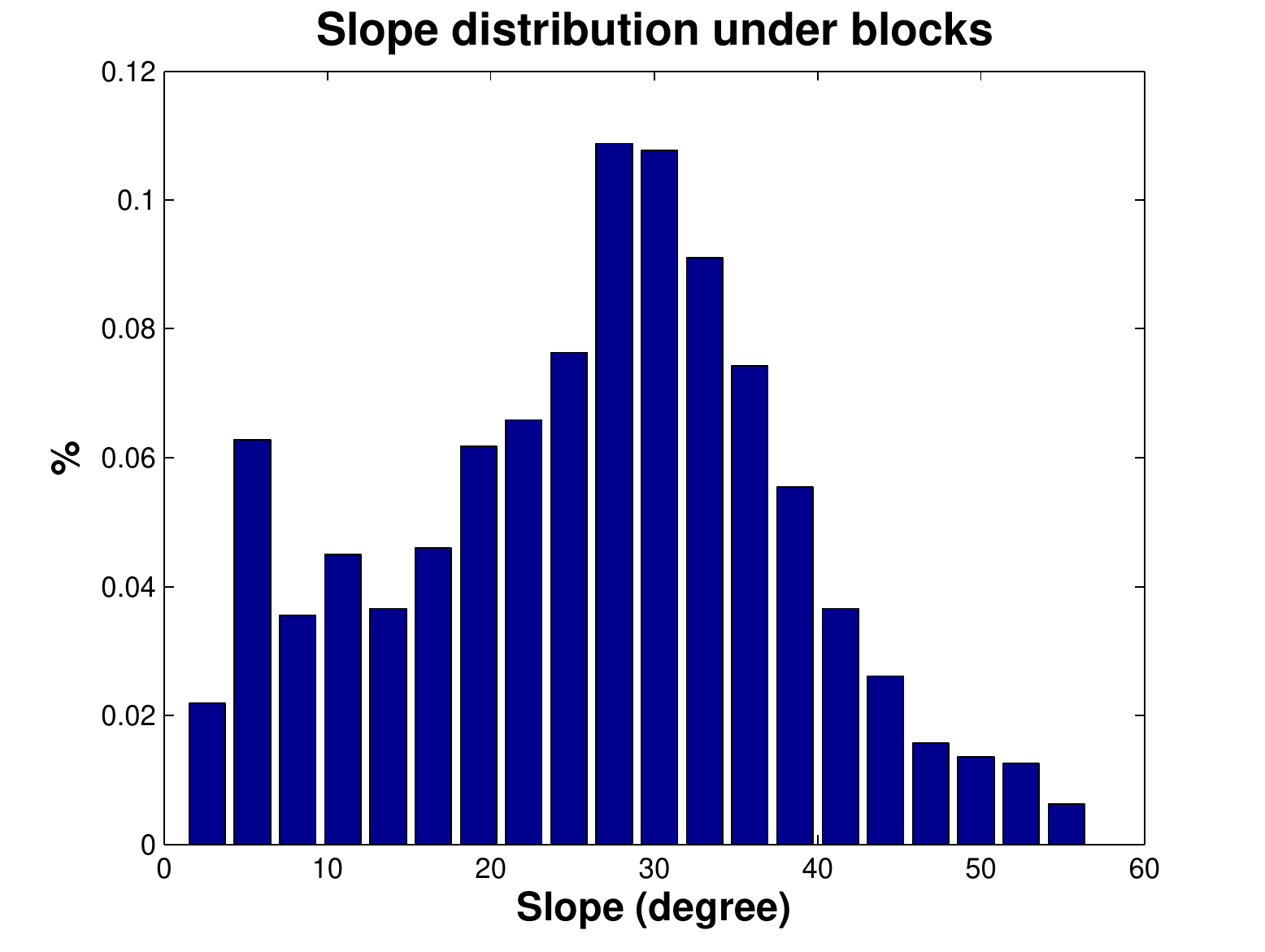}
\caption{\label{slope}Distribution of the slope of the bedrock at the position of the blocks.}
 \end{figure}

The spatial extensions of the glacier as well as the block heights in 1999 and 1982 is given by the respective DEM.
The slope $\phi$ of the bedrock ranges from $0^\circ$ to $45^\circ$
(see Fig \ref{slope}).

\subsection{Extension of \citet{Faillettaz&al2010} model: accounting the effect of subglacial water flow}
\label{improvement}

In our initial model, the effect of subglacial water flow on basal motion was not accounted. Basal processes play a key role in the instability (see \ref{activephase}). 
Melt water flow at the glacier bed influences the glacier dynamics \citep{Bell2008} in two ways: (i) it lubricates the bed and (ii), if water becomes pressurized, the glacier can be partially decoupled from its bed.
The lubrication of the bedrock leading to a progressive decrease of the ice-bedrock frictional resistance is modeled by decreasing the friction coefficient $\mu_0$ (in Eq. \ref{frictionlaw}) \citep{Faillettaz&al2011b}.
The following assumptions are made:
\begin{enumerate}
\item The friction coefficient depends on the subglacial water pressure \citep{Schweizer&Iken1992,Jay&al2011}. 
\item The subglacial water pressure and discharge variations at the daily time scale are similar \citep{Boulton&al2007}.
\item As a consequence of the two aformationned assumptions, the decrease of the friction coefficient can be assumed to be proportional to the discharge of subglacial water flow.
\item The ice thickness does not vary drastically along the glacier tongue. Hence the gradient of the hydraulic potential and therefore the flow path of water is directly given by the topography of the bedrock \citep{Flowers&Clarke1999}.
\end{enumerate}

\begin{figure}

\begin{tabular}{@{}c@{}}

\noindent\includegraphics[width=20pc]{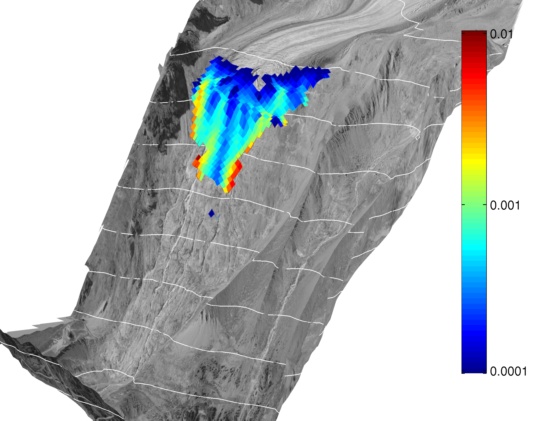}(a)\\
\noindent\includegraphics[width=20pc]{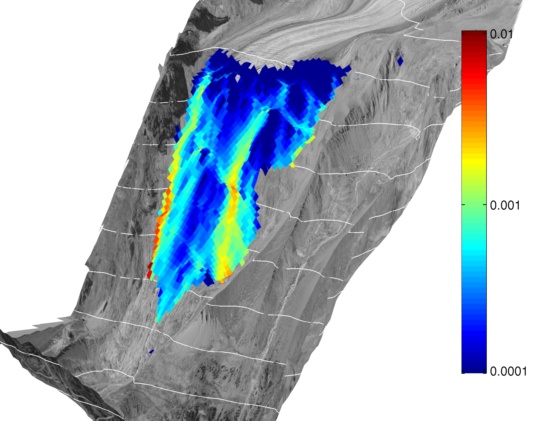}(b)
\end{tabular}
\caption{Relative contribution of each block to the total subglacial water discharge ($\rm Z_{ev}$) for 1999 (a) and 1982 (b). Red zones indicate where discharge of subglacial water is large.}
\label{drainage}
\end{figure}

At a first step, the amount of basal water flowing under each block has to be evaluate. 
We used a toolbox developped by \citet{Schwanghart&Kuhn} to assess the subglacial drainage network with a DEM of the glacier bed. Most commonly DEMs are represented as rectangular grids where an elevation value is
assigned to each cell. Two steps are needed to evaluate the basal water flow under each cell of the DEM. First, the flow direction is evaluated via a transfer matrix M. Elements in M contain the relative
amount of discharge Mij transferred from one cell i to a maximum number of eight downward neighbors with the index j. Thereby the transfer ratios are proportional to the downward slope to the respective neighbor.
Second, by counting the number of cells draining in each grid cell, it calculates the upstream contributing drainage area. Finally, this toolbox gives the relative contribution from each cell of the DEM to the global runoff at the outlet. 
The amount of water flowing through a particular cell of the DEM is then obtained by multiplying the total runoff at the outlet by its relative contribution.
Results are shown in Fig. \ref{drainage}.

In a second step, the friction coefficient is adapted according to the basal water discharge under each block with the following relationship:
\begin{equation}
\label{frictionQdep}
\mu_o(i,j)=\mu_1 - Z_{ev}(i,j) \cdot Q \cdot c_p,
\end{equation}
where $\mu_o$ is the friction coefficient of a given block (i,j), $\mu_1$ a constant friction coefficient, $Z_{ev}(i,j)$ is the relative contribution of the block, 
$Q$ is the daily runoff at the glacier terminus and $c_p$ is a parameter describing the relative influence of the basal water on the friction coefficient.
In this way, the friction coefficient of a block is assumed to decrease when the local subglacial water discharge (and therefore the local subglacial water pressure) is increasing.

\subsubsection{New algorithm accounting for subglacial basal water flow}

\begin{figure}

\noindent{\includegraphics[width=20pc]{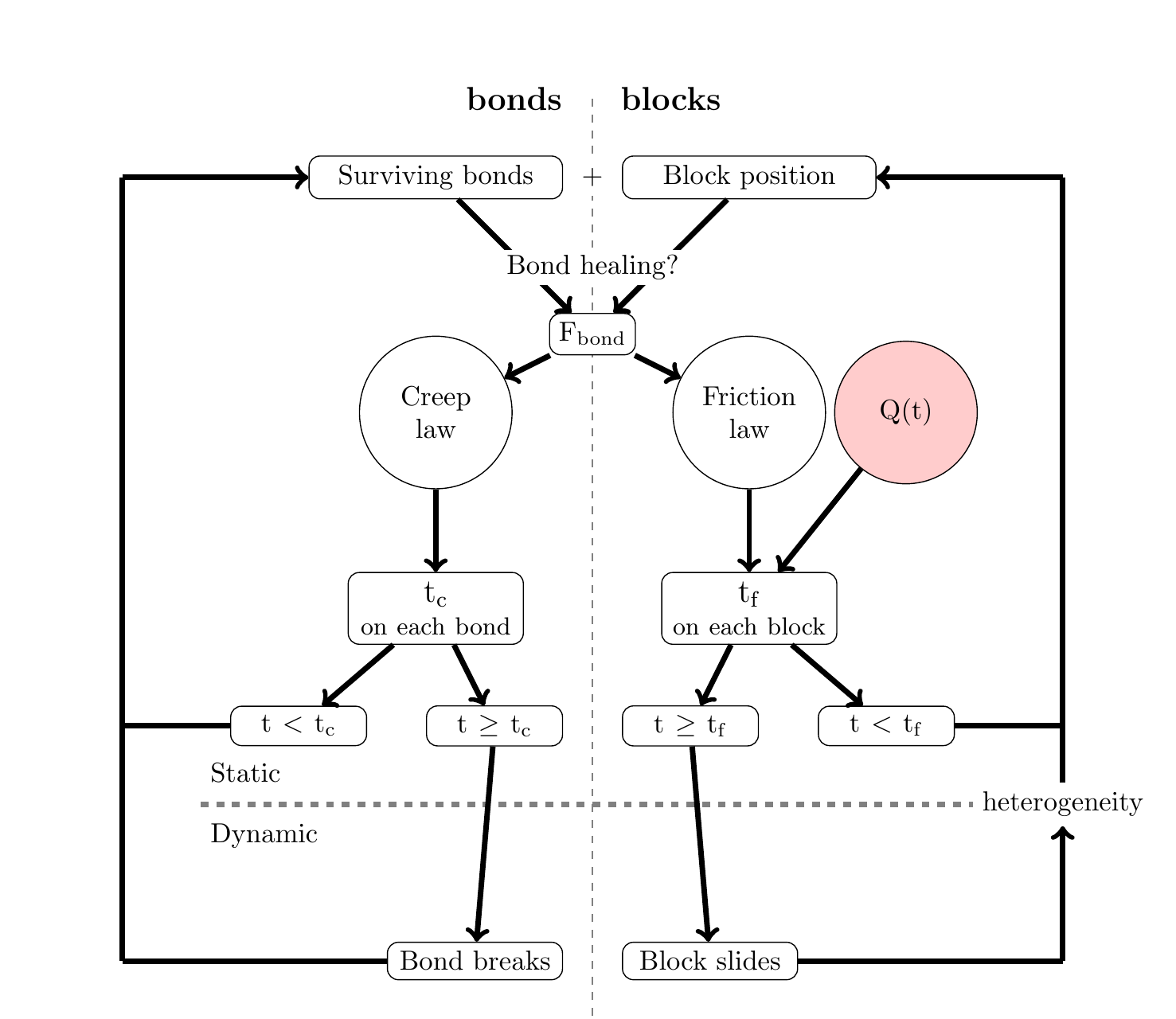}}
\caption{Schematic flowchart of this {\bf modified spring-block} model accounting for subglacial water flow.
}
\label{Newalgor}
\end{figure}

The different steps describing how the instability is modeled are plotted
in Fig. \ref{Newalgor} and developped in Annexe 2.
As explained previously, two phases have to be distinguished:
\begin{itemize} 
\item[(i) ]A quasi-static (quiescent) phase corresponding to the nucleation of block
  sliding and bond rupture.
\item[(ii) ]A dynamical (active) phase corresponding to the sliding phase of the blocks
  and the failure of bonds.
\end{itemize}

The new algorithm takes into account the subglacial water flow. At each time step of the run, the friction coefficient under each block is modified according to the local subglacial discharge. It is evaluated with Eq. \ref{frictionQdep} based on the runoff at the outlet given by the model of \citet{Farinotti&al2011} and the matrix $Z_{ev}$ calculated from the DEM of the bedrock topography.

\section{Numerical results}
\label{numres}

The aim of the numerical simulations is to better understand the development of a glacier sliding instability, considering Allalingletscher as an example.
In particular, we intend to provide answers to the following questions:

\begin{itemize}
\item Why did the glacier only broke off in 1965 and 2000?
\item Does the glacier geometry play a role in the occurrence of the break-off?
\item What is the influence of subglacial water on the glacier dynamic and on the break-off?
\item Does the subglacial drainage network efficiency play a role in the onset of the instability?
\item Is it possible to find precursory signs of the break-off?
\item Would it be possible to predict the break-off?
\end{itemize}

\subsection{Tuning the creep parameters}
\label{tuningcreep}
The first step of the numerical simulation is to find plausible creep parameters able to reproduce the observed glacier behavior.
Creep parameters turn out to be crucial.
As explained in section \ref{creeplaw}, the creep parameters are $\beta$ and $C$.
\citet{Faillettaz&al2010} showed that three different regimes can be expected from the model, namely a fragmentation regime, slab regime or stick and slip regime.
As showed in section \ref{glaciological investigations}, the glacier broke off in 2000 as a slab avalanche.
So we have to find a parameter set able to reproduce the observed behavior.

\begin{figure}

\begin{tabular}{@{}c@{}c@{}}

\noindent\includegraphics[width=9.5pc]{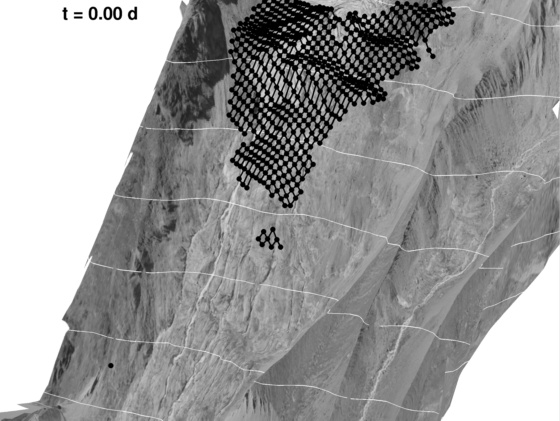}(a)&
\noindent\includegraphics[width=9.5pc]{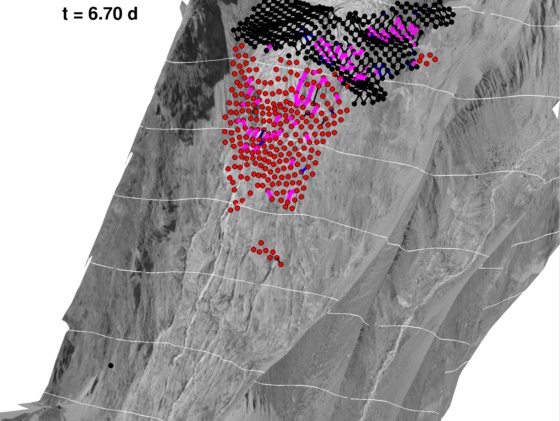}(b)\\
\noindent\includegraphics[width=9.5pc]{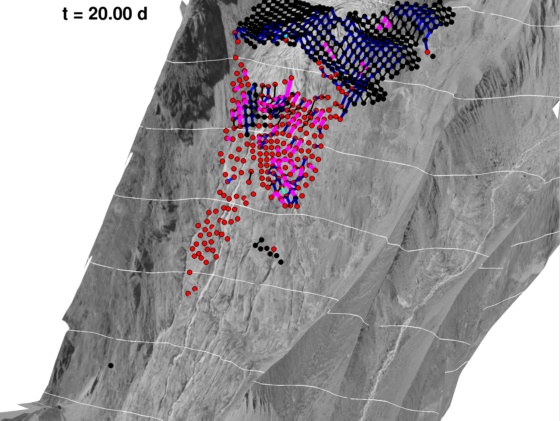}(c)&
\noindent\includegraphics[width=9.5pc]{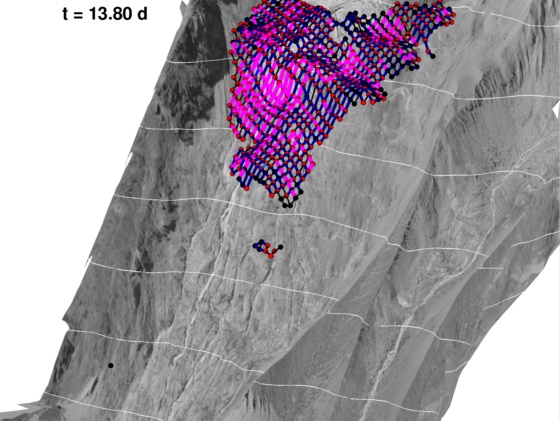}(d)\\
\end{tabular}
\caption{Snapshots illustating the three possible regimes: fragmentation (b), slab (c) and stick-slip(d)}
\label{behavior}
\end{figure}

In this part, we will use the 1999 surface topography for our simulation. 
 At each time step, we evaluate, for each block, the basal slope (see Fig. \ref{geometric_evolution}).
We performed different runs with the 1999 glacier geometry by varying creep parameters.
Results are shown in Fig. \ref{behavior}.

In the fragmentation regime, creep occurs before the nucleation of sliding could develop, so that bonds break and the bottom part undergoes a fragmentation process with the creation of a heterogeneous number of sliding blocks (Fig. \ref{behavior}(b)).
In the stick-slip regime, the time needed for damage initiation is so great that the whole glacier tongue undergoes a series of internal stick slip events, associated with an initial slow average downward motion. After a sufficient time, all blocks slide (Fig. \ref{behavior}(d)).
In the slab regime, in which neither damage nor frictional sliding dominates, the occurrence of a macrocrack propagating roughly along the location of the largest curvature associated with the change of slope from the stable frictional regime in the upper part to the unstable frictional state in the lower part (Fig. \ref{behavior}(c)).

In the following, we used the parameter set that reproduces the last regime.

\subsection{Influence of the glacier geometry on the instability}
\label{geometry}
To investigate the contribution of the glacier geometry on the onset of the instability, we performed numerical simulation on two different glacier geometries in year 2000 (when a break-off occurred) and year 1982 (when active phase could be evidenced without break-off).
We initialized our simulation as described in the previous section.
Table \ref{creeplawtab} summarizes the 
parameters used in our simulations.

 \begin{table}
\caption{Numerical values of parameters used in our simulations. }
\begin{center}
\begin{tabular}{cccccc}
\hline
\multicolumn{6}{c}{{\bf Friction parameters}}\\
A  & $\theta_0$ & $\mu_0$ & $\dot{\delta_0}$\\
 - &    d       & -       &$m.d^{-1}$\\
 0.1 & 100 & 0.8 & $10^{-3}$  \\
\hline
\multicolumn{6}{c}{{\bf Creep parameters}}\\ 
$E$&$\beta$ & $C \sim 1/K$  & $\xi$ &$e_{01}$&$e_{02}$\\
Pa&$\rm Pa^{-1}$ & $s$& -     & -   &   -    \\
 $10^9$   &$10^{-7}$ & $10^3$ & 10 & 0.003&0.003\\
\hline
\multicolumn{6}{c}{{\bf Runoff parameters}}\\ 
$\mu_1$&$c_p$ \\
&\\
 0.8   &6\\
\hline\end{tabular}
\end{center}

\label{creeplawtab}
\end{table}

\begin{figure}

\begin{tabular}{@{}c@{}c@{}}

\noindent\includegraphics[width=9.5pc]{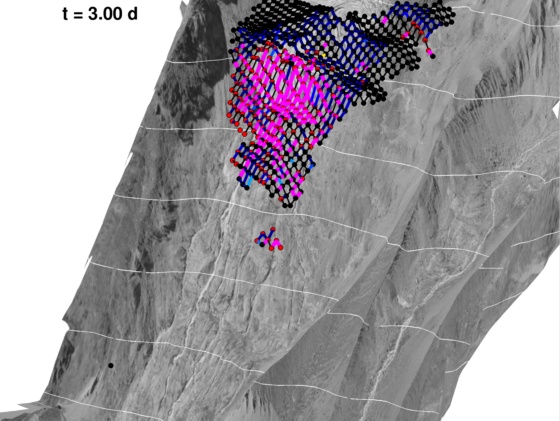}(a1)&
\noindent\includegraphics[width=9.5pc]{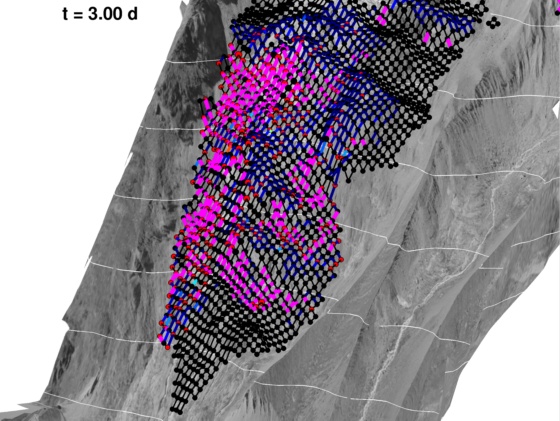}(a2)\\
\noindent\includegraphics[width=9.5pc]{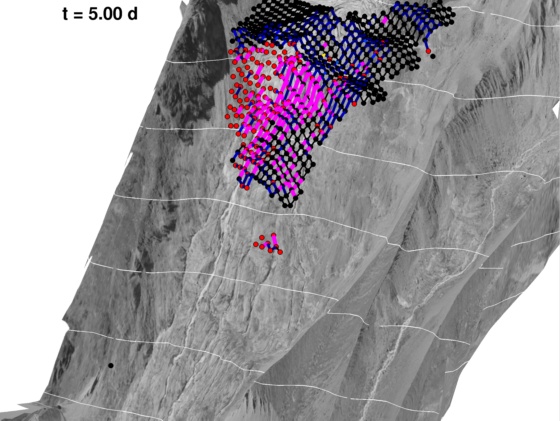}(b1)&
\noindent\includegraphics[width=9.5pc]{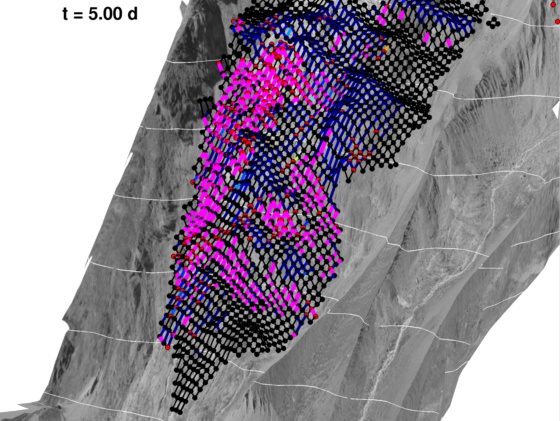}(b2)\\
\noindent\includegraphics[width=9.5pc]{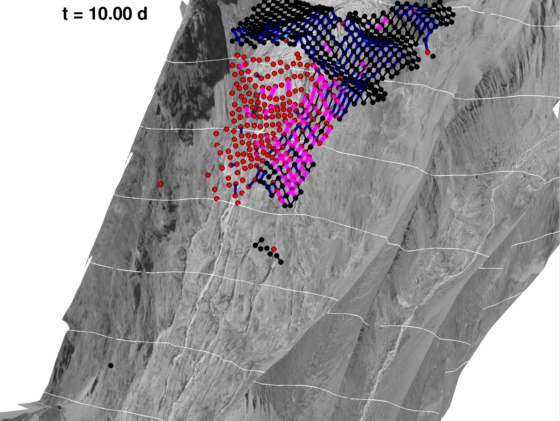}(c1)&
\noindent\includegraphics[width=9.5pc]{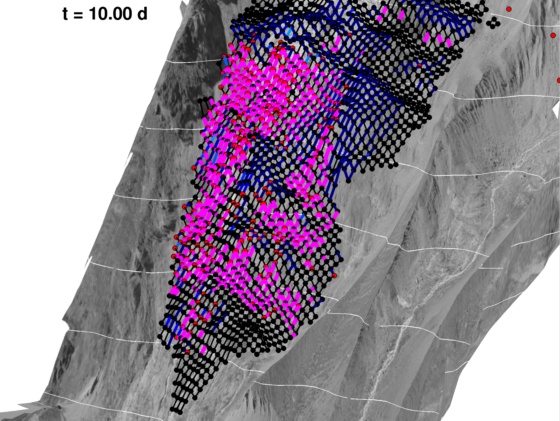}(c2)\\
\noindent\includegraphics[width=9.5pc]{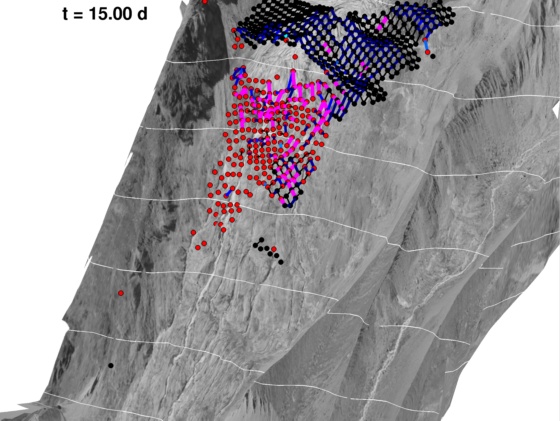}(d1)&
\noindent\includegraphics[width=9.5pc]{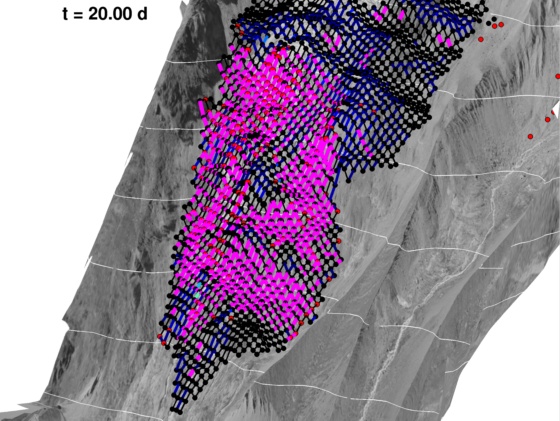}(d2)\\
\noindent\includegraphics[width=9.5pc]{Figure14c}(e1)&
\noindent\includegraphics[width=9.5pc]{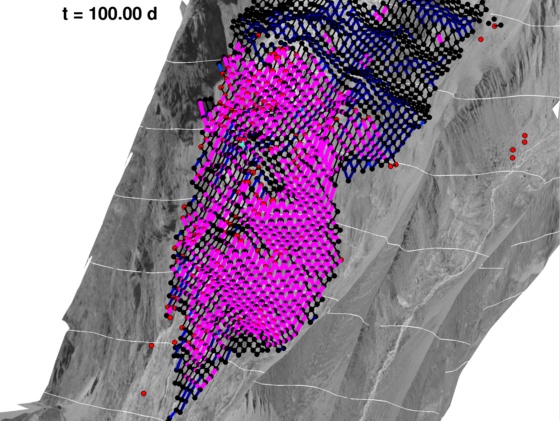}(e2)\\
\end{tabular}
\caption{Snapshot describing the rupture progression for the glacier geometry in the years 2000 and 1982, with a constant friction coefficient.}
\label{diffgeo}
\end{figure}

Fig. \ref{diffgeo} shows the results of simulations for both geometries of 2000 and 1982 with the same creep and friction parameters.
As the terminal part of the glacier lies on a average slope of $30^0$, we imposed the friction coefficient to be a bit lower than its mean slope, i.e. $\mu_0=0.5$.

With this set of parameters, the initiation of the instability could be reproduced.
At the beginning of both simulations, blocks situated in the steepest part start to slide, inducing a zone with high tension at the slope changes.
Then, blocks located downstream start to slide resulting in the apparition of a compression zone directly downstream of this process zone (Fig. \ref{diffgeo}(b)).
This compression zone grows as the instability develops.
In the case of the 1982 geometry, the compression zone continues to propagate downwards and reaches the terminus where the glacier is less steep. The whole glacier is then supported by the lower part of the tongue, which stabilizes the glacier and stops the development of the instability (Fig. \ref{diffgeo}(e2)).
On the contrary, in the case of the 2000 geometry, the compression zone reaches the terminus where the slope is still large. Nothing supports the advance of the glacier, resulting in a global break-off of the tongue (Fig. \ref{diffgeo}(e1)).

The geometrical configuration of the glacier tongue plays therefore a major role in the development of the instability. In the case of Allalingletscher, the glacier tongue where the terminus is located beyond the position of 1965 is stabilized as the slope near the terminus is lower.

\subsection{Role of subglacial water: Channelized vs. distributed subglacial drainage network}
\label{rolewater}

\subsubsection{Channelized subglacial drainage network}
We pointed out that the subglacial water and drainage network play a role in the initiation and the development of the instability.
The presence of subglacial water could be modeled by decreasing the friction coefficient as a function of discharge (see \ref{improvement}). 

To investigate the relative contribution of the drainage network at the onset of the instability, we performed different type of simulations.
We have already shown in section \ref{tuningcreep} that a global decrease of the friction (i.e. modeling a distributed subglacial drainage network) could qualitatively reproduce the break-off event in the year 2000. 

\begin{figure}

\begin{tabular}{@{}c@{}c@{}}
\noindent\includegraphics[width=9.5pc]{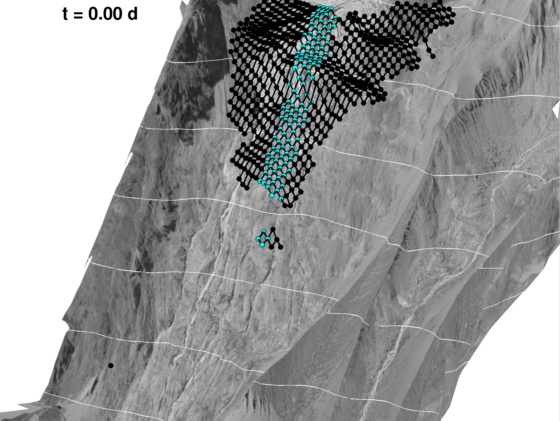}(a)&
\noindent\includegraphics[width=9.5pc]{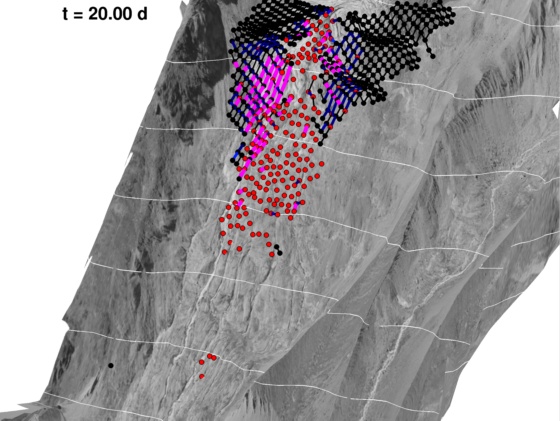}(b)\\
\end{tabular}
\caption{(a) Blue blocks indicate where friction coefficient was changed. (b) Snapshot at the end of the run.}
\label{bande}
\end{figure}

To study the effect of a channelized drainage network, we performed a simulation where the friction coefficient was decreased locally, in a band roughly situated at the center of the glacier (Fig \ref{bande}a). The band width was set to 3 blocks, i.e. 60 m. This allows to reproduce the effect of a flow path of subglacial water in a restricted area.
Results are shown in Fig. \ref{bande}.
After the development of the crown crevasse, the instability develops, leading to a localized break-off with a triangular shape. Moreover, to obtain such a rupture, the friction coefficient had to be set to a very low value ($\mu_0=0.2$), indicating a stronger stability in the case of a channelized network than with a distributed one.
A channelized network could not reproduce, even qualitatively, the break-off event of the year 2000.

\subsubsection{Distributed subglacial drainage network}

\begin{figure}

\begin{tabular}{cc}
\noindent\includegraphics[width=20pc]{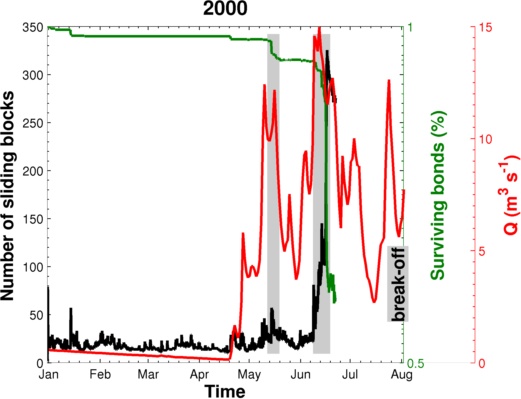}(a)\\
\noindent\includegraphics[width=20pc]{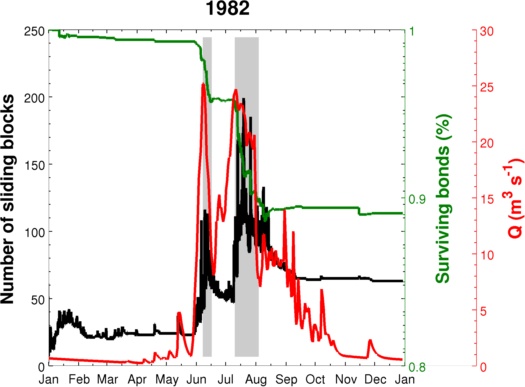}(b)
\end{tabular}
\caption{Runoff, number of sliding blocks and surviving bonds as a function of time for the situation in 2000 (a) and 1982 (b)}
\label{comp}
\end{figure}

We performed two runs with the same creep and friction parameters but for two different years (1982 and 2000).
For these 2 years, modeled daily runoff at the outlet of the glacier are available (red lines in Fig. \ref{comp}). The model allows to redistribute the water under each block according to Eq. \ref{frictionQdep} (Fig. \ref{drainage}).

\begin{figure}
\begin{tabular}{cc}
\noindent\includegraphics[width=9.5pc]{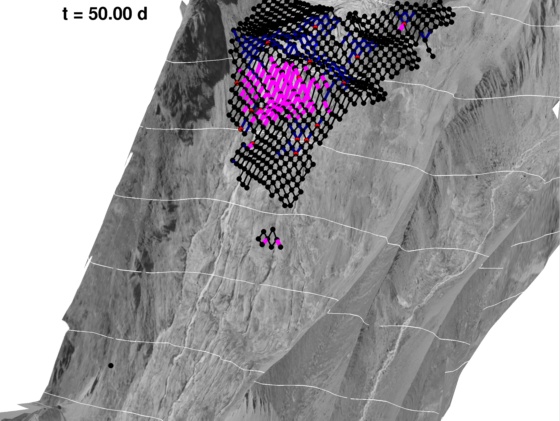}(a1)&
\noindent\includegraphics[width=9.5pc]{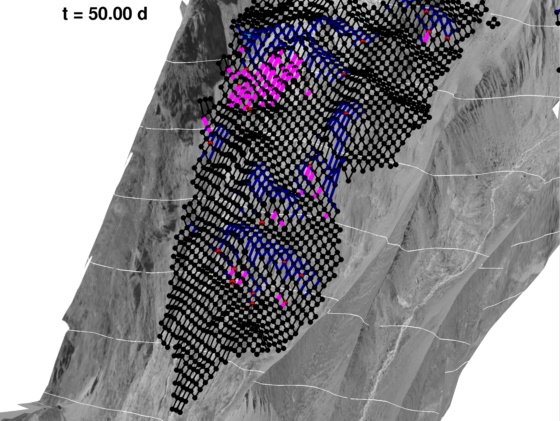}(a2)\\
\noindent\includegraphics[width=9.5pc]{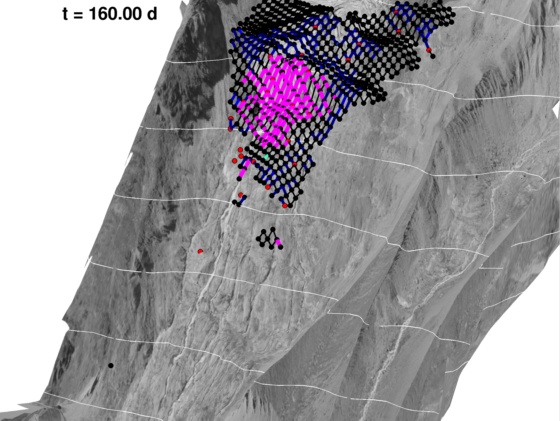}(b1)&
\noindent\includegraphics[width=9.5pc]{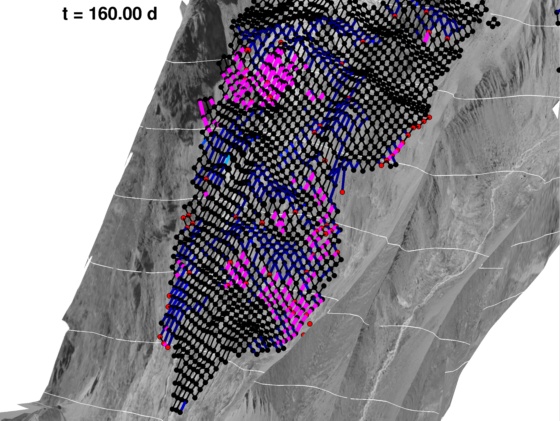}(b2)\\
\noindent\includegraphics[width=9.5pc]{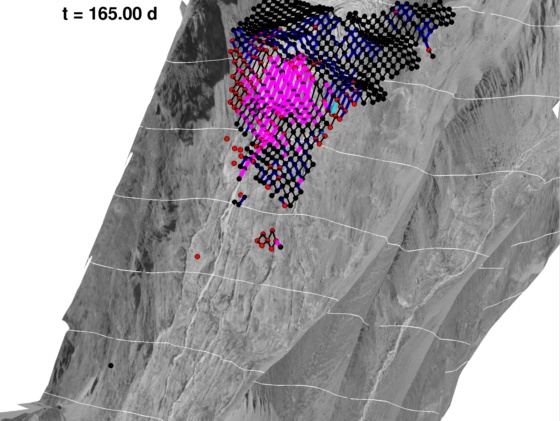}(c1)&
\noindent\includegraphics[width=9.5pc]{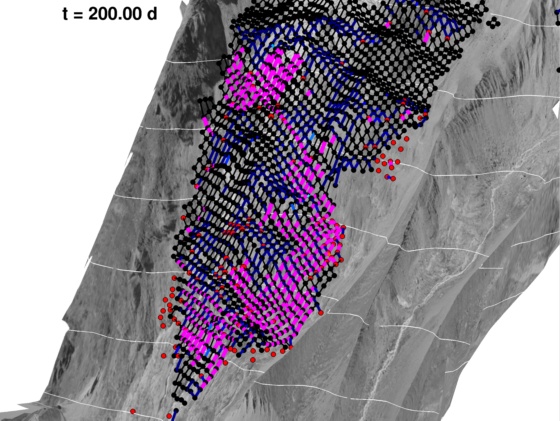}(c2)\\
\noindent\includegraphics[width=9.5pc]{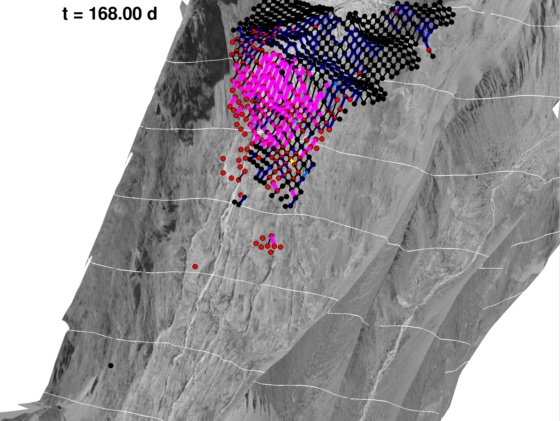}(d1)&
\noindent\includegraphics[width=9.5pc]{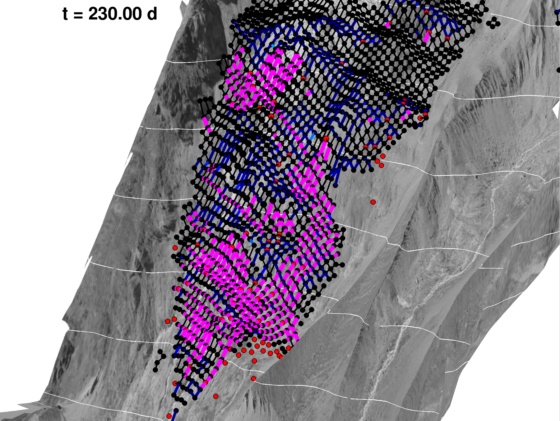}(d2)\\
\noindent\includegraphics[width=9.5pc]{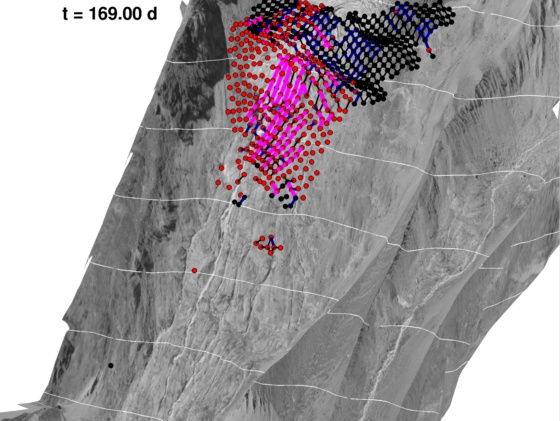}(e1)&
\noindent\includegraphics[width=9.5pc]{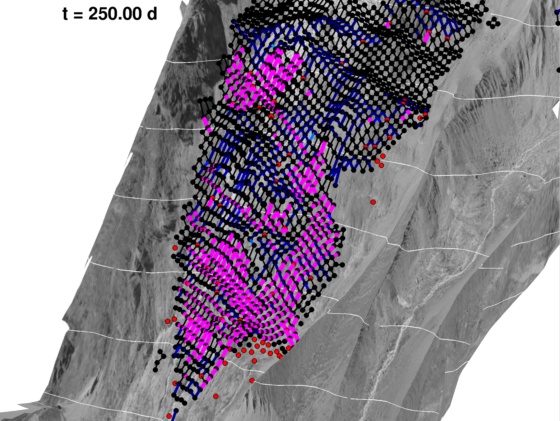}(e2)\\
\noindent\includegraphics[width=9.5pc]{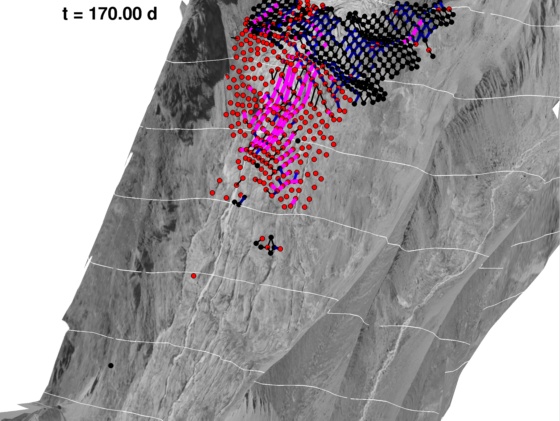}(f1)&
\noindent\includegraphics[width=9.5pc]{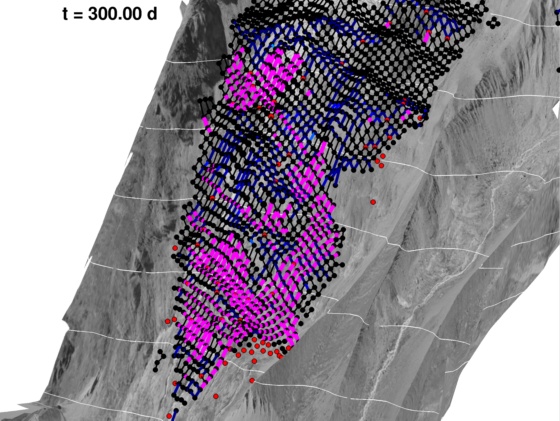}(f2)\\
\end{tabular}
\caption{\label{compwater}Snapshots describing the rupture progression and sliding instability in the block lattice for the years 2000 and 1982.}
\end{figure}

Fig. \ref{compwater} shows the evolution of the instability for the years 2000 and 1982. 
In 2000, after the opening of the crown crevasse and the formation of a compressive zone at the glacier tongue (Fig. \ref{comp}(a)), the glacier enters in a quiescent regime. Even the increase in runoff in May does not affect the dynamic of the glacier tongue (Fig. \ref{comp}(a)). Suddenly, after 165 days of simulation (in mid-June), the tongue starts to be unstable, resulting in a fragmentation process that propagates over the whole tongue in a few days.
We stopped the simulation when the glacier tongue was completely disaggregate. 
In 1982, the situation is different as the tongue extension is larger. The compression zone propagates downstream until the lowest glacier tongue, where the bed slope decreases. An increase in the number of sliding blocks can be noticed between mid July and beginning of September (Fig. \ref{comp}(b)), indicating an ongoing active phase. Interestingly, this active phase is not directly correlated with the runoff magnitude, as it drops in August in the course of the ongoing active phase. It also appears that once triggered, the active phase needs some time to stop.

\section{Discussion}
With the results obtained with 
our model, we could highlight the mechanisms leading to the sliding instability. 
The main result is that the active phase is triggered by an increase of subglacial water flow, its duration depending on the extension of the glacier tongue.

\subsection{Existence of active phases and prediction}

An ongoing active phase does not necessarily lead to a catastrophic break-off. 
Some further conditions are required to trigger the large scale rupture:
\begin{itemize}
\item[(i)] A critical geometric configuration of the glacier tongue,
\item[(ii)] A distributed drainage network.
\end{itemize}
This explains why such instabilities are so rare.

Our model is able to explain and predict the occurrence of the active phases observed on Allalingletscher, but not the final catastrophic break-off. For the situation in year 2000, the model could reproduce with a good timing the active phase appearing before the final rupture.
In the following, as the glacier is completely disaggregated, the simulation could not be pursued and the occurrence of the break-off event could not be predicted.
Moreover, the rapid motion of the disagregated glacier likely affected the efficiency of the subglacial drainage network. This process could lead to a positive feedback: an enhanced basal motion affects the efficiency of subglacial drainage network leading to increase subglacial water pressure and basal motion further.
This positive feedback induced by the onset of the active phase is not modeled.

\subsection{Impact of the runoff evolution on the break-off}

Our results show for both years 2000 and 1982 that fragmentation phases (drop of surviving bonds highlighted by gray bands in Fig. \ref{comp}) coincide with periods where the daily runoff is decreasing. On the contrary, when the runoff increases, the bond breaking process is suddenly stopped.
Moreover, the glacier motion reacts to runoff variations with a delay depending on its size. Fig. \ref{comp} indicates a clear correlation between runoff magnitude and the number of sliding blocks with a larger delay in 1982 than 2000.

\begin{figure}
\noindent\includegraphics[width=20pc]{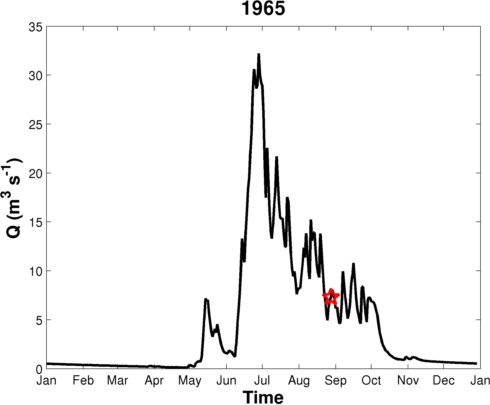}
\caption{\label{1965} Modeled runoff regime in the year 1965 (The red star represents the break-off).}
\end{figure}

These results provide new insights on the maturation process of the instability. 
They suggest the following sequence of ongoing processes:
\begin{enumerate}
\item The onset of the active phases is induced by an increase of subglacial water flow  with a delay depending on its size.
\item The initiation of the fracturation process starts in a period of decreasing runoff, i.e. during a rapid recoupling phase of the glacier onto its bedrock. During this phase, an intensive fracturation of the glacier tongue is initiated.
\item A catastrophic break-off requires the combination of two opposite phenomena: first, the glacier needs to be in an active phase with a strong enhanced basal motion, and second, this active phase must be stopped abruptly with a rapid recoupling of the glacier to its bed.
\end{enumerate}
The likelihood of this process chain can be verified with observations during the years 2000 (Fig. \ref{comp}(a)) and 1965 (Fig. \ref{1965}) when the glacier broke off. It appears that, for both years,
one month before the rupture, the runoff drastically increases, which could led to an active phase.
During the 6 days prior to the break-off, the runoff dropped from 13 to 5~$\rm m^3~s^{-1}$ in 2000 and from 14 to 5~$\rm m^3~s^{-1}$ in 1965.
It appears that the process chain leading to the final break-off in 1965 and 2000 inferred by our modeling results is confirmed by these observations.

With these new insights explaining the sliding instabilities for Allalingletscher, we can analyse other situations.

\subsection{Analysis of observed sliding instabilities of other glaciers}

\subsubsection{Break-off of Le glacier du Tour (France)}
\begin{figure}
\begin{tabular}{cc}
\noindent\includegraphics[width=9.5pc]{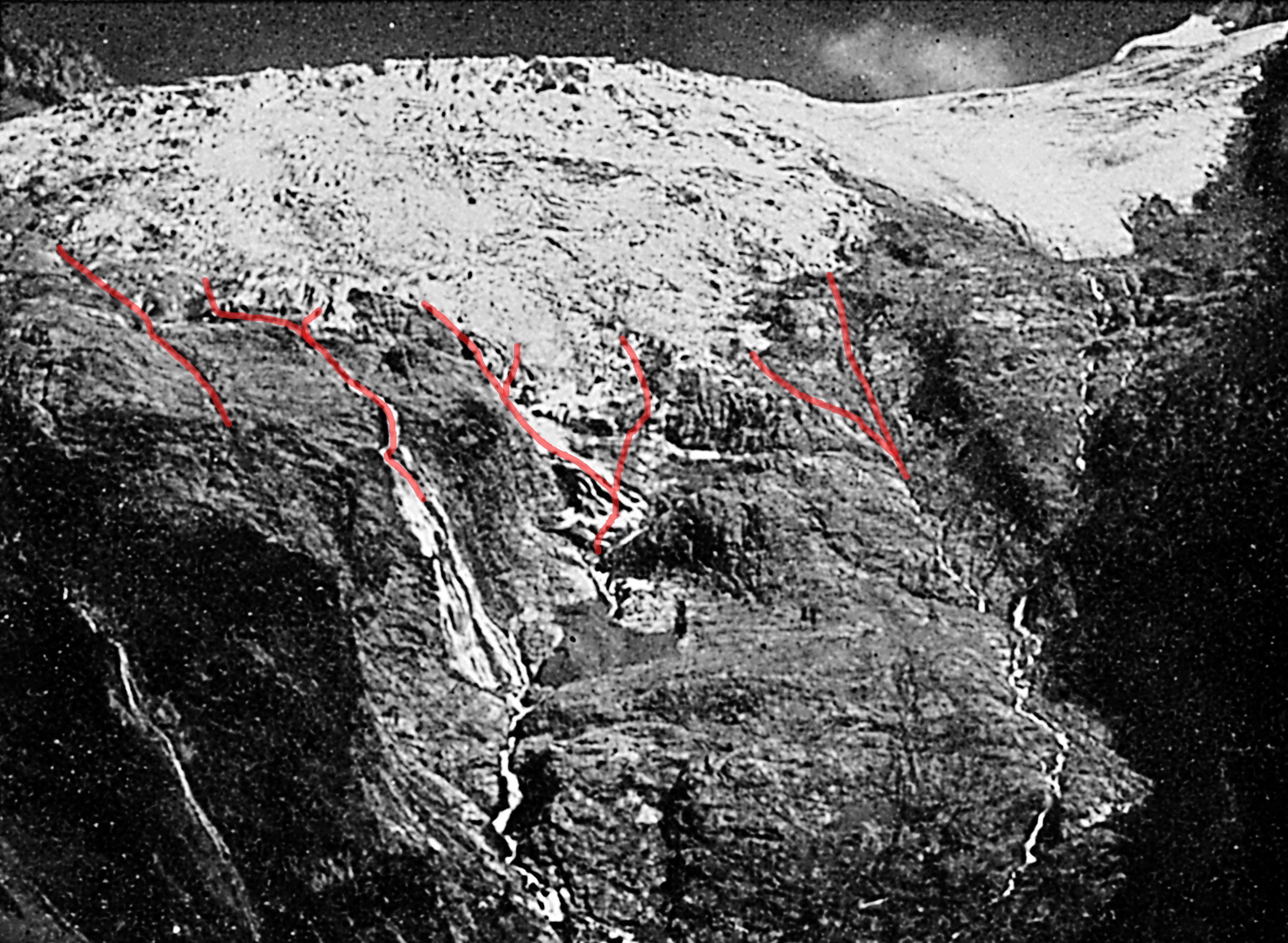}(a)&
\noindent\includegraphics[width=9.5pc]{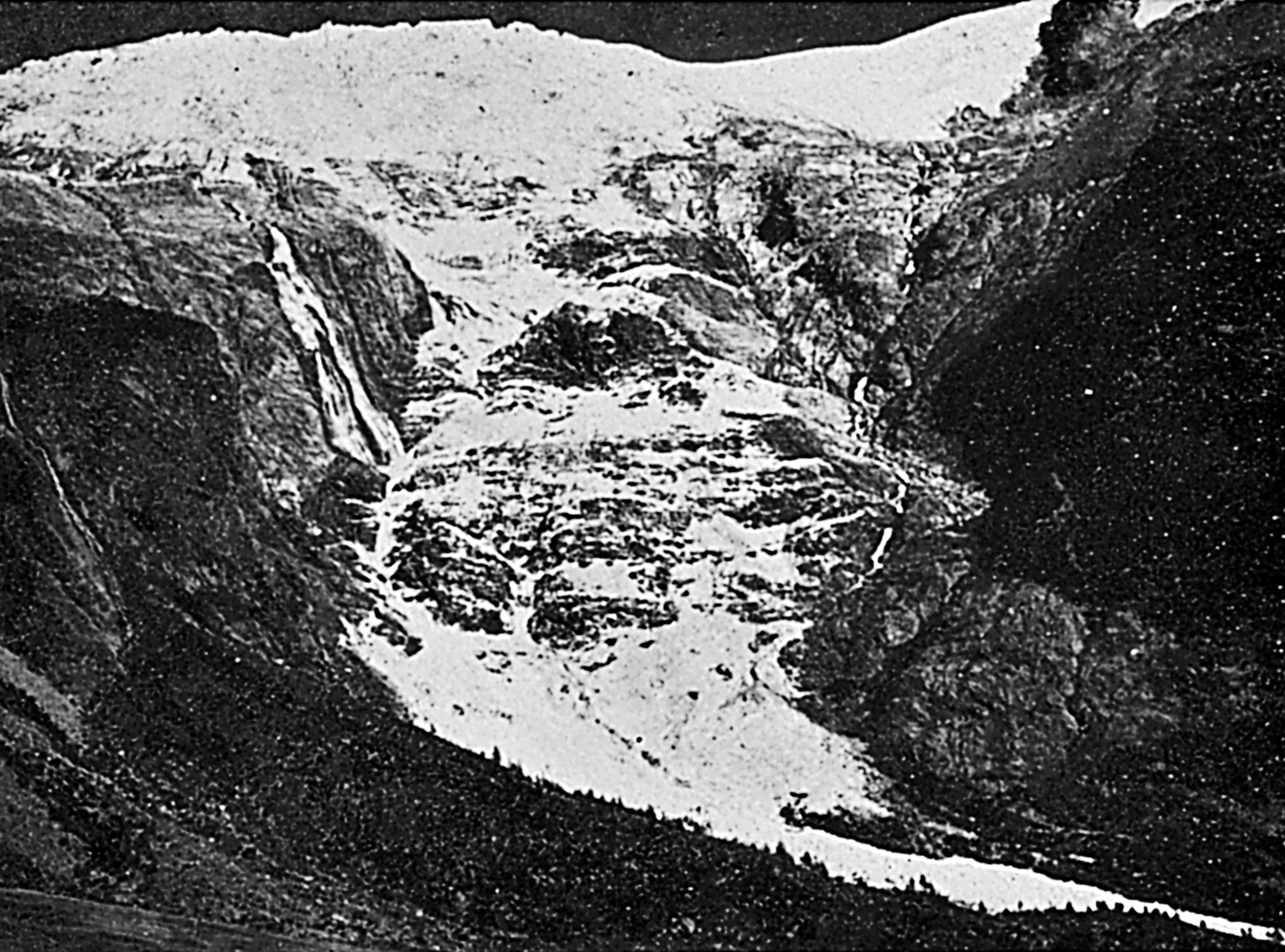}(b)
\end{tabular}
\caption{\label{Tourav}Le glacier du Tour in 1949 (a) before the break-off (the possible drainage network highlighted in red (b) after the break-off}
\end{figure}

\begin{figure}

\noindent\includegraphics[width=20pc]{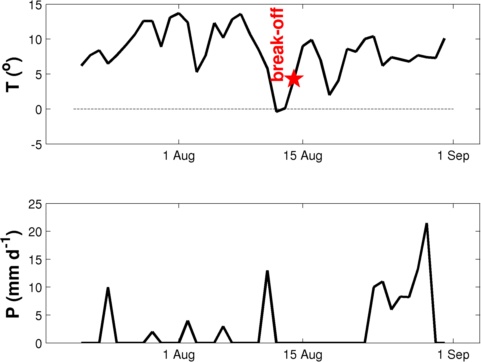}
\caption{\label{Tour} Temperature and precipitation in summer 1949 at the Col du Grand Saint Bernard (2472 m a.s.l.), located about 20 km South West from the Glacier du Tour. The break-off is indicated with a red star.}
\end{figure}

The terminal tongue of the glacier du Tour (Mont Blanc, France) broke off on August 14th 1949. The estimation of the volume of the break-off ranges from 0.5 to 2 millions cubic meters \citep{Glaister1951}. At that time, the glacier geometry was similar to Allalingletscher. The upper part of the glacier was lying on a flat zone, whereas its tongue ended in a very steep terrain (about 40 degrees) that is not able to support the glacier tongue. 
The geometrical configuration of the glacier tongue was also in a critical configuration.
Moreover, different subglacial streams emerging from the glacier could be distinguished, indicating a distributed subglacial drainage network (see Fig. \ref{Tourav}). 
The two main criteria for a break-off in the case of a sliding instability were likely met.
Meteorological conditions prevailing in this area were analyzed (Fig. \ref{Tour}) to infer a runoff history prior to the break-off event.
This temperature time series indicates that in the month preceding the break-off, temperatures were high (about 10 $^\circ$C at 2472 m). Five days before the break-off, the temperatures dropped from 14 to 0 $^\circ$C. 
Subglacial runoff magnitude should have drastically decreased during this period.
This confirms that the recoupling of the glacier on its bed played a decisive role in the break-off event.

\subsubsection{Triftgletscher (Bernese Alps, Switzerland)}
The Triftgletscher is situated in the Bernese Alps in Switzerland. The glacier experienced a rapid retreat during the last 20 years. After the year 2000, the tongue ended in a steep terrain (about 35 degree) and its stability was questionable \citep{Dalban&al2011}. In the following summer surface velocities increase from 1 to 4 $\rm m d^{-1}$, similar to a mini-active phase.
However, up to now, the glacier did not break-off. 

All the basic conditions for the glacier to break-off were fullfilled, except the existence of a distributed subglacial drainage network. 
This might be the reason for the absence of the initiation of active phase here.

\subsubsection{Giesengletcher (Bernese Alps, Switzerland)}

\begin{figure}

\begin{tabular}{@{}c@{}}
\noindent\includegraphics[width=20pc]{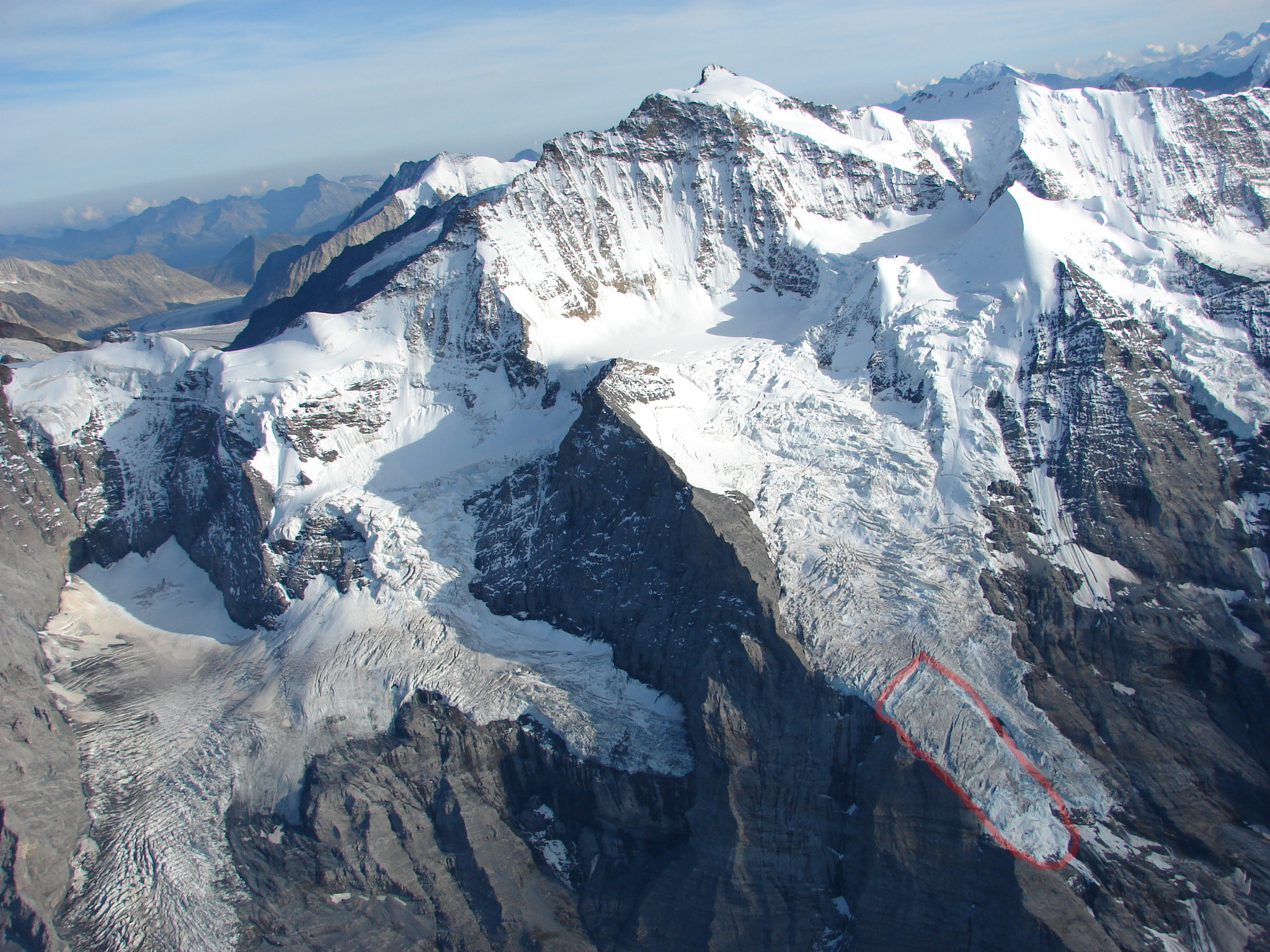}(a)\\
\noindent\includegraphics[width=20pc]{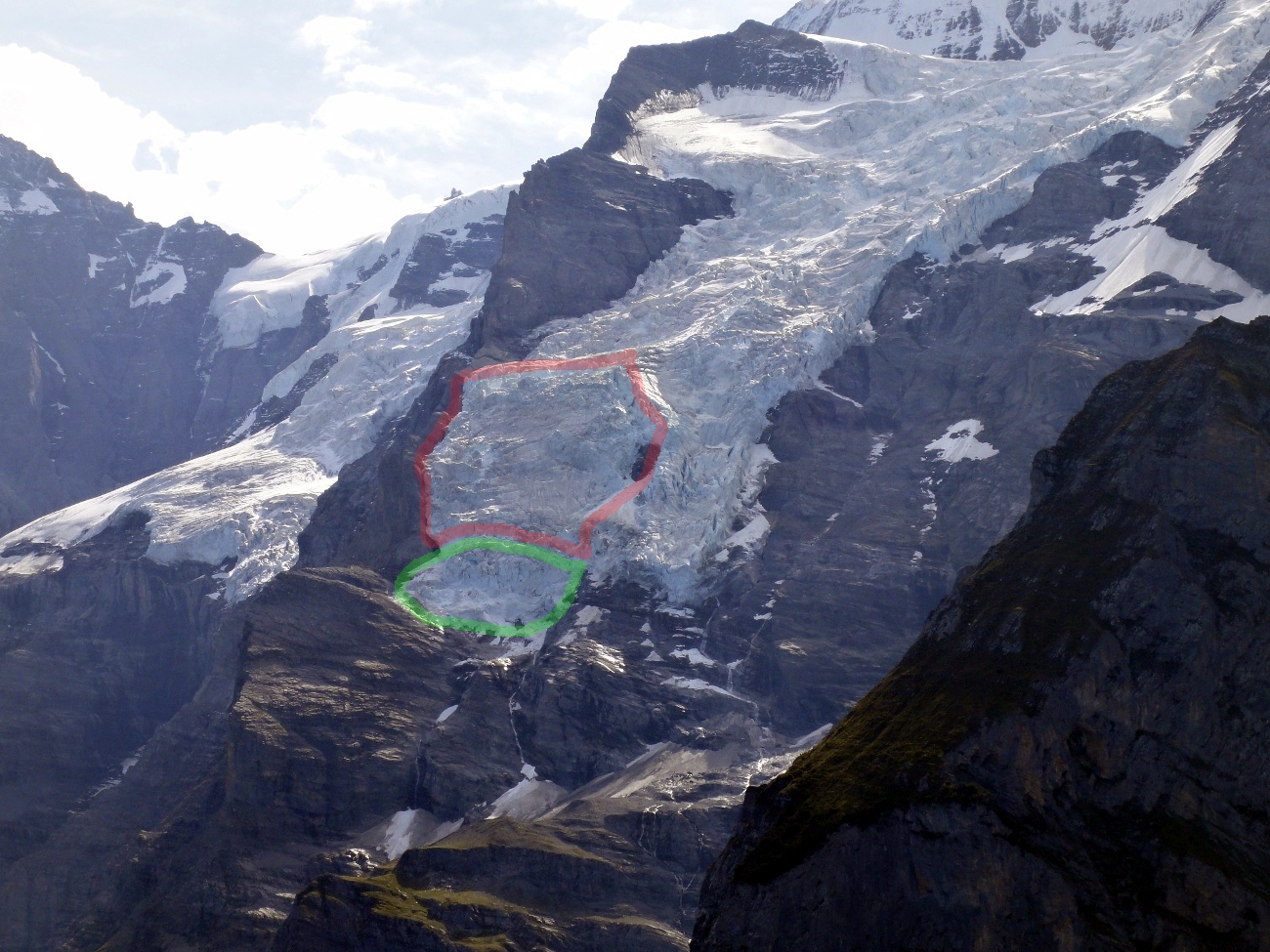}(b)
\end{tabular}
\caption{(a) Situation of Giesengletscher in 2008 (highlighted in red). (b) Situation in 2011. Green zone indicates the stable part of the tongue, in red the possible unstable zone.}
\label{Giesen}
\end{figure}

In a more general context, climate change may affect the stability of sliding glaciers.
As a general glacier retreat in the Alps is observed, the geometry of some glaciers might potentially evolve toward a critical situation.
As an example, Giesengletscher (Bernese Alps, Switzerland) would be a suitable candidate to a future catastrophic break-off. The terminus of Giesengletscher is located at about 2500 m a.s.l. in the Bernese Alps (Fig. \ref{Giesen}). In 2008, a crevasse spanning the whole glacier was observed on the glacier tongue indicating an ongoing active phase on the steepest part (about 35 degree, Fig. \ref{Giesen}(a)). The situation is nevertheless not critical yet, as the glacier terminus is resting on a moderate slope, which stabilizes the glacier tongue (green zone in Fig. \ref{Giesen}(b)).
Moreover, its bedrock is likely smooth, so that the drainage network is expected to be distributed.
Except for the geometrical configuration of the tongue, all conditions for the glacier to break-off are fulfilled, indicating that the disappeance of the supporting glacier terminus could lead to a critical situation, as the tongue geometry would become critical.

\section{Conclusions}

Glacier sliding instabilities may occur on temperate glacier tongues. Such instabilities are strongly affected by the subglacial hydrology: infiltrated melt water may indeed cause (i) a lubrication of the bed and (ii) a decrease of the effective pressure at the glacier bed and consequently a decrease of basal friction.
Available data from Allalingletscher (Valais, Switzerland) indicate that the glacier tongue experienced an active phase during 2-3 weeks in summer in most years between 1965-2000 with strongly enhanced basal motion. This glacier broke-off twice in 1965 and 2000. In order to scrutinize in more detail the processes governing the sliding instabilities, a numerical model developed to investigate gravitational instabilities in heterogeneous media was applied to Allalingletscher. This modified spring-block model enables to account for various geometrical configurations of the glacier and also for interactions between basal sliding and tension cracking. The impact of subglacial water flow on basal motion was included in the model.

Our results indicate that basically three conditions have to be fulfilled for a break-off to occur in the case of a sliding instability:
(i) a critical geometrical configuration of the glacier tongue is needed, where the glacier terminus rests on a steep slope.
(ii) the glacier has to experience an ``active phase" during which basal motion drastically increases,
(iii) the subglacial drainage network has to be distributed.
Moreover, our modeling results point out that the break-off always occurs after a sudden phase of recoupling of the glacier on its bedrock, characterized by a drop of subglacial water flow.
This new precursory sign was observed for both 1965 and 2000 break-off event.
We also showed that all these conditions were fulfilled in 1949 before the break-off of the glacier du Tour, and moreover, we demonstrated that the same drop of subglacial water runoff occurred five days before the catastrophic break-off, indicating that this new precursory sign might be decisive.
In a more general context, climate change may affect the stability of steep glacier tongues, as they retreat and may evolve toward a critical geometry.
This model casts a gleam of hope for a better understanding of the ultimate rupture of such glacier sliding instabilities.

\appendix

\section*{Initiation of sliding for a single block}

This appendix complements 
Section \ref{frictionlawchap} by providing details of the calculation
of the critical time at which unstable sliding of a given block occurs.
Section \ref{frictionlawchap} describes the 
sub-critical sliding process of a given block interacting via state-and-velocity
solid friction with its inclined basal surface. When the sub-critical
sliding velocity $d\delta/dt$ diverges (we refer to the time
when this occurs as the ``critical time'' $t_f$ for the frictional sliding instability), 
this indicates a change towards a dynamical sliding regime where inertia (the block mass and its acceleration
in the Newton's law) has to be taken into account.

Let us calculate explicitly how the critical time $t_f$ is obtained and define its dependence on the parameters and boundary conditions.
Let us call $T \equiv \Arrowvert \sum \vec F_{\rm bond}\;-\;T_{\rm weight} \vec x \Arrowvert$ 
(or $N \equiv N_{\rm weight}$) the total shear (or normal) force exerted
on a given block, where $\vec F_{\rm bond}$ is the force exerted by a neighboring 
spring bond, and $N_{\rm weight}$ and $T_{\rm weight}$ are the 
normal and tangential forces due to the weight of the block. We then have
\begin{equation}
\label{hjgjhnt'en}
\mu = \frac{T}{N}=\tan \phi ~,
\end{equation}
where $\phi$ is the angle of the basal surface supporting the block.
Therefore, 
\begin{equation}
\mu_0 + A\;
\ln \frac{\dot{\delta}}{\dot{\delta}_0}+B\; \ln \frac{\theta}{\theta_0} =  \tan \phi ~.
\end{equation}
As explained in Section \ref{frictionlawchap}, $A -B$ is usually very small for natural material, of the order of $A-B
\approx \pm 0.02$.
For the sake of simplicity, we assume $A=B$. As recalled
in Section \ref{frictionlawchap}, this choice is not restrictive as it recovers
the two important regimes \citep{Helmstetter&al2004}. This leads to
\begin{equation}
\ln \; \frac{\dot{\delta}}{\dot{\delta_0}}\frac{\theta}{\theta_0}\;=\; \frac{\tan\phi-\mu_0}{A}~,
\end{equation}
whose solution is
\begin{equation}
\dot{\delta}.\theta\;=\;(\dot{\delta}_0\;\theta_0)\;\exp \left(\frac{\tan\phi-\mu_0}{A}\right)~.
\label{kgkgtw}
\end{equation}

Combining Eqs. \ref{qwertyu} and \ref{kgkgtw}, we obtain
\begin{equation}
\dot{\theta}=1-\frac{\theta \;\dot{\delta}}{D_c}=1- {\;\dot{\delta}_0\;\theta_0 \over D_c}~
\exp \left(\frac{\tan\phi-\mu_0}{A}\right)~.
\end{equation}
After integration it reads
\begin{equation}
\label{thetaev}
\theta\;=\; \theta_0 + \Big( 1-\frac{\;(\dot{\delta}_0\;\theta_0)\;\exp \left(\frac{\tan\phi-\mu_0}{A}\right)}{D_c}\Big) t~,
\end{equation}
and using Equation (\ref{kgkgtw}), we obtain
\begin{equation}
\dot{\delta} ~=~ { \;(\dot{\delta}_0\;\theta_0)\;\exp \left(\frac{\tan\phi-\mu_0}{A}\right)
\over
\theta_0 + \Big( 1-{\;\dot{\delta}_0\;\theta_0 \over D_c}\;\exp \left(\frac{\tan\phi-\mu_0}{A}\right)\Big) t}~.
\label{ghnt}
\end{equation}
This expression exhibits the usual regimes: a
finite time singularity is obtained for $\;(\dot{\delta}_0\;\theta_0)\;\exp \left(\frac{\tan\phi-\mu_0}{A}\right)>D_c$. 
In this case, Expression (\ref{ghnt}) can be re-written as
\begin{equation}
\label{tfev}
\dot{\delta}\;= {D_c\; \dot{\delta}_0\;\theta_0\;\exp \left(\frac{\tan\phi-\mu_0}{A}\right)
\over D_c - \dot{\delta}_0\;\theta_0\;\exp \left(\frac{\tan\phi-\mu_0}{A}\right)} \cdot {1 \over t_f-t} 
\end{equation}
with
\begin{equation}
t_f\;=\; {D_c \theta_0
\over D_c - \dot{\delta}_0\;\theta_0\;\exp \left(\frac{\tan\phi-\mu_0}{A}\right)}~.
\end{equation}

We can simplify Expression (\ref{tfev}) by using the condition that, for $\mu=\tan \phi=\mu_0$, we should have $t_f \rightarrow \infty$.
But, for $\mu=\mu_0$, $\exp\left(\frac{\mu_s-\mu_0}{A}\right)=1$ and thus, 
for the condition $t_f \to \infty$ to hold, we need 
\begin{equation}
{\dot{\delta}_0\;\theta_0 \over  D_c} = 1~.
\label{hntbkwg}
\end{equation} 
The final expression for the critical time $t_f$ signaling the transition from 
a subcritical sliding to the dynamical inertial sliding is, for $\mu>\mu_0$,
\begin{equation}
t_f\;=\;\frac{\theta_0}{\exp(\frac{\mu_s-\mu_0}{A})-1}~,
\label{tkhntrk}
\end{equation}
while $t_f \to \infty$ for $\mu \leq \mu_0$. Note that
the dependence on $\dot{\delta_0}$ has disappeared due to the 
relation (\ref{hntbkwg}).

To summarize, a given configuration of blocks and spring tensions determines
the value of  $T \equiv \Arrowvert \sum \vec F_{\rm bond}\;-\;T_{\rm weight} \vec x \Arrowvert$ 
and $N \equiv N_{\rm weight}$ and therefore of $\mu$ via 
(\ref{hjgjhnt'en}). Knowing $\mu$ and given the other material parameters
$\theta_0, \mu_0$ and $A$, we determine the time $t_f$ for the transition to the dynamical
regime for that block via Equation (\ref{tkhntrk}). 

\section*{General Algorithm}

The simulation of the frictional process for each given block proceeds as follows:
\begin{enumerate}
\item A given configuration of blocks and spring tensions determines
the value of  $T \equiv \Arrowvert \sum \vec F_{\rm bond}\;-\;T_{\rm weight} \vec x \Arrowvert$ 
and $N \equiv N_{\rm weight}$ for each block,
and therefore their solid friction coefficient $\mu$ corresponds to the ratio $T/N$.

\item  At a given time, knowing the runoff at the outlet, $\mu_0$ is evaluated for a given block with Eq. \ref{frictionQdep}. 

\item Knowing $\mu$ for a given block and with the other material parameters
$\theta_0, \mu_0$ and $A$ for that block, the time $t_f$  for the transition
to the dynamical sliding regime is calculated with
expression~(\ref{frictionlaw}). Eq.~\ref{frictionlaw} for 
$t_f$ gives the waiting time until the next block starts to slide. 

\item When the block undergoes a transition into the dynamical sliding regime at time $t_f$, 
its subsequent dynamics should obey Newton's law.

\item The dynamical slide of the block goes on as long as the velocity
of the block remains positive. When its velocity reaches zero, we assume
that the block is no more sliding. 
To account for the heterogeneity and roughness of the sliding surface,
we assume that the state variable $\theta_0$ is reset to a new random value after
the dynamical sliding stops. This random value
is taken to reflect the characteristics of the new asperities 
constituting the fresh surface of contact.

\item After a dynamical slide, the forces exerted by the springs that connect
the block to its neighbors are updated, as is the new gravitational
force (if the basal surface has a curvature), the new value of $\mu$ is 
obtained, the time counter for frictional creep is reset to zero, and a new process of slow frictional creep
develops over the new waiting time $t_f$, that is, in general, different from the 
previous one.
\end{enumerate}

In summary, simulation of the damage process leading to bond rupture between
blocks proceeds as follows.
\begin{enumerate}
\item Given an initial configuration of all the blocks within the network, the
elastic forces exerted by all bonds can be calculated from their extension/compression.

\item For each bond $i$ subjected to an initial stress $s_0(i)$, we calculate the corresponding 
critical time $t_{c,0}(i)$ at which it would rupture if neither of the two blocks
connected to it moved in the meantime.
For those bonds where $s_0(i)<s^*$ defined in Equation (\ref{tcs}), $t_{c,0}(i)$ is infinite.

\item Some bonds will eventually fail, modifying the force balance on their
blocks and accelerating the transition to the sliding regime, after which 
the stresses in the bonds connected to the same blocks are modified.

\end{enumerate}


\end{document}